\documentclass[12pt]{article}  	% use "amsart" instead of "article" for AMSLaTeX format
\usepackage{geometry}                		% See geometry.pdf to learn the layout options. There are lots.
\usepackage{graphicx}				% Use pdf, png, jpg, or eps§ with pdflatex; use eps in DVI mode
\usepackage{amssymb,amsthm,xcolor,caption,subcaption}					
\usepackage[square,sort,comma,numbers]{natbib}
\usepackage{amsmath}
\usepackage{epstopdf}
\usepackage{bbm}
\usepackage{mathtools}
\usepackage{algorithm}
\usepackage{times}
\usepackage{algpseudocode}
\usepackage{chngcntr}
\usepackage{url}
\usepackage{JASA_manu}

\newcommand{\mA}{\mathcal{A}}

\newcommand{\abs}[1]{\left\vert#1\right\vert}
\newcommand{\aabs}[1]{\Big\vert#1\Big\vert}

\newcommand{\norm}[1]{\big\Vert#1\big\Vert}

\newcommand\wt[1]{{ \widetilde{#1} }}

\newcommand \bbP{\mathbb{P}}
\newcommand \bbE{\mathbb{E}}

\def\T{{ \mathrm{\scriptscriptstyle T} }}

\def\m{\mathcal}
\def\mb{\mathbb}

\def\ind{\mathbbm{1}}
\def\up{{ \mathrm{\scriptscriptstyle \uparrow} }}
\def\down{{ \mathrm{\scriptscriptstyle \downarrow} }}

\newtheorem{theorem}{Theorem}[section]
\newtheorem{lemma}{Lemma}[section]
\newtheorem{proposition}[theorem]{Proposition}
\newtheorem{corollary}[theorem]{Corollary}
\newtheorem{rem}{Remark}[section]

\usepackage{xr}
\begin{document}
%\begin{center}
%{\Large {\bf Simplex factor models for multivariate unordered categorical data}}
%\end{center}

\title{Probabilistic community detection with unknown number of communities\footnote{To appear in JASA 2018}}

\author{ {\bf Junxian Geng} \\ Department of Statistics, Florida State University, Tallahassee, FL, \\email: jgeng@stat.fsu.edu \\
{\bf Anirban Bhattacharya}\\
Department of Statistics, Texas A\&M University, College Station, TX, \\email: anirbanb@stat.tamu.edu\\
{\bf Debdeep Pati} \\Department of Statistics, Texas A\&M University, College Station, TX, \\email: debdeep@stat.tamu.edu\\
}

\maketitle

\begin{abstract}
 A fundamental problem in network analysis is clustering the nodes into groups which share a similar connectivity pattern.  Existing algorithms for community detection assume the knowledge of the number of clusters or estimate it a priori using various selection 
criteria and subsequently estimate the community structure.  Ignoring the uncertainty in the first stage may lead to erroneous clustering, particularly when the community structure is vague. We instead propose a coherent probabilistic framework for simultaneous estimation of the number of communities and the community structure, adapting 
recently developed Bayesian nonparametric techniques to network models. An efficient Markov chain Monte Carlo (MCMC) algorithm is proposed which obviates the need to perform reversible jump MCMC on the number of clusters. The methodology is shown to outperform recently developed community detection algorithms in a variety of synthetic data examples and in benchmark real-datasets.  Using an appropriate metric on the space of all configurations, we develop non-asymptotic Bayes risk bounds even when the number of clusters is unknown.  Enroute, we develop concentration properties of non-linear functions of Bernoulli random variables, which may be of independent interest in analysis of related models. 
  
\end{abstract}
%\vspace{-0.05in}

\section{Introduction}
Data available in the form of networks are increasingly becoming common in modern applications ranging from brain remote activity, protein interactions, web applications, social networks to name a few.  Accordingly, there has been an explosion of activities in the statistical analysis of networks in recent years; see \cite{goldenberg2010survey} for a review of various application areas and statistical models. Among various methodological \& theoretical developments, the problem of community detection has received widespread attention. Broadly speaking, the aim there is to cluster the network nodes into groups which share a similar connectivity pattern, with sparser inter-group connections compared to more dense within-group connectivities; a pattern which is observed empirically in a variety of networks \cite{goldenberg2001talk}. Various statistical approaches has been proposed for community detection and extraction. These include hierarchical clustering (see \cite{newman2004detecting} for a review), spectral clustering \cite{white2005spectral,zhang2007identification,rohe2011spectral}, and algorithms based on optimizing a global criterion over all possible partitions, such as normalized cuts \cite{shi2000normalized} and network modularity \cite{newman2004finding}. 

\begin{figure}[h]
	\centering
	\includegraphics[width=.4\textwidth]{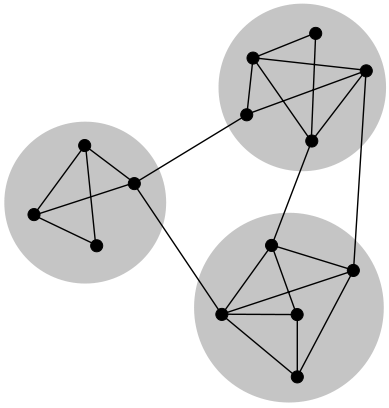}
	\caption{\label{fig:Community Structure}{\em A sketch of a network displaying community structure, with three groups of nodes with dense internal edges and sparser edges among groups.}	}
	\label{fig:sbm}
\end{figure}

From a model-based perspective, the stochastic block model (SBM; \cite{holland1983stochastic}) and its various extensions \cite{karrer2011stochastic,airoldi2009mixed} enable formation of communities in networks. A generic formulation of an SBM starts with clustering the nodes into groups, with the edge probabilities $\bbE A_{ij} = \theta_{ij}$ solely dependent on the cluster memberships of the connecting nodes.  A realization of a network from an SBM is shown in Figure \ref{fig:sbm}; formation of a community structure is clearly evident.  
This clustering property of SBMs has inspired a large literature on community detection \citep{bickel2009nonparametric,newman2012communities,zhao2012consistency,karrer2011stochastic,zhao2011community,amini2013pseudo}.  

A primary challenge in community detection  is the estimation of both  the number of communities and the clustering configurations. 
Essentially all existing community detection 
algorithms assume the knowledge of the number of communities \citep{airoldi2009mixed,bickel2009nonparametric,amini2013pseudo} or estimate it {\em a priori} using either of cross-validation, hypothesis testing, BIC or spectral methods \citep{daudin2008mixture,latouche2012variational,wang2015likelihood,le2015estimating}. 
Such two stage procedures ignore uncertainty in the first stage and are prone to increased erroneous cluster assignments when there is inherent variability in the number of communities. Although  model based methods  are attractive for inference and quantifying uncertainty,  fitting block models from a frequentist point of view, even with the number of communities known, is a non-trivial task especially for large networks, since in principle the problem of optimizing over all possible label assignments is NP-hard. 

Bayesian inference offers a natural solution to this problem by providing a probabilistic framework for simultaneous inference of the number of clusters and the clustering configurations.  However,  the case of unknown number of communities  poses a stiff computational challenge even in a fully Bayes framework.   \cite{snijders1997estimation,nowicki2001estimation} developed  a MCMC algorithm to estimate the parameters in a SBM for a given number of communities.  Often, a frequentist estimate of $k$ is first determined through a suitable criterion; e.g., integrated likelihood \cite{daudin2008mixture,zanghi2008fast,latouche2012variational}, composite likelihood BIC \cite{saldana2015many} etc., with a subsequent Bayesian model fitted with the estimated number of components.  In a fully Bayesian framework,  a prior distribution is assigned on the number of communities which is required to be updated at each iteration of an MCMC algorithm.  This calls for complicated search algorithms in variable dimensional parameter space such as the reversible jump MCMC algorithm \cite{green1995reversible}, which are difficult to implement and automate, and are known to suffer from lack of scalability and mixing issues. \cite{mcdaid2013improved} proposed an algorithm by `collapsing' some of the  nuisance parameters which allows them to implement
an efficient algorithm based on the allocation sampler of \cite{nobile2007bayesian}.  However, 
the parameter ($k$) indicating the number of components still cannot be marginalized out within the Gibbs sampler requiring complicated Metropolis moves to simultaneously update the clustering configurations and $k$.

In this article, we consider a Bayesian formulation of an SBM \cite{snijders1997estimation,nowicki2001estimation,mcdaid2013improved} with standard conjugate Dirichlet-Multinomial prior on the community assignments and Beta priors on the edge probabilities.
%and a prior on $k$ supported on integers.     
 Our contribution is two-folds. First, 
we allow simultaneous learning of the number of communities and the community memberships via a prior on the number of communities $k$.  
%We exhibit through our simulation examples that such two stage procedure which ignores uncertainty in the first stage may be prone to increased misclassification when there is inherent variability in the number of communities or the community structure is vague. 
A seemingly automatic choice to allow uncertainty in the number of communities is to use a Bayesian nonparametric approach such as the Chinese restaurant process (CRP) \citep{pitman1995exchangeable}. While it has been empirically observed that CRPs often have the tendency to create tiny extraneous clusters, it has only been recently established that CRPs lead to inconsistent estimation of the number of clusters in a fairly general setting \citep{miller2015mixture}. We instead adapt the mixture of finite mixture (MFM) approach of \cite{miller2015mixture} which alleviates the drawback of CRP by automatic model-based pruning of the tiny extraneous clusters leading to consistent estimate of the number of clusters. Moreover, MFM admits a clustering scheme similar to the CRP which is exploited to develop an efficient MCMC algorithm. In particular, we analytically marginalize over the number of communities to obtain an efficient Gibbs sampler and avoid resorting to complicated reversible jump MCMC algorithms or allocation samplers. We exhibit the efficacy of our proposed MFM-SBM approach over existing two-stage approaches and the CRP prior through various simulation examples. We envision simple extensions of MFM-SBM to degree corrected SBM \cite{karrer2011stochastic} and mixed membership block model \cite{airoldi2009mixed}, which will be reported elsewhere. 

Our second contribution is to develop a framework for consistent community detection, where we derive non-asymptotic bounds on the posterior probability of the true configuration. As a consequence, we can show that the marginal posterior distribution on the set of community assignments increasingly concentrates (in an appropriate sense) on the true configuration with increasing number of nodes. This is a stronger statement than claiming that the true configuration is the {\em maximum a posteriori} model with the highest posterior probability.  
%We are not aware of any similar theoretical investigations in Bayesian network models.  
Although there is now a well-established literature on posterior convergence in density estimation and associated functionals in Bayesian nonparametric mixture models (see for example, \cite{kruijer2010adaptive} and references therein), there are no existing results on clustering consistency in network models or beyond to best of our knowledge. In fact, the question of consistency of the number of mixture components has only been resolved very recently \cite{miller2015mixture,rousseau2011asymptotic}. Clustering consistency is clearly a stronger requirement and significantly more challenging to obtain than consistency of the number of mixture components.  
%\cite{nguyen2013convergence} recently studied convergence of mixing measures 
%
%More recently, convergence of the mixing measure has been studied by \cite{nguyen2013convergence}. However, to best of our knowledge, there are no existing results on clustering consistency. 
We exploit the conjugate nature of the Bayesian SBM to obtain the marginal likelihoods for each cluster configuration, and subsequently use probabilistic bounds on the log-marginal likelihood ratios to deliver our non-asymptotic bound. 
We hope our results on selection consistency has a broader appeal to the Bayesian model selection community; see in particular the second paragraph in \S \ref{sec:theory} for a detailed discussion.  

The rest of the paper is organized as follows. We start with a brief review of the SBM in 
\S \ref{sec:sbm}.  The Bayesian methods for simultaneous inference on the number of clusters and the clustering configurations are discussed in \S \ref{sec:cd} and the Gibbs sampler is provided in \S \ref{ssec:GS}.  The theory for consistent community detection is developed in \S \ref{sec:theory}. Simulation studies and comparisons with existing methods are provided in \S \ref{sec:sim} and illustration of our method on a benchmark real dataset is in \S \ref{sec:real}. Additional simulations exploring sensitivity, convergence diagnostics, and robustness, and proofs of all technical results, are provided in a separate supplemental document. The supplemental document additionally contains a second real data example.  

\section{Stochastic Block models} \label{sec:sbm}

We use $\m A  = (A_{ij}) \in \{0, 1\}^{n \times n}$ to denote the adjacency matrix of a network with $n$ nodes, with $A_{ij} = 1$ indicating the presence of an edge from node $i$ to node $j$ and $A_{ij} = 0$ indicating a lack thereof. We consider undirected networks without self-loops so that $A_{ij} = A_{ji}$ and $A_{ii} = 0$. The sampling algorithms presented here can be trivially modified to directed networks with or without self-loops. The theory would require some additional work in case of directed networks though conceptually a straightforward modification of the current results should go through. 

The probability of an edge from node $i$ to $j$ is denoted by $\theta_{ij}$, with $A_{ij} \sim \mbox{Bernoulli}(\theta_{ij})$ independently for $1 \le i < j \le n$. In a $k$-component SBM, the nodes are clustered into communities, with the probability of an edge between two nodes solely dependent on their community memberships.  Specifically,  
\begin{align}\label{eq:sBm}
A_{ij} \mid Q, k \sim \mbox{Bernoulli}(\theta_{ij}), \quad \theta_{ij} = Q_{z_i z_j}, \quad 1 \leq i < j \leq n, 
\end{align}
where $z_i \in \{1, \ldots, k\}$ denotes the community membership of the $i$th node and $Q = (Q_{rs}) \in [0, 1]^{k \times k}$ is a symmetric matrix of probabilities, with $Q_{rs} = Q_{sr}$ indicating the probability of an edge between any node $i$ in cluster $r$ and any node $j$ in cluster $s$.

Let $\m Z_{n, k} = \big\{(z_1, \ldots, z_n) : z_i \in \{1, \ldots, k\}, 1 \le i \le n \big\}$ denote all possible clusterings of $n$ nodes into $k$ clusters.  
Given $z \in \m Z_{n, k}$, let $A_{[rs]}$ denote the $n_r \times n_s$ sub matrix of $A$ consisting of entries $A_{ij}$ with $z_i = r$ and $z_j = s$. The joint likelihood of $A$ under model \eqref{eq:sBm} can be expressed as 
{\small \begin{align}\label{eq:like}
P(A \mid z, Q, k) = \prod _{1 \leq r\leq s \leq k} P(A_{[rs]} \mid z, Q), \quad P(A_{[rs]} \mid z, Q, k) = \prod_{1\leq i < j \leq n:  z_i = r, z_j = s} Q_{rs}^{A_{ij}} (1 - Q_{rs})^{1 - A_{ij}}.
\end{align}}
A common Bayesian specification of the SBM when $k$ is given can be completed by assigning independent priors to $z$ and $Q$. We generically use $p(z, Q) = p(z) p(Q)$ to denote the joint prior on $z$ and $Q$. When $K$ (the true number of clusters) is unknown, a natural Bayesian solution is to place a prior on $k$. This is described in \S \ref{sec:cd}. 

\section{Bayesian community detection in SBM} \label{sec:cd}
A natural choice of a prior distribution on $(z_1, z_2, \ldots, z_n)$ that allows automatic inference on the number of clusters $k$ is the CRP  \citep{aldous1985exchangeability,pitman1995exchangeable,Neal:2000}.  A CRP is described through the popular Chinese restaurant metaphor:  imagine customers arriving at a Chinese restaurant with infinitely many tables with the index of the table having a one-one correspondence with the cluster label.  The first customer is seated at the first table, so that $z_1 =1$. Then  $z_i, i=2, \ldots, n$ are defined through the following conditional distribution  (also called a P\'{o}lya urn scheme \cite{blackwell1973ferguson}) 
\begin{eqnarray}\label{eq:crp}
P(z_{i} = c \mid z_{1}, \ldots, z_{i-1})  \propto   
\begin{cases}
\abs{c}  , \quad  \text{at an existing table labeled}\, c\\
\alpha,  \quad \quad \quad \, \text{if} \, $c$\,\text{is a new table}.  
\end{cases}
\end{eqnarray}
The above prior for $\{z_i\}$ can also be defined through a stochastic process where 
at any positive-integer time $n$,  the value of the process is a partition $\mathcal{C}_n$ of the set $\{1, 2, 3, \ldots,n\}$, whose probability distribution is determined as follows. At time $n = 1$, the trivial partition $\{ \{1\} \}$ is obtained with probability $1$. At time $n + 1$ the element  $n + 1$ is either  i) added to one of the blocks of the partition $\mathcal{C}_n$, where each block is chosen with probability  $\abs{c}/(n + 1)$ where $\abs{c}$ is the size of the block, or ii) added to the partition $\mathcal{C}_n$ as a new singleton block, with probability $1/(n + 1)$.  Marginally, the distribution of $z_i$ is given by the stick-breaking formulation of a Dirichlet process \cite{sethuraman1994constructive}:   
 \begin{eqnarray}\label{eq:DPM}
 z_i \sim \sum_{h=1}^{\infty}  \pi_h \delta_h,  \quad \pi_h = \nu_h \prod_{l < h} (1- \nu_l), \quad \nu_h \sim \mbox{Beta}(1, \alpha).
 \end{eqnarray}
%where $\alpha$ is assigned a prior  $p_{\alpha}(\cdot)$. 
% The formulation \eqref{eq:DPM} is commonly used as a prior for the locations of Gaussian observations to induce clusters in continuous measurements.   Under \eqref{eq:crp} and \eqref{eq:DPM},  the probability of a partition $\mathcal{C}_n$ is given by 
% \begin{eqnarray} \label{eq:partitioncrp}
% p_{\mathrm{DP}} (\mathcal{C}_n) =  V_n^{\mathrm{DP}}(t) \prod_{c \in \mathcal{C}}  ( \abs{c} -1)!
%\end{eqnarray}
% where $t = \abs{\mathcal{C}_n}$ and $V_n^{\mathrm{DP}}(t) =  \int \frac{\alpha^t}{\alpha^{(n)}} p_\alpha(\alpha) d\alpha$.   
Let $t = \abs{\mathcal{C}_n}$ denote the number of blocks in the partition $\m C_n$. Under \eqref{eq:crp}, one can obtain the probability of block-sizes $s= (s_1, s_2, \ldots, s_t)$ of a partition $\mathcal{C}_n$ as
\begin{eqnarray} \label{eq:sizeprobcrp}
p_{\mathrm{DP}}(s) \propto  \prod_{j=1}^t  s_j^{-1}.  
\end{eqnarray}
It is clear from \eqref{eq:sizeprobcrp} that CRP assigns large probabilities to clusters with relatively smaller size. A striking consequence of this has been recently discovered \citep{miller2015mixture} where it is shown that the CRP produces extraneous clusters in the posterior leading to inconsistent estimation of the {\em number of clusters} even when the sample size grows to infinity. \cite{miller2015mixture} proposed a modification of the CRP based on a mixture of finite mixtures (MFM) model to circumvent this issue:   %The modified CRP is induced by   on $z_i, i=1,2, \ldots, n$:
\begin{eqnarray}\label{eq:MFM}
k \sim p(\cdot), \quad (\pi_1, \ldots, \pi_k) \mid k \sim \mbox{Dir}(\gamma, \ldots, \gamma), \quad z_i \mid k, \pi \sim \sum_{h=1}^k  \pi_h \delta_h,\quad  i=1, \ldots, n, 
\end{eqnarray}
 where $p(\cdot)$ is a proper p.m.f on $\{1, 2, \ldots, \}$ and $\delta_h$ is a point-mass at $h$.  
%Under \eqref{eq:MFM},  the p.m.f of $\mathcal{C}_n$ can be written as \footnote{Let $x^{(m)} = x(x+1) \cdots (x + m -1)$ and $x_{(m)} =x(x-1) \cdots (x - m +1)$, 
%with $x^{(0)} = x_{(0)} =1$}
%\begin{eqnarray*}
%p_{\mathrm{MFM}}(\mathcal{C}_n) = V^{\mathrm{MFM}}_n(t) \prod_{c \in \mathcal{C}_n} \gamma^{(\abs{c})}
%\end{eqnarray*}
%where $t = \abs{\mathcal{C}_n}$ is again the number of blocks in the partition, and
%$V^{\mathrm{MFM}}_n(t) = \sum_{k=1}^{\infty} \frac{k_{(t)}}{(\gamma k)^{(n)}} p(k)$.  \
\cite{miller2015mixture} showed that the joint distribution of $(z_1, \ldots, z_n)$ under \eqref{eq:MFM} admit a P\'{o}lya urn scheme akin to CRP:
%which generates partitions $\mathcal{C}_1, \mathcal{C}_2, \ldots, $ as follows: 
 \begin{enumerate}
 \item Initialize with a single cluster consisting of element 1 alone:  $\mathcal{C}_1 = \{\{1\} \}$, 
 \item For $n=2, 3, \ldots, $ place element $n$ in 
 \begin{enumerate}
 \item an existing cluster $c \in \mathcal{C}_{n-1}$ with probability $\propto \abs{c} + \gamma$
 \item a new cluster with probability $\propto \frac{V_n(t+1)}{V_n(t)} \gamma$
 \end{enumerate}
 where $t = \abs{\mathcal{C}_{n-1}}$. 
 \end{enumerate}
$V_n(t)$ is a coefficient of partition distribution that need to be precomputed in this model, 
\begin{align*} 
\begin{split}
V_n(t) &= \sum_{n=1}^{+\infty}\dfrac{k_{(t)}}{(\gamma k)^{(n)}} p(k)
\end{split}					
\end{align*} 
where $k_{(t)}=k(k-1)...(k-t+1)$, and $(\gamma k)^{(n)} = {\gamma k}(\gamma k+1)...(\gamma k+n-1)$. (By convention, $x^{(0)} = 1$ and $x_{(0)}=1$).

%This restaurant process generated by the P\'{o}lya urn scheme bears close resemblance to the CRP, 
Compared to the CRP, the introduction of new tables is slowed down by the factor $V_n(\abs{\mathcal{C}_{n-1}}  +1)/ V_n(\abs{\mathcal{C}_{n-1}})$, thereby allowing a model-based pruning of the tiny extraneous clusters.   An alternative way to understand this is to look at  the probability of block-sizes $s= (s_1, s_2, \ldots, s_t)$ of a partition $\mathcal{C}_n$  with $t= |\m C_n|$ under MFM.  As opposed to \eqref{eq:sizeprobcrp},  the probability of the cluster-sizes $(s_1, \ldots, s_t)$ under MFM is 
\begin{eqnarray} \label{eq:sizeprob}
p_{\mathrm{MFM}}(s) \propto  \prod_{j=1}^t  s_j^{\gamma-1}.  
\end{eqnarray} 
From \eqref{eq:sizeprobcrp} and \eqref{eq:sizeprob}, it is easy to see that MFM assigns comparatively smaller probability to clusters with small sizes.  The parameter $\gamma$ controls the relative size of the clusters;  small $\gamma$ favors lower entropy $\pi$'s, while large $\gamma$ favors higher entropy $\pi$'s.   
   
Adapting MFM to the SBM setting, our model and prior can be expressed hierarchically as: 
\begin{align}\label{eq:MFMSBM}
\begin{split}
& k \sim p(\cdot), \text{where $p(\cdot)$ is a p.m.f on \{1,2, \ldots\} }\\
& Q_{rs} = Q_{sr} \stackrel{\text{ind}} \sim \mbox{Beta}(a, b),  \quad r, s = 1, \ldots, k, 
\\
& \mbox{pr}(z_i = j \mid \pi, k) = \pi_j, \quad j = 1, \ldots, k, \, i = 1, \ldots, n,  \\
& \pi \mid k \sim \mbox{Dirichlet}(\gamma, \ldots, \gamma),\\
& A_{ij} \mid z, Q, k \stackrel{\text{ind}} \sim \mbox{Bernoulli}(\theta_{ij}), \quad \theta_{ij} = Q_{z_i z_j}, \quad 1 \le  i < j \le n. 
\end{split}
\end{align}
A default choice of $p(\cdot)$ is a $\mbox{Poisson}(1)$ distribution truncated to be positive \cite{miller2015mixture}, which is assumed through the rest of the paper. We refer to the hierarchical model above as MFM-SBM. While MFM-SBM admits a CRP representation, an important distinction from infinite mixture models hinges on the fact that for any given prior predictive realization, one draws a value of $k$ and as $n$ grows the individuals are distributed into the $k$ clusters. On the other hand, the number of clusters keeps growing with $n$ for the infinite mixture models. 

\subsection{ {\bf Gibbs sampler }} \label{ssec:GS}
Our goal is to sample from the posterior distribution of the unknown parameters $k, z = (z_1, \ldots, z_n) \in \{1, \ldots, k\}^n$ and $Q = (Q_{rs}) \in [0, 1]^{k \times k}$. \cite{miller2015mixture} developed the MFM approach for clustering in mixture models, where their main trick was to analytically marginalize over the distribution of $k$. While MFM-SBM is different from a standard Bayesian mixture model, we could still exploit the P\'{o}lya urn scheme for MFMs to analytically marginalize over $k$ and develop an efficient Gibbs sampler. The sampler is presented in Algorithm \ref{algorithm} in Appendix \ref{alg:1} of the supplemental document, which efficiently cycles through the full conditional distribution of $Q$ and $z_i \mid z_{-i}$ for $i=1, 2, \ldots, n$, where $z_{-i} = z \backslash \{z_i\}$. The marginalization over $k$ allows us to avoid complicated reversible jump MCMC algorithms or even allocation samplers.   In practice, one way to initialize the number of clusters is to use a frequentist approach (e.g. \cite{le2015estimating}).  
 For the initialization of cluster configurations, we randomly assign all observations into those clusters.

\section{Consistent community detection}\label{sec:theory}

In this section, we provide theoretical justification to the proposed approach by showing that marginal posterior distribution on the space of community assignments concentrates on the truth exponentially fast as the number of nodes increases. At the very onset, some clarification is required regarding the mode of convergence, since the community assignments are only identifiable up to arbitrary labeling of the community indicators within each community.  For example, in a network of $5$ nodes with $2$ communities, consider two community assignments $z$ and $z'$, with $z_1 = z_3 = z_5 = 1$ \& $z_2 = z_4 = 2$; and $z_1' = z_3' = z_5' = 2$ \& $z_2' = z_4' = 1$. Clearly, although $z$ and $z'$ are different as $5$-tuples, they imply the same community structure and the posterior cannot differentiate between $z$ and $z'$. To bypass such {\em label switching} issues, we consider a permutation-invariant Hamming distance introduced in \cite{zhang2015minimax} as our loss function and bound the posterior expected loss (equivalently, the Bayes risk) with large probability under the true data generating mechanism. The concentration of the posterior on the true community assignment (up to labeling) follows as a straightforward corollary of the Bayes risk bound. 

Consistency results for our Bayesian procedure complements a series of recent frequentist work on consistent community detection \cite{bickel2009nonparametric,zhao2012consistency,zhang2015minimax,gao2015achieving,abbe2015community,abbe2015recovering,abbe2015detection} among others.  From a Bayesian viewpoint, our result contributes to a growing literature on consistency of Bayesian model selection procedures when the number of competing models grow exponentially relative to the sample size \cite{johnson2012bayesian,narisetty2014bayesian,castillo2015bayesian,shin2017scalable}. Our present problem has two key distinctions from these existing results which primarily focus on variable selection in (generalized) linear models: (a) the model space does not have a natural nested structure as in case of (generalized) linear models, which requires additional care in enumeration of the space of community assignments; and (b) the log-marginal likelihood differences between a putative community assignment and the truth is not readily expressible as a $\chi^2$-statistic, necessitating careful analysis of such objects.

\subsection{ {\bf Preliminaries }}

We introduce some basic notations here that are required to state our main results. Notations that only appear in proofs are introduced at appropriate places in the supplemental document. 

Throughout $C, C'$ etc denote constants that are independent of everything else but whose values may change from one line to the other. $\ind(B)$ denotes the indicator function of set $B$. 
%Let $\m A = (a_{ij}) \in \{0, 1\}^{n \times n}$ denote the $n \times n$ adjacency matrix of the network. 
For two vectors $x = \{x_i\}$ and $y = \{y_i\}$ of equal length $n$, the Hamming distance between $x$ and $y$ is $d_H(x, y) = \sum_{i=1}^n \ind(x_i \ne y_i)$. For any positive integer $m$, let $[m] := \{1, \ldots, m\}$. A community assignment of $n$ nodes into $K < n$ communities is given by $z = (z_1, \ldots, z_n)^{\T}$ with $z_i \in [K]$ for each $i \in [n]$. Let $\m Z_{n, K}$ denote the space of all such community assignments. For a permutation $\delta$ on $[K]$, define $\delta \circ z$ as the community assignment given by $\delta \circ z(i) = \delta(z_i)$ for $i \in [n]$. Clearly, $\delta \circ z$ and $z$ provide the same clustering up to community labels. 
Define $\langle z \rangle$ to be the collection of $\delta \circ z$ for all permutations $\delta$ on $[K]$; we shall refer to $\langle z \rangle$ as the equivalence class of $z$. 
%For example, $z$ and $z'$ in the previous page belong to the same equivalence class, with $z' = \delta \circ z$, where $\delta(1) = 2$ and $\delta(2) = 1$.  
Define a permutation-invariant Hamming distance (c.f. \cite{zhang2015minimax})
\begin{align}\label{eq:perm_inv}
d(z, z') = \inf_{\delta} d_H(\delta \circ z, z')
\end{align}
where the infimum is over all permutations of $[K]$. Note that $d(z, z') = 0$ if and only if $z$ and $z'$ are in the same equivalence class, i.e., $\langle z \rangle = \langle z' \rangle$. 
%\\[1ex]
%Let $\bbP$ denote probability under the true data generating mechanism. 

\subsection{ {\bf Homogeneous SBMs }}
To state our theoretical result, we restrict attention to {\em homogeneous SBMs}.   
%It is well-known that MFMs lead to consistent estimation of the number of clusters \cite{miller2015mixture} in a Gaussian mixture model.    
%In our context,  a straightforward application of Doob's Theorem (refer to \S 7.1.4 of \cite{miller2015mixture}) implies $\Pi( k =2 \mid \m A) \to 1$ a.s. $\bbP_0$.   Hence,  we do not impose a prior on $k$ and fix $k=2$ in the subsequent analysis.
An SBM is called homogeneous when the $Q$ matrix in \eqref{eq:sBm} has a compound-symmetry structure, with $Q_{rs} = q + (p-q) I(r=s)$, so that all diagonal entries of $Q$ are $p$ and all off-diagonal entries are $q$. Thus, the edge probabilities 
\begin{align*}
\theta_{ij} = 
\begin{cases}
p & \text{if $z_i = z_j$}, \\
q & \text{if $z_i \ne z_j$}.
\end{cases}
\end{align*}
For a homogeneous SBM, the likelihood function for $p, q, z,k$ assumes the form
\begin{align}
f(\m A \mid z, p, q,k) 
& = \prod_{i < j} \theta_{ij}^{ a_{ij} } \, (1 - \theta_{ij})^{1 - a_{ij}} \notag \\
& = p^{A_{\up}(z)} (1 - p)^{n_{\up}(z) - A_{\up}(z)} q^{A_{\down}(z) } (1 - q)^{n_{\down}(z) - A_{\down}(z)},  \label{eq:lik}
\end{align}
where 
\begin{alignat}{4}
n_{\up}(z) &= \sum_{i < j} \ind(z_i = z_j), \quad A_{\up}(z) &&= \sum_{i < j} a_{ij} \ind(z_i = z_j), \label{eq:nup}\\ 
n_{\down}(z) &= \sum_{i < j} \ind(z_i \ne z_j), \quad A_{\down}(z) &&= \sum_{i < j} a_{ij} \ind(z_i \ne z_j). \label{eq:ndown}
\end{alignat}
Clearly, $n_{\down}(z) = {n \choose 2} - n_{\up}(z)$. 
\\[1ex]
As in \S 3, we consider independent $\mbox{U}(0, 1)$ priors on $p$ and $q$. %We comment here that our results continue to hold for the uniform prior on $\m Z_{n,2}$. 
A key object is the {\em marginal likelihood} of $z$, denoted $\m L(\m A \mid z, k)$, obtained by integrating over the priors on $p$ and $q$. Exploiting Beta-binomial conjugacy, we have,
\begin{align}
\m L(\m A \mid z, k) 
& = \bigg\{ \int_{0}^1 p^{A_{\up}(z)} (1 - p)^{n_{\up}(z) - A_{\up}(z)}  dp \bigg\} \, \bigg\{ \int_{0}^1 q^{A_{\down}(z) } (1 - q)^{n_{\down}(z) - A_{\down}(z)} dq \bigg\} \notag \\
%& = \frac{\Gamma\{ A_{\up}(z) + 1 \} \Gamma\{ n_{\up}(z) - A_{\up}(z) +1 \} }{ \Gamma\{ n_{\up}(z) + 2\} } \, \frac{\Gamma\{ A_{\down}(z) + 1 \} \Gamma\{ n_{\down}(z) - A_{\down}(z) +1 \} }{ \Gamma\{ n_{\down}(z) + 2\} } \\
& = \frac{1}{n_{\up}(z) + 1} \, \frac{1}{ { n_{\up}(z) \choose A_{\up}(z) } } \, \frac{1}{n_{\down}(z) + 1} \, \frac{1}{ {n_{\down}(z) \choose A_{\down}(z)} }. \label{eq:marglik}
\end{align} 
Letting $\Pi(z\mid k)$ denote the prior probability of the community assignment $z$ conditional on $k$, its posterior probability $\Pi(z \mid k, \m A) \propto \m L(\m A \mid z, k) \Pi(z\mid k)$. Observe that each one of $n_{\up}(z), n_{\down}(z), A_{\up}(z)$ and $A_{\down}(z)$ are labeling invariant, i.e., they assume a constant value on $\langle z \rangle$, and hence so is $\m L(\m A \mid z, k)$. Hence, as long as the prior $\Pi(\cdot \mid k)$ is labeling invariant, the same can thus be concluded regarding the posterior $\Pi(\cdot \mid k,  \m A)$. For example, the Dirichlet-multinomial prior (conditional on $k$) in \eqref{eq:MFM} in \S 3 is labeling invariant. 

%For this reason, we focus on $\Pi( \langle z_0 \rangle \mid \m A)$ as our object of study, where $z_0$ is (one representation of ) the true community assignment. 

\subsection{ \bf{Main result for known $K$ case} }

Our first set of results pertain to the case when the number of communities $K$ is fixed and known. We assume the true network-generating model is a homogeneous SBM with $K$ communities, and true within- and between-community edge probabilities $p_0$ and $q_0$ respectively. We note that unlike several existing results, we do not assume knowledge of $p_0$ and $q_0$. Let $z_0$ denote the true community assignment. 
\\
We state our assumptions on these quantities below.  \\[1ex]
{\bf (A1)} Assume the number of nodes $n$ is an integer multiple of $K$, with each community having an equal size of $n/K$. Without loss of generality, we assume that $z_{0i} = \lfloor (i-1)/K \rfloor + 1$ for $i = 1, \ldots, n$.  \\[2ex]
{\bf (A2)} The true edge probabilities $p_0 \ne q_0$ satisfy $n \bar{D}(p_0, q_0)/K \to \infty$ as $n \to \infty$, where 
\begin{align}\label{eq:bar_D}
\bar{D}(p_0, q_0) := \frac{ (p_0 - q_0)^2}{ (p_0 \vee q_0) \{1 - (p_0 \wedge q_0)\}}.
\end{align}
with $\vee$ and $\wedge$ denoting maximum and minimum respectively. \\
{\bf (A1)} assumes a balanced network which is fairly common in the literature; see for example, \cite{zhang2015minimax}. Extension to the case where the community sizes are unequal but of the same order can be accomplished, albeit with substantially more tedious counting arguments. Condition {\bf (A2)} is automatically satisfied if $p_0$ and $q_0$ do not vary with $n$. However, ${\bf (A2)}$ is much stronger in that one can accommodate {\em sparse networks} where $p_0$ and $q_0$ decay to zero. Indeed, parameterizing $p_0 = a/n$ and $q_0 = b/n$, the condition in {\bf (A2)} amounts to $(a - b)^2/(a \vee b) \to \infty$. Recent information-theoretic results [Theorem 1.1 of \cite{zhang2015minimax}, equation (16) in \cite{abbe2015community}] show that the condition $(a-b)^2/(a \vee b) \to \infty$ is necessary for complete recovery of the community assignments. The quantity $\bar{D}(p_0, q_0)$ is closely related to Renyi divergence measures between $\mbox{Bernoulli}(p_0)$ and $\mbox{Bernoulli}(q_0)$ distributions that appear in the information-theoretic lower bounds. 

We next state a Lipschitz-type condition on the log-prior mass on the community assignments. \\
{\bf (P1) } Assume $z_0$ satisfies {\bf (A1)}. The prior $\Pi$ on $\m Z_{n, K}$ satisfies 
\begin{align}\label{eq:lip}
\abs{ \log \Pi(z) - \log \Pi(z_0) } \le C K d(z, z_0),
\end{align}
for all $z \in \m Z_{n, K}$. 

\begin{rem}
{\bf (P1)} requires $\log \Pi(\cdot)$ to be Lipschitz continuous with respect to the distance $d$, with Lipschitz constant bounded by a multiple of $K$. {\bf (P1)} is satisfied by the Dirichlet-multinomial prior in \S 3. Straightforward calculations yield, for the Dirichlet-multinomial prior with Dirichlet concentration parameter $\gamma$, 
$$
\frac{ \Pi(z) }{ \Pi(z_0) } = \prod_{h=1}^K \frac{\Gamma(n_h(z) + \gamma)}{ \Gamma(n/K + \gamma)},
$$
where, recall $n_h(z) = \sum_{i=1}^n \ind(z_i = h)$. The inequality \eqref{eq:lip} follows from an application of the following two-sided bound for the gamma function: for any $x >0$, $\log \Gamma(x) = (x - 1/2) \log x - x + R(x)$, with $0 < R(x) < (12 x)^{-1}$.
\end{rem}

%In Theorem \ref{thm:clus_cons} below, we state a non-asymptotic probabilistic bound on the posterior probability  
Let $\bbP$ denote probability under the true data generating mechanism. We now provide a bound to the posterior expected loss of $d(z, z_0)$, i.e., $E [d(z, z_0) \mid \m A]$, that holds with large $\bbP$-probability (w.r.t. $\m A$), in Theorem \ref{thm:clus_cons} below. The proof is deferred to Appendix \ref{sec:pfmt}  of the supplemental document. 
\begin{theorem}\label{thm:clus_cons}
Recall the permutation-invariant Hamming distance $d(\cdot, \cdot)$ from \eqref{eq:perm_inv}. Assume the true cluster assignment $z_0$ satisfies {\bf (A1)}, and the true within \& between edge probabilities $p_0$ and $q_0$ satisfy {\bf (A2)}. Also, assume that the prior $\Pi$ on $\m Z_{n, K}$ satisfies {\bf (P1)}. Then, %$\Pi( \langle z_0 \rangle \mid \m A) \to 1$ a.e. $[\bbP_0]$. 
\begin{align*}
E [ d(z, z_0) \mid \m A] \le \exp \bigg\{ - \frac{ C n \bar{D}(p_0, q_0) }{K} \bigg\}, 
\end{align*}
holds with $\bbP$-probability at least $1 - e^{- C (\log n)^{\nu}}$ for some $\nu > 1$. 
\end{theorem}
%\noindent Although Theorem \ref{thm:clus_cons} is stated for a fixed $K$, an inspection of the proof reveals that $K$ can be allowed to grow with $n$ as long as $K^2 = o(n)$. 
An immediate corollary of Theorem \ref{thm:clus_cons} is that the posterior almost surely concentrates on the true configuration $z_0$. To see this, let $\m C$ denote the large $\bbP$-probability set in Theorem \ref{thm:clus_cons}. We have, inside $\m C$, 
\begin{align*}
\Pi[ \langle z \rangle =  \langle z_0 \rangle \mid \m A] = \Pi[d(z, z_0) = 0 \mid \m A] = 1 - \Pi[ d(z, z_0) > 1 \mid \m A] \ge 1 -  \exp \bigg\{ - \frac{ C n \bar{D}(p_0, q_0) }{K} \bigg\},
\end{align*} 
where the penultimate inequality follows from Markov's inequality. We summarize in the following Corollary which is a straightforward application
 of the first Borel-Cantelli Lemma. 
\begin{corollary}\label{cor:clus_cons}
Suppose the conclusion of Theorem \ref{thm:clus_cons} holds. Then, 
\begin{align*}
\Pi[ \langle z \rangle =  \langle z_0 \rangle \mid \m A]   \geq 1 -  \exp \bigg\{ - \frac{ C n \bar{D}(p_0, q_0) }{K} \bigg\}  \quad \text{almost surely 
$\mathbb{P}$ as} \,\,  n \to \infty.
\end{align*}
\end{corollary}
Corollary \ref{cor:clus_cons} ensures that as  $n \to \infty$, for almost every network sampled from $\mathbb{P}$,  $\Pi[ \langle z \rangle =  \langle z_0 \rangle \mid \m A]$ is close to $1$ at the same rate obtained in  Theorem \ref{thm:clus_cons}. This is possible since $\mathbb{P}(\m C^c)$ decreases sufficiently fast to $0$ as $n \to \infty$. 

The proof of Theorem \ref{thm:clus_cons} is lengthy and thus provided in Appendix \ref{sec:pfmt} of the supplemental document. We briefly comment on some of the salient aspects here. The key ingredient in proving Theorem \ref{thm:clus_cons} is to uniformly bound from below the difference in log-marginal marginal likelihood between the true community assignment $z_0$ and a putative community assignment $z$ with $d(z, z_0) = r$. As a first step, we approximate the log-marginal likelihood $\log \m L(\m A \mid z)$ by $\wt{\ell}(z) := n_{\up}(z) h\{ A_{\up}(z)/n_{\up}(z) \} + n_{\down}(z) h\{ A_{\down}(z)/n_{\down}(z) \}$, where $h(x) = x \log x + (1-x) \log(1-x)$ for $x \in (0, 1)$. This is essentially a Laplace approximation of the log-marginal likelihood and the error in approximation can be bounded appropriately. We construct a set $\m C$ with $\bbP(\m C) \ge 1 - e^{- C (\log n)^{\nu}}$ in Proposition \ref{prop:main} stated in the supplemental document such that within $\m C$, 
\begin{align}\label{eq:marglik_diff_main}
\wt{\ell}(z_0) - \wt{\ell}(z) \ge \frac{ C \bar{D}(p_0, q_0) \ n \ d(z, z_0)} {K},
\end{align}
for all $z \in \m Z_{n, K}$. Equation \eqref{eq:marglik_diff_main} combined with the prior mass condition {\bf (P1)} essentially delivers the proof of Theorem \ref{thm:clus_cons}. 

A couple of intertwined technical challenges show up in obtaining a concentration bound of the form \eqref{eq:marglik_diff_main}. First, the random quantities $\wt{\ell}(z_0)$ and $\wt{\ell}(z)$ can be highly dependent, particularly when $d(z, z_0)$ is small, which rules out separately analyzing the concentration of each term around its expectation. However, a combined analysis of the difference is complicated by the presence of the non-linear function $h$. We note that $h$ is non-Lipschitz, and hence standard concentration inequalities for Lipschitz functions of several independent variables cannot be applied. We crucially exploit convexity of $h$ to analyze 
the difference $\wt{\ell}(z_0) - \wt{\ell}(z)$. A careful combinatorial analysis of terms arising inside the bounds (Lemma \ref{lem:mota_n} in the supplemental document) along with concentration inequalities for sub-Gaussian random variables \cite{vershynin2010introduction} deliver the desired bound.

\subsection{ \bf{Main result for unknown $K$ case} }

We now partially aim to answer the question: if the true $K$ is unknown and a prior is imposed on $k$ which assigns positive mass to the true $K$, can we recover $K$ and the true community assignment $z_0$ from the posterior? To best of our knowledge, this question hasn't been settled even for usual mixture models, and a complete treatment for SBMs is beyond the scope of this paper. An inspection of the proof of Proposition \ref{prop:main} in the supplemental document will reveal that the only place where the fact that both $z$ and $z_0$ lie in $\m Z_{n, K}$ has been used in Lemma \ref{lem:mota_n}. The primary difficulty in extending the theoretical results in the previous subsection to the variable $k$ case precisely lie in generalizing the combinatorial bounds in Lemma \ref{lem:mota_n}. Recall the metric $d$ in \eqref{eq:perm_inv} is defined on $\m Z_{n, K}$. To define $d(z_1, z_2)$ for $z_1 \in \m Z_{n, K_1}$ and $z_2 \in \m Z_{n, K_2}$, an option is to embed all the $Z_{n, k}$s inside $\cap_{k=1}^{K_{max}} Z_{n, k}$, where $K_{\max}$ is an upper bound on the number of communities. This substantially complicates the analysis as one now has to take into account zero counts for one or more communities in obtaining the combinatorial bounds. 

We consider the following simplified setting. Suppose the true $K$ can be either $2$ or $3$. Given $K$, the network is generated exactly as in the previous subsection, i.e., according to a homogeneous SBM with equal-sized communities satisfying {\bf (A1)} and {\bf (A2)}. We  do not assume knowledge of the true $K$, and use a MFM-SBM model with a prior on $k$ supported on $\{2, 3\}$.  We only require $\Pi(k)$ to have positive probability on both $2$ and $3$. We show below that the posterior of $k$ concentrates on the true $K$, characterizing the rate of concentration. 
\begin{theorem}\label{thm:clus_consk}
Assume the true cluster assignment $z_0$ satisfies {\bf (A1)} with $K \in \{2, 3\}$, and the true within \& between edge probabilities $p_0$ and $q_0$ satisfy {\bf (A2)}. Also, assume that the prior $\Pi$ on $\m Z_{n, k}$ satisfies {\bf (P1)} conditional on $k$ and $\Pi(k) >0$ for $k \in  \{2, 3\}$. Then, %$\Pi( \langle z_0 \rangle \mid \m A) \to 1$ a.e. $[\bbP_0]$. 
\begin{align*}
\Pi(k = K \mid \m A) \geq 1 - \exp\{-c n^q \},
\end{align*}
for some constant $c > 0$, with $\bbP$-probability at least $1 - e^{-t_n}$ for $t_n \to \infty$ where $q=1$ and $t_n =o(\sqrt{n})$ for $K=2$ and $q=2$ and $t_n =o(n)$ for $K=3$.  
\end{theorem}

\begin{figure}[htp!]
	\centering
	\includegraphics[width=0.9\textwidth]{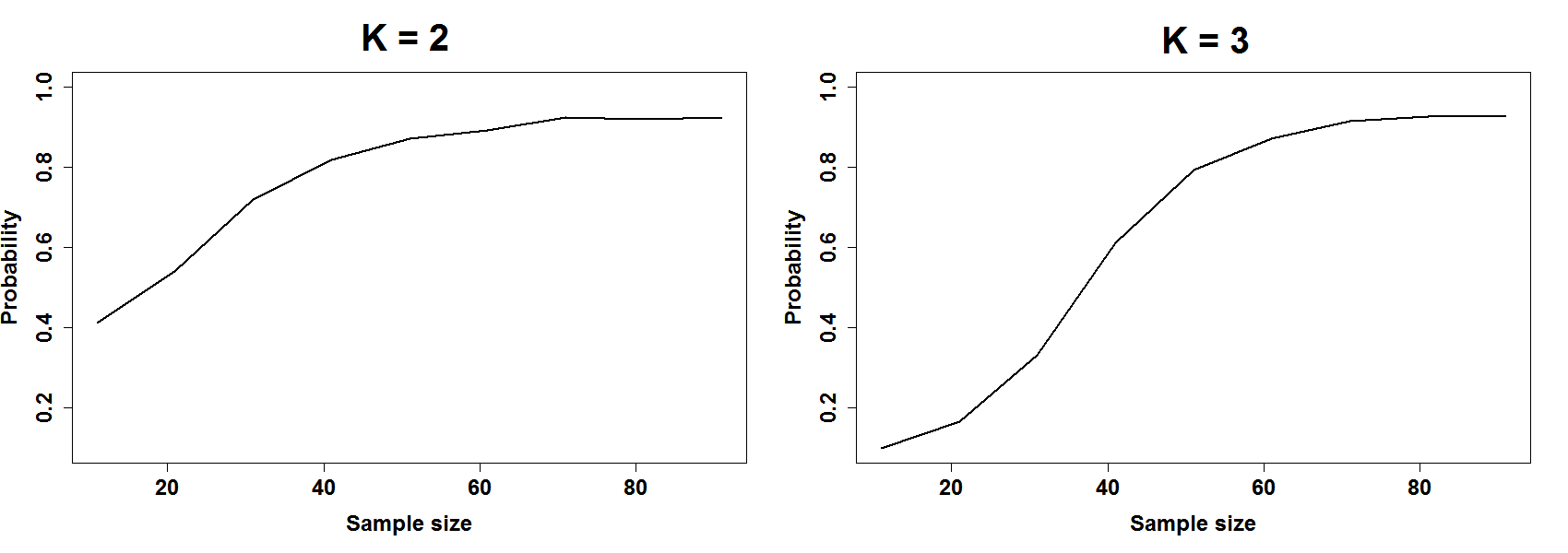}
	\caption{ {\em Growth rate of the posterior probability of the true number of components, $\Pi(k = K \mid \m A)$, as sample size $n$ increases, under the setup of Theorem  \ref{thm:clus_consk}. Left panel corresponds to the case when $K=2$, while the right panel corresponds to $K = 3$.}}
\label{fig:aveplot}
\end{figure}

The proof is deferred to Appendix \ref{sec:pfcor}  of the supplemental document. 
Theorem \ref{thm:clus_consk} is an illustration of model-selection consistency when the goal is to identify the number of clusters $K$. 
%  The likelihood of the SBM can potentially derive strength from $O(n^2)$ edges as opposed to $O(n)$ data points in standard regression and mixture models.   In the overfitted case when $K=2$ and the model is fitted with 
%$k=3$,  the  marginal likelihood ratio corresponding to a given configuration $z$ against the null $z_0$ becomes the weakest  when the Rand index between $z$ and $z_0$ is close to $1$.  Even in this case, the ratio is exponentially small in $n$ as opposed to standard mixture or regression models (typically polynomial in $n$ in such cases).  Since the number of such configurations $z$ is  sub-exponential, the Bayes factor becomes exponentially small. In the underfited case, since the Rand index between the fitted and the true configuration never approaches $1$, the  marginal likelihood ratio derives strength from all the $O(n^2)$ edges and  is of the order $e^{-n^2}$. 
In the overfitted case when $K=2$ and the model is fitted with 
$k=3$, the posterior can successfully ``empty-out'' the extraneous cluster and recover the true number of clusters.  The likelihood of the SBM can potentially derive strength from $O(n^2)$ edges as opposed to $O(n)$ data points in standard regression and mixture models.   In the overfitted case when $K=2$ and the model is fitted with 
$k=3$,  the  marginal likelihood ratio corresponding to a given configuration $z$ against the null $z_0$ becomes the weakest  when the Rand index between $z$ and $z_0$ is close to $1$. In this case, the marginal likelihood ratio corresponding to $k=3$ and $K=2$ is only exponentially small ($e^{-n}$) when the rand-index between the true configuration and fitted configuration is close to $1$.  Apparently, this may appear to impede model selection consistency since the model complexity is exponential in $n$.  However, it turns out that the number of configurations for which the rand-index is sufficiently close to $1$ is only polynomial in $n$.  This is also aided by the Dirichlet-Multinomial formulation which restricts $\Pi(z\mid k) / \Pi(z_0 \mid K)$ for configurations close to $z_0$ to be at most polynomial in $n$.  Hence  the Bayes factor is exponentially small in $n$ delivering an exponential concentration of the posterior of $k$.   This is a clear distinction with  standard mixture or regression models (typically polynomial in $n$ in such cases \cite{rousseau2011asymptotic,drton2017bayesian}). In the underfited case,  the Rand-Index between the true and the fitted configuration can never be close to $1$ which makes separation between  the log-marginal likelihoods of the order of $n^2$.   This is strong enough to offset the exponential model complexity as well as the prior ratio leading to a posterior concentration rate of $e^{-n^2}$.  

To empirically demonstrate the posterior probability bounds for  $K=2$ and $K =3$ in Theorem \ref{thm:clus_consk}, we conduct a small simulation study under the setup of the theorem. Figures \ref{fig:aveplot} displays $\Pi(k = K \mid \m A)$ averaged over $100$ replicated datasets plotted against $n$ when $K = 2$ and $K=3$ respectively and $(p_0, q_0) = (0.5, 0.1)$.  
%The definitions of $p$ and $q$ can be found in Section \ref{sec:sim}. 
It is evident that $\Pi(k = K \mid \m A)$ approaches $1$ at a faster rate for $K=3$ than for $K=2$.  
%\begin{proof}
%Define $\m D_n$ to be the set inside $\bbP_0$ in Theorem \ref{thm:clus_cons}. 
%%\begin{eqnarray*}
%%\mathcal{D}_n =  \bigg\{ \abs{ \frac{1}{\Pi( \langle z_0 \rangle \mid \m A)} - 1 } \le e^{- C n \log n} \bigg\}.    
%%\end{eqnarray*}
%Since $\sum_{n=1}^{\infty} \bbP_0 (\mathcal{D}_n^c) < \infty$, by the first Borel-Cantelli Lemma,  
%$\bbP_0 (\mbox{lim sup} \, \, \mathcal{D}_n^c)  = 0$, implying $\bbP_0 (\mbox{lim inf} \, \, \mathcal{D}_n)  = 1$. Since $\{\mbox{lim inf} \,\,  \mathcal{D}_n\}  \subset \{ \Pi( \langle z_0 \rangle \mid \m A) \to  1 \}$, the conclusion follows. 
%\end{proof}

\section{Simulation studies} \label{sec:sim}

In this section, we investigate the performance of the proposed MFM-SBM approach from a variety of angles. At the very onset, we outline the skeleton of the data generating process followed throughout this section. \\
\underline{{\bf Step 1:}} Fix the number of nodes $n$ \& the true number of communities $K$. \\
\underline{{\bf Step 2:}} Generate the true clustering configuration $z_0 = (z_{01}, \ldots, z_{0n})$ with $z_{0i} \in \{1, \ldots, K\}$. To this end, we fix the respective community sizes $n_{01}, \ldots, n_{0K}$, and without loss of generality, let $z_{0i} = l$ for all $i = \sum_{j<l}n_{0,j}+1, \ldots, \sum_{j<l}n_{0,j} + n_{0l}$ and $l = 1, \ldots, K$. We consider both balanced (i.e., $n_{0l} \sim \lfloor n/K \rfloor$ for all $l$) and unbalanced networks. In the unbalanced case, the community sizes are chosen as $n_{01}:\cdots :n_{0K} = 2: \cdots:K+1$. \\
\underline{{\bf Step 3:}} Construct the matrix $Q$ in \eqref{eq:sBm} with $q_{rs} = q + (p-q) I(r=s)$, so that all diagonal entries of $Q$ are $p$ and all off-diagonal entries are $q$. We fix $q = 0.10$ throughout and vary $p$ subject to $p > 0.10$. Clearly, smaller values of $p$ represent weaker clustering pattern.  \\
\underline{{\bf Step 4:}}  Generate the edges $A_{ij} \sim \mbox{Bernoulli}(Q_{z_{0i} z_{0j}})$ independently for $1 \leq i < j \leq n$.

The Rand index \cite{rand1971objective} is used to measure the accuracy of clustering. Given two partitions $\mathcal{C}_1 = \{X_1, \ldots, X_r\}$ and $\mathcal{C}_2 = \{Y_1, \ldots, Y_s\}$ of $\{1, 2, \ldots, n\}$, let $a, b, c$ and $d$ respectively denote the number of pairs of elements of $\{1, 2, \ldots, n\}$ that are (a) in a same set in $\m C_1$ and a same set in $\m C_2$, (b) in different sets in $\m C_1$ and different sets in $\m C_2$, (c) in a same set in $\m C_1$ but in different sets in $\m C_2$, and (d) in different sets in $\m C_1$ and a same set in $\m C_2$. The Rand index $\mathrm{RI}$ is 
\begin{eqnarray*}
 \mathrm{RI} = \frac{a+b}{a+b+c+d} = \frac{a+b}{{n \choose 2 }}. 
\end{eqnarray*}
Clearly, $0 \leq \mathrm{RI} \leq 1$ with a higher value indicating a better agreement between the two partitions. In particular, $\mathrm{RI} = 1$ indicates $\m C_1$ and $\m C_2$ are identical (modulo labeling of the nodes). 

We also briefly discuss the estimation of $k$ from the posterior. In our collapsed Gibbs sampler, $k$ is marginalized out and hence we do not directly obtain samples from the posterior distribution of $k$. However, we can still estimate $k$ based on the posterior distribution of $|z|$, the number of unique values (occupied components) in $(z_1, \ldots, z_n)$. This is asymptotically justified for mixtures of finite mixtures as in \S 4.3.2 of \citep{Miller:Thesis} who  showed that the (prior) posterior distribution of $|z|$ behaves very similarly to that for the number of components $k$ when $n$ is large. This approach also works well in finite samples as demonstrated below.   %We obtain $\mbox{M}$ posterior samples and obtain posterior summary measures based on samples post burn-in. Inference on the number of clusters and clustering configurations is obtained employing the modal clustering method of \cite{dahl2009modal}.

In all the simulation examples considered below, we employed Algorithm \ref{algorithm} with $\gamma =1$ and $a=b=1$ to fit the MFM-SBM model; we shall henceforth refer to this as the MFM-SBM algorithm.  For all simulations, a truncated Poisson prior with mean $1$ is assumed on $k$.  We arbitrarily initialized our algorithm with $9$ clusters and randomly allocated the cluster configurations in all the examples. We experimented with various other choices and did not find any evidence of sensitivity to the initialization; a detailed sensitivity analysis can be found in Appendix \ref{sec:cond} of the supplemental document. In more complex real networks,  a practical guideline for the truncated Poisson mean is to take an empirical Bayes approach and set it to the estimated number of clusters from a frequentist algorithm (such as BHM considered in the paper).

%\subsection{{\bf Convergence analysis}}

\subsection{ {\bf  Estimation performance }} \label{sec:k-est}

We now study the accuracy of MFM-SBM in terms of estimating the number of communities as well as the community memberships. As benchmark for comparison, we consider two modularity based methods available in the R Package \texttt{igraph} which first estimate the number of communities by some model selection criterion and subsequently optimize a modularity function to obtain the community allocations. The first competitor, called the leading eigenvector method (LEM: \cite{newman2006finding}), finds densely connected subgraphs by calculating the leading nonnegative eigenvector of the modularity matrix of the graph. The second competitor, called the hierarchical modularity measure (HMM; \cite{blondel2008fast}), implements a multi-level modularity optimization algorithm for finding the community structure. Our experiments suggests that these two methods have the overall best performance among available methods in the R Package \texttt{igraph}. In addition to LEM and HMM, we also consider a couple of very recent spectral methods which have been developed solely for estimating the number of communities and have been shown to outperform a wide variety of existing approaches  based on BIC, cross-validation etc. These methods are based on the spectral properties of certain graph operators, namely the non-backtracking matrix (NBM) and the Bethe Hessian matrix (BHM).  We also compare our algorithm to trans-dimensional MCMC algorithms like reversible jump MCMC or allocation samplers \cite{nobile2007bayesian} that also allow the number of components to be inferred from data. We found the very recent preprint \citep{newman2016estimating} (MH-MCMC) that came out (\texttt{C} code publicly available) while this article was in submission which implements a similar idea to update $k$ using Metropolis--Hastings moves and also uses a Dirichlet-multinomial prior.

%\textcolor{red}{We also compare to the Bayesian competitor (MH-MCMC).  }

\begin{figure}[htbp]
	\centering
	\includegraphics[width=0.90\textwidth]{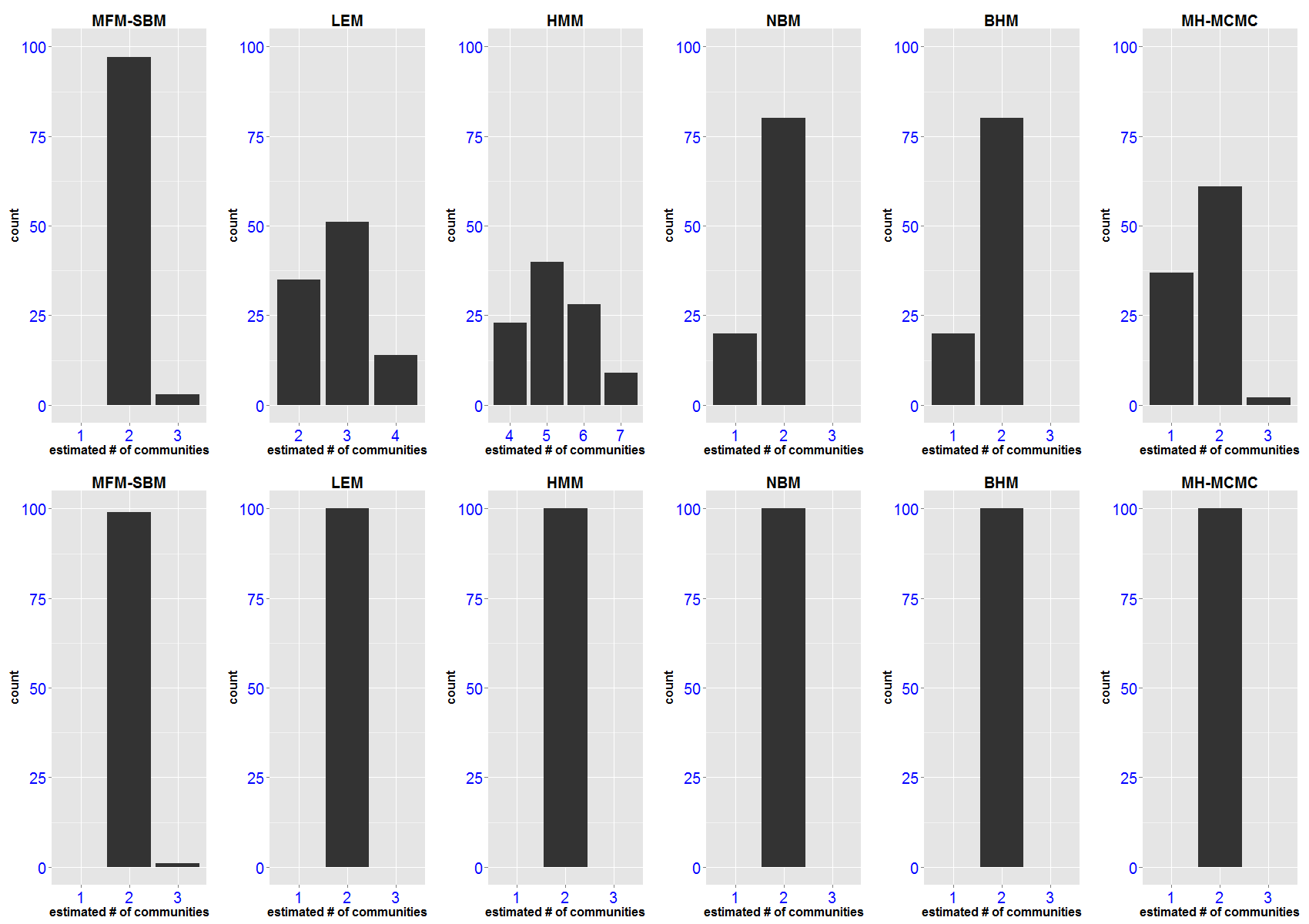}
	\caption{ {\em Balanced network with $100$ nodes and $2$ communities. Histograms of estimated number of communities across $100$ replicates. The lower panel is the case when the community structure in the network is prominent (p = 0.5); the top panel is for a vague block structure (p = 0.24).  From left to right:  our method (MFM-SBM), leading eigenvector method (LEM), hierarchical modularity measure (HMM), non back-tracking matrix (NBM), Bethe Hessian matrix (BHM) \& MH-MCMC. }}
	\label{fig:6}	
\end{figure}

\begin{figure}[htbp]
	\centering
	\includegraphics[width=0.90\textwidth]{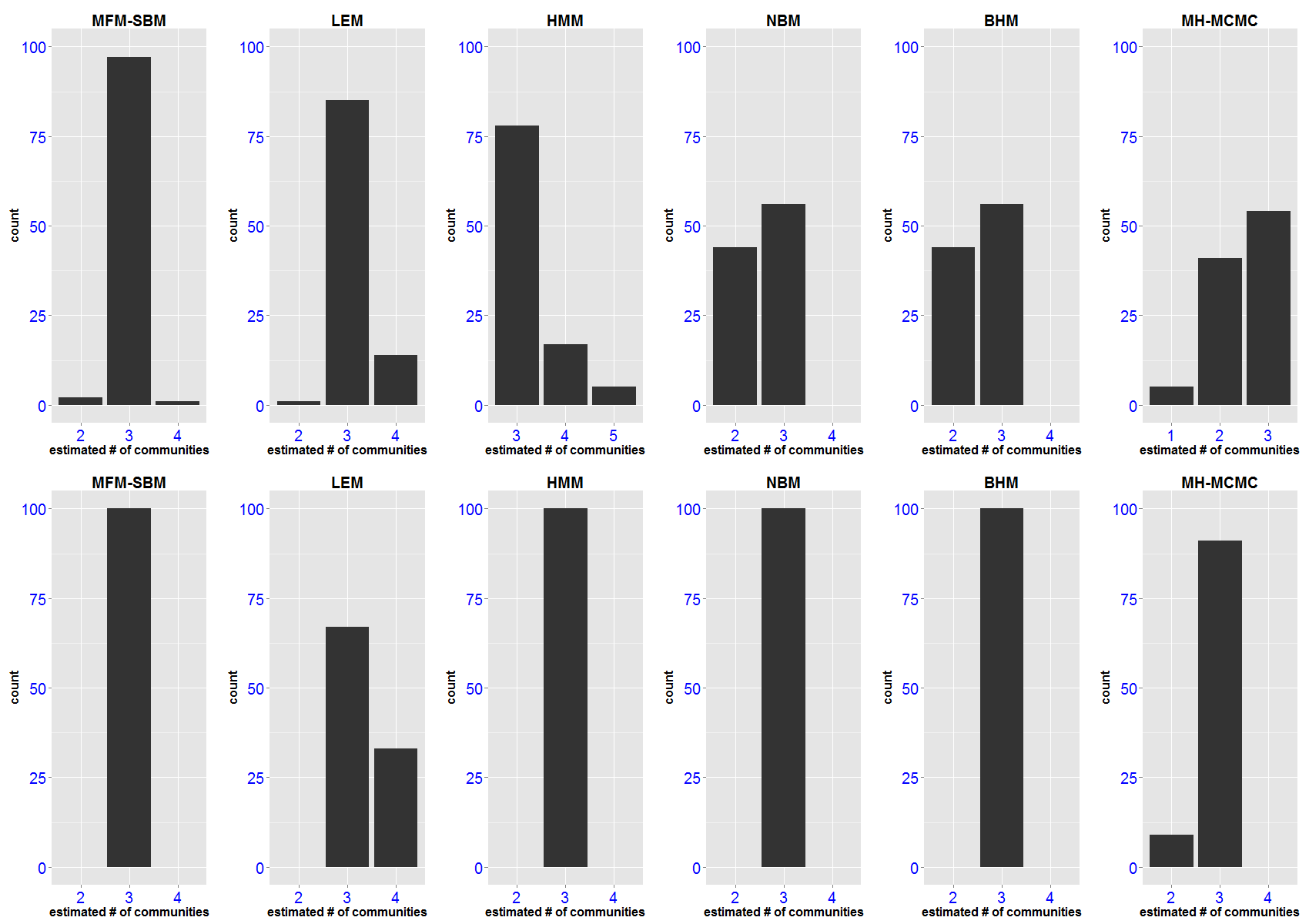}
	\caption{{\em Balanced network with $100$ nodes and $3$ communities. Histograms of estimated number of communities across $100$ replicates. The lower panel is the case when the community structure in the network is prominent (p = 0.5); the top panel is for a vague block structure (p = 0.33). From left to right:  our method (MFM-SBM), leading eigenvector method (LEM), hierarchical modularity measure (HMM), non back-tracking matrix (NBM) , Bethe Hessian matrix (BHM) \& MH-MCMC.} }
\label{fig:7}	
\end{figure}

We consider balanced networks with 100 nodes and different choices of $K$ and $p$. We generate $100$ independent datasets using the steps outlined at the beginning of the section and compare the different approaches based on the proportion of times the true $K$ is recovered among the $100$ replicates. For MFM-SBM, we used random initializations to run 10 MCMC chains in parallel for 250 iterations each, and took majority voting among the posterior modes of $k$ from each chain to arrive at a final point estimate.  
The summaries from the 100 replicates are provided in Figures \ref{fig:6} and \ref{fig:7}. 

From the lower panels of Figures \ref{fig:6} and \ref{fig:7}, we can see that when the community structure in the network is prominent ($p = 0.5$), all three methods have $100 \%$ accuracy. However, the situation is markedly different when the block structure is vague, as can be seen from the top panels of the respective figures. When the true number of communities is $2$ and $p = 0.24$ (top panel of Figure \ref{fig:6}), MFM-SBM comprehensively outperforms the competing methods. When $p = 0.33$ with $3$ communities (top panel of Figure \ref{fig:7}), our method continues to have the best performance.

We next proceed to compare the estimation performance in recovering the true community memberships using the Rand index as a discrepancy measure.    For MFM-SBM,  inference on the clustering configurations is obtained employing the modal clustering method of \cite{dahl2009modal}. Comparisons with LEM,HMM and MH-MCMC are summarized in Table \ref{tab:1}; NBM and BHM are excluded since they only estimate the number of communities. When the block structure is more vague (small $p$), MFM-SBM provides more accurate estimation of the community memberships. 
% \textcolor{red}{From Table \ref{tab:1} and Table \ref{tab:2}, we can also see that when the block structure is more vague (small $p$), MFM-SBM provides more accurate estimation of the community memberships than MH-MCMC.} 

\begin{table}[htbp]
\begin{center}
\begin{tabular}{|c|cccc|}
\hline
$(K, p)$ & {\bf MFM-SBM} & {\bf LEM} &  {\bf HMM} &  {\bf MH-MCMC}\\ \hline
$K=2, p=0.50$ & 0.99 (1.00) & 1.00 (0.99) & 1.00 (1.00) & 1.00 (1.00)  \\ \hline
$K=2, p=0.24$ & 0.97 (0.84) & 0.35 (0.79) & NA (NA) & 0.61 (0.78) \\ \hline
$K=3, p=0.50$ & 1.00 (1.00) & 0.67 (0.96) & 1.00 (0.99) & 0.91 (0.99) \\ \hline
$K=3, p=0.33$ & 0.97 (0.93) & 0.85 (0.79) & 0.78 (0.89) & 0.54 (0.93) \\ \hline 
\end{tabular}
\end{center}
\caption{ {\em The value outside the parenthesis denotes the proportion of correct estimation of the number of clusters out of 100 replicates. The value inside the parenthesis denotes the average Rand index value when the estimated number of clusters is true. NA's indicate no correct estimation of the number of clusters out of all replicates.   }}
\label{tab:1}
\end{table}
We also conducted a thorough simulation study to assess robustness of the method to misspecification in Appendix \ref{sec:k-est_mis} of the supplemental document.

\section{Benchmark real datasets} \label{sec:real}
We consider two real-datasets popularly considered in the literature i) the dolphin social network data and the ii) US political books network. Both can be found in 
\url{http://www-personal.umich.edu/~mejn/netdata/}.  We mention analysis of the first dataset in \S \ref{sec:dsn} and the defer the analysis of the second dataset to Appendix \ref{sec:uspb} of the supplemental document.

\subsection{{\bf Community detection in dolphin social network data}} \label{sec:dsn}
\begin{figure}[htp!]
	\centering
	\includegraphics[width=1\textwidth]{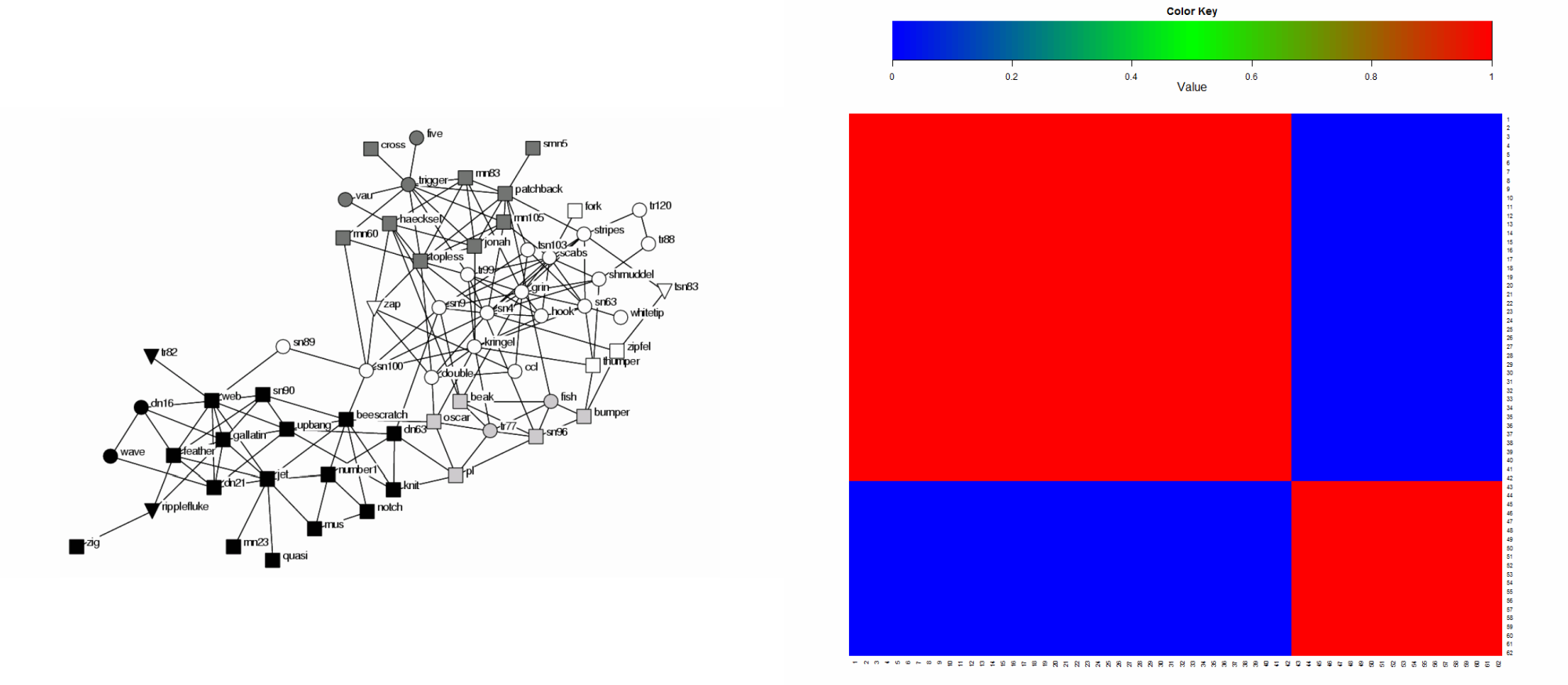}
	\caption{{\em Reference configuration for the dolphin network. Left panel: Vertex color indicates community membership: black and non-black vertices represent the principal division into two communities. Shades of grey represent sub-communities. Females are represented with circles, males with squares and individuals with unknown gender with triangles. Right panel: Heatmap of the membership matrix  $B$ of the reference configuration $z^0$ defined as $B_{ij}= \ind(z_i^0=z_j^0)$.  } }
	\label{fig:dolphin}
\end{figure}

We consider the social network dataset \citep{Lusseau2003dolphin} obtained from a community of 62 bottlenose dolphins (Tursiops spp.) over a period of seven years from 1994 to 2001. The nodes in the network represent the dolphins, and ties between nodes represent associations between dolphin pairs occurring more often than by random chance.  A reference clustering of this undirected network with $62$ nodes is in Figure \ref{fig:dolphin} (Refer to Figure 1 in \citep{Lusseau2004dolphin}).  The reference clustering shows several sub-communities based on gender, age and other demographic characteristics. There are 58 ties between males and males, 46
between females and females, and 44 between males and females, for a total of 159 ties
altogether.  We are interested in recovering the principal division  into two communities as indicated by the black and the non-black vertices just from the adjacency matrix itself. 

 \begin{table}[htp!]
\begin{center}
\begin{tabular}{|c|c|c|c|c|c|c|c|}
\hline
Method & {\bf MFM-SBM} & \bf{NBM} & {\bf BHM} &  {\bf LEM} & {\bf HMM} & {\bf MH-MCMC}\\ \hline
 Number of clusters & 2 & 2  & 2  & 5 & 5 & 3  \\ \hline

\end{tabular}
\end{center}
\caption{ {\em Estimated number of clusters for dolphin data}}
\label{tab:dolphin}
\end{table}
Results from our method (MFM-SBM) is based on 10,000 MCMC iterations leaving out a burn-in of 4,000, initialized at a randomly generated configuration with 9 clusters.  The elements of probability matrix $Q$ are assigned independent $\text{Beta}(1,1)$ priors.
From Table \ref{tab:dolphin}, it is evident that our method (MFM-SBM), NBM and BHM  provide consistent estimate of the number of clusters (being same as the reference clustering), while the other three overestimated the number of clusters.  

\begin{figure}[htp!]
	\centering
	\includegraphics[width=1\textwidth]{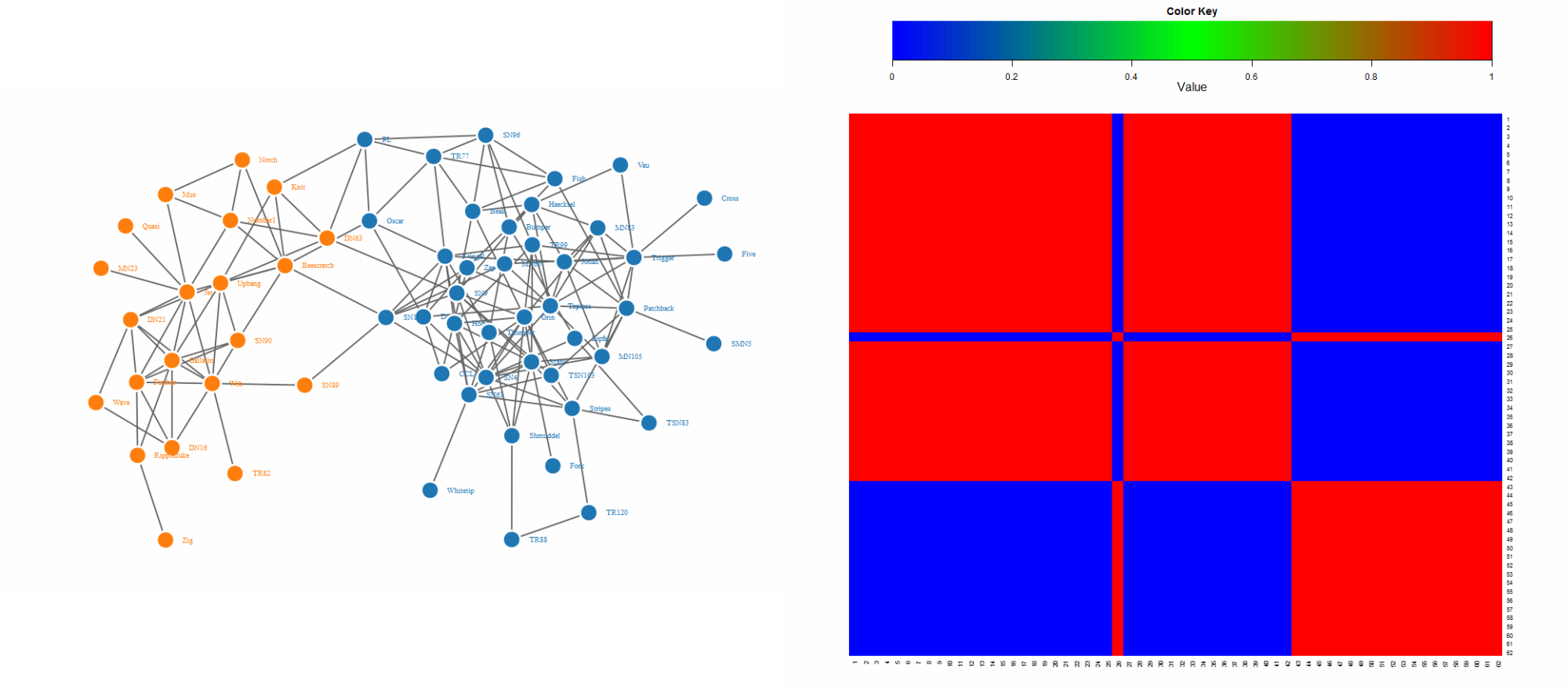}
	\caption{{\em Estimated configuration for the dolphin network using MFM-SBM. Left panel: Vertex color indicates community membership. Right panel: Heatmap of the membership matrix  $\hat{B}$ of the estimated  configuration $\hat{z}$. Perfect concordance with the reference configuration except for the assignment of the 8$^{th}$ subject. }}
	\label{fig:dol2}
\end{figure}

\begin{figure}[htp!]
	\centering
	\includegraphics[width=1\textwidth]{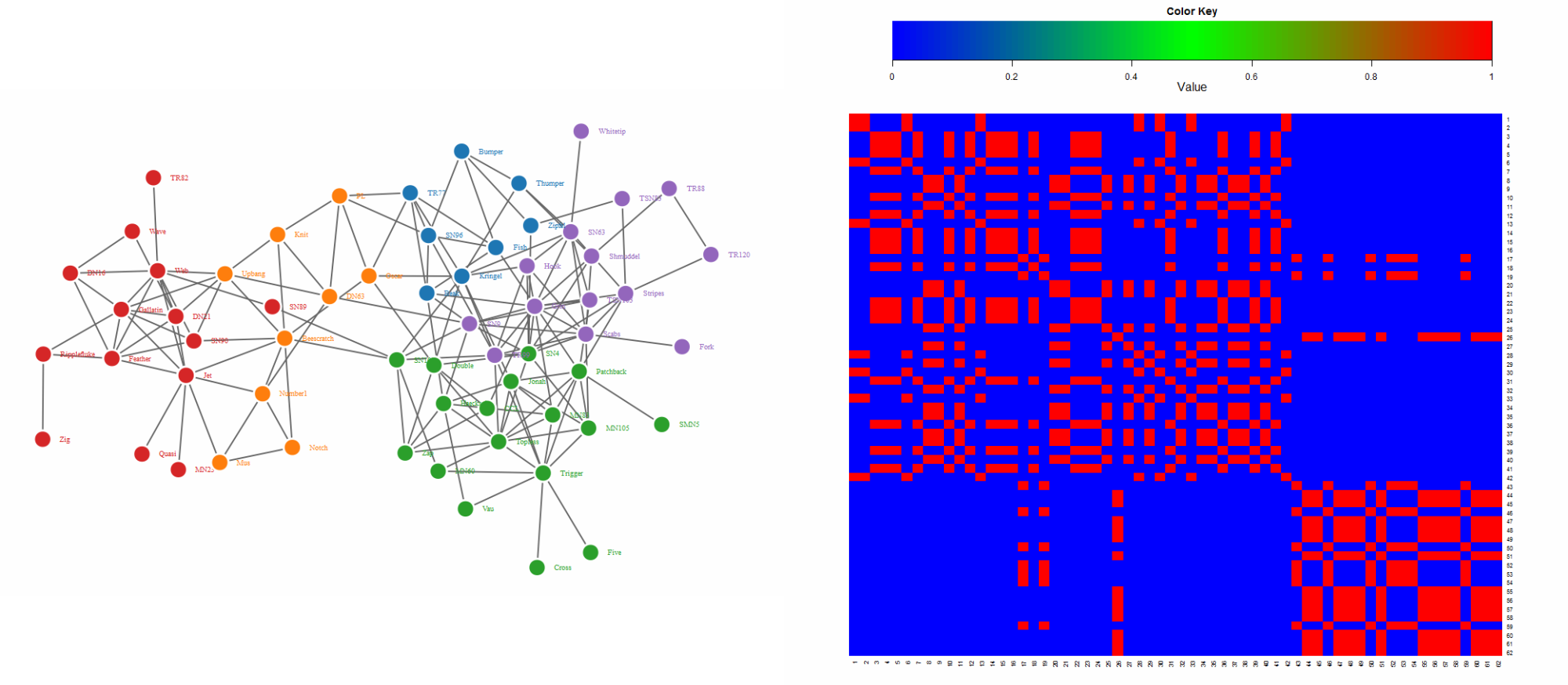}
	\caption{ {\em Estimated configuration for the dolphin network using LEM. Left panel: Vertex color indicates community membership. Right panel: Heatmap of the membership matrix  $\hat{B}$ of the estimated  configuration $\hat{z}$. The number of clusters is estimated to be 4.    Aside from cluster splitting, the assignment of $3$ subjects are different from the that in reference configuration.}
}	
\label{fig:dolcomp1}
\end{figure}

\begin{figure}[htp!]
	\centering
	\includegraphics[width=1\textwidth]{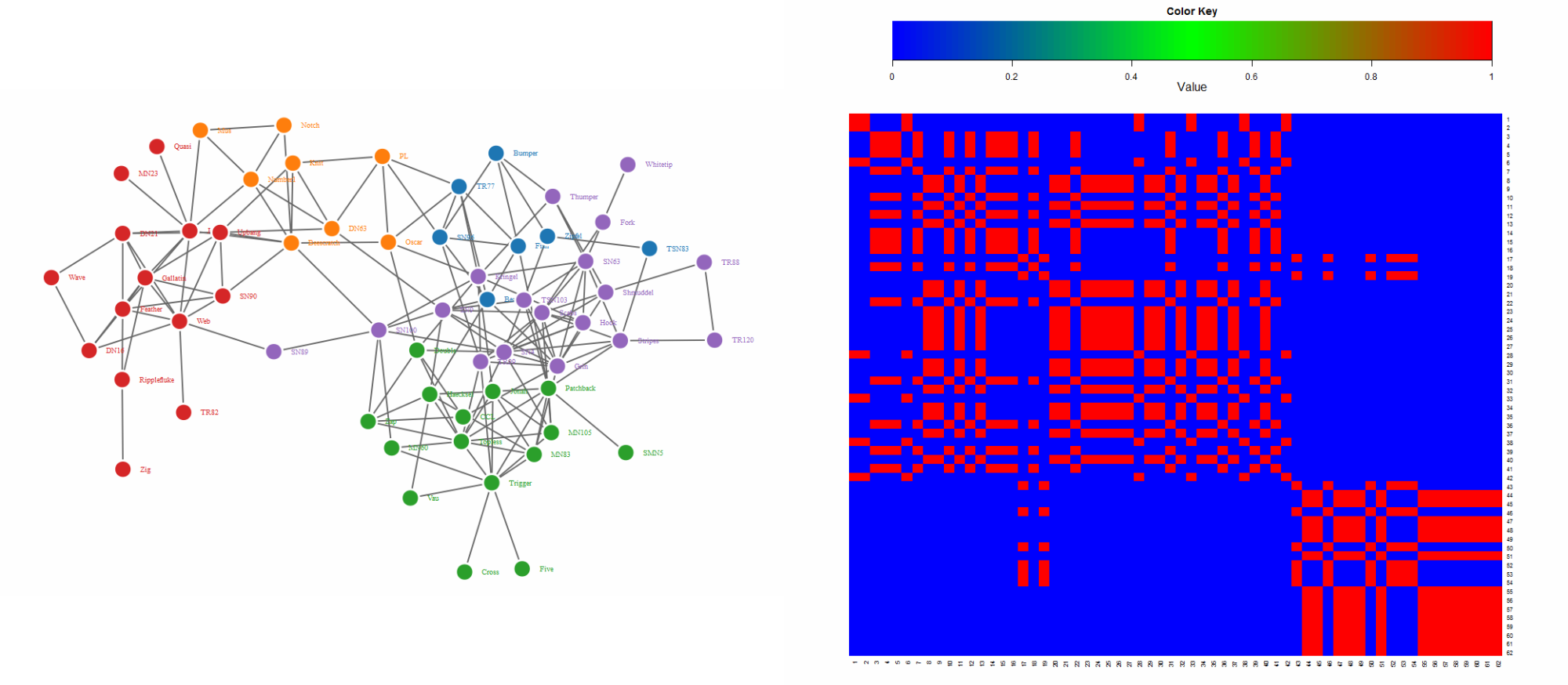}
	\caption{ {\em Estimated configuration for the dolphin network using HMM. Left panel: Vertex color indicates community membership. Right panel: Heatmap of the membership matrix  $\hat{B}$ of the estimated  configuration $\hat{z}$.  The number of clusters is estimated to be  4 and the assignment of $2$ subjects are different from that in reference configuration aside from cluster splitting. }}
	\label{fig:dolcomp2}
\end{figure}

\begin{figure}[htp!]
	\centering
	\includegraphics[width=1\textwidth]{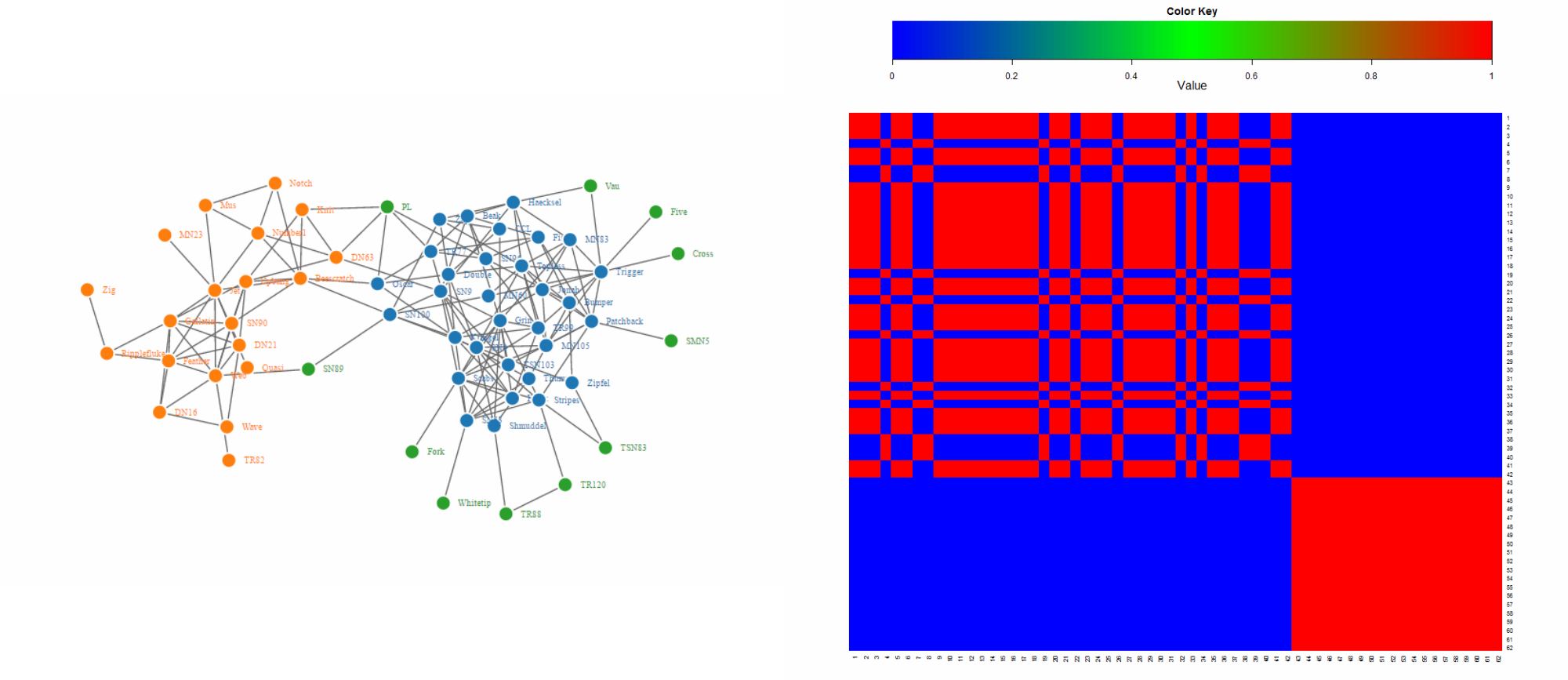}
	\caption{ {\em Estimated configuration for the dolphin network using MH-MCMC. Left panel: Vertex color indicates community membership. Right panel: Heatmap of the membership matrix  $\hat{B}$ of the estimated  configuration $\hat{z}$. The number of clusters is estimated to be 3.} }
	\label{fig:dolBSBM}
\end{figure}

%Unlike the previous study, we do not need any constraint on the diagonal and off diagonal values of the edge probability matrix $Q$.  
From Figure \ref{fig:dol2}, we see that the estimated configuration from MFM-SBM is very similar to the reference clustering (the only difference is in the assignment of the 8th subject). 
%As shown in Table \ref{tab:dolphin},the modularity based approaches (LEM and HMM) in the \texttt{igraph} package overestimate the number of clusters as shown in respectively. 
The heatmaps in Figures \ref{fig:dolcomp1}-\ref{fig:dolcomp2} show both LEM and HMM incur a few missclassified nodes. 
Figure \ref{fig:dolBSBM} shows MH-MCMC overestimate the number of clusters, with the larger cluster corresponding to the reference configuration split into two smaller clusters indicating that the mixing of the MCMC has been affected by the trans-dimensional moves. 

\section{Discussion}
We proposed a Bayesian approach for discovering the number of communities as well as the groups in a network, which has excellent performance in both simulation and real data examples. The contribution of the article is learning the number of communities and the configurations simultaneously in a coherent probabilistic framework. 
The approach is also proved to yield consistent detection of the number of communities, which is to the best of our knowledge the first such result in a Bayesian paradigm.  As an intermediate result, we developed concentration inequalities for non-linear functions of Bernoulli random variables (refer to Proposition \ref{prop:main} in the supplemental document) which may be useful in analysis of related network models.  The method can be extended easily to numerous modification of stochastic block models including the degree-corrected, mixed membership and the covariate adjusted versions. 

\section{Acknowledgement}
Dr. Bhattacharya acknowledges NSF DMS 1613156 and NSF CAREER (DMS 1653404)
and Dr. Pati acknowledges NSF DMS 1613156 for supporting this research.

\newpage
\appendix
\section*{Appendices}
\counterwithin{lemma}{section}

\section{{\bf Algorithm 1}} \label{alg:1}
We present the details of the Gibbs sampling algorithm mentioned in \S 3.1 of the main document. 

\begin{algorithm}[htbp!]
\caption{Collapsed sampler for MFM-SBM}
\label{algorithm}
\begin{algorithmic}[1]
\Procedure{c-MFM-SBM} {}
\\ Initialize $z = (z_1, \ldots, z_n)$ and $Q = (Q_{rs})$.
\For{each iter $=1$ to $\mbox{M}$ }
\\ Update $Q = (Q_{rs})$ conditional on $z$ in a closed form as
\begin{align*} 
\begin{split}
p(Q_{rs} \mid A) & \sim \mbox{Beta}(\bar{A}_{[rs]}+a,n_{rs}-\bar{A}_{[rs]}+b)
\end{split}			
\end{align*} 
Where $\bar{A}_{[rs]}=\sum_{z_i=r,z_j=s,i\neq j}A_{ij}$, $n_{rs} = \sum_{i \neq j}I(z_i=r,z_j=s),  r=1,\ldots,k; s=1,\ldots,k$.  Here $k$ is the number of clusters formed by current $z$.
\\ Update $z = (z_1, \ldots, z_n)$ conditional on $Q = (Q_{rs})$, for each $i$ in $(1,...,n)$, we can get a closed form expression for $P(z_i = c \mid z_{-i}, A, Q)$:
\[\propto \left\{
\begin{array}{ll}
      [\abs{c} + \gamma] [\prod_{j > i} Q_{cz_j}^{A_{ij}}
(1- Q_{cz_j})^{(1 - A_{ij})}] [\prod_{k < i} Q_{z_kc}^{A_{ki}}
(1- Q_{z_kc})^{(1 - A_{ki})}] & \text{at an existing table c} \\
      \frac{V_n(\abs{\mathcal{C}_{-i}}  +1)}{V_n(\abs{\mathcal{C}_{-i}}} \gamma  m(A_i) & \text{if c is a new table} \\
\end{array} 
\right. \]
where $\mathcal{C}_{-i}$ denotes the partition obtained by removing $z_i$ and 
\small
\begin{eqnarray*}
m(A_i) =  \prod_{t= 1}^{\abs{\mathcal{C}_{-i}}} \big[\mbox{Beta}(a, b)\big]^{-1}  
\mbox{Beta}\bigg[\sum_{j \in \mathcal{C}_t,j>i} A_{ij} + \sum_{j \in \mathcal{C}_t,j<i} A_{ji} + a, \abs{\mathcal{C}_t}  - 
\sum_{j \in \mathcal{C}_t,j>i} A_{ij} - \sum_{j \in \mathcal{C}_t,j<i} A_{ji} + b\bigg].
\end{eqnarray*}
\normalsize
\EndFor
\EndProcedure
\end{algorithmic}
\end{algorithm}

\section{ {\bf  Estimation performance under model misspecification}} \label{sec:k-est_mis}

\begin{figure}[htbp]
	\centering
	\includegraphics[width=1\textwidth]{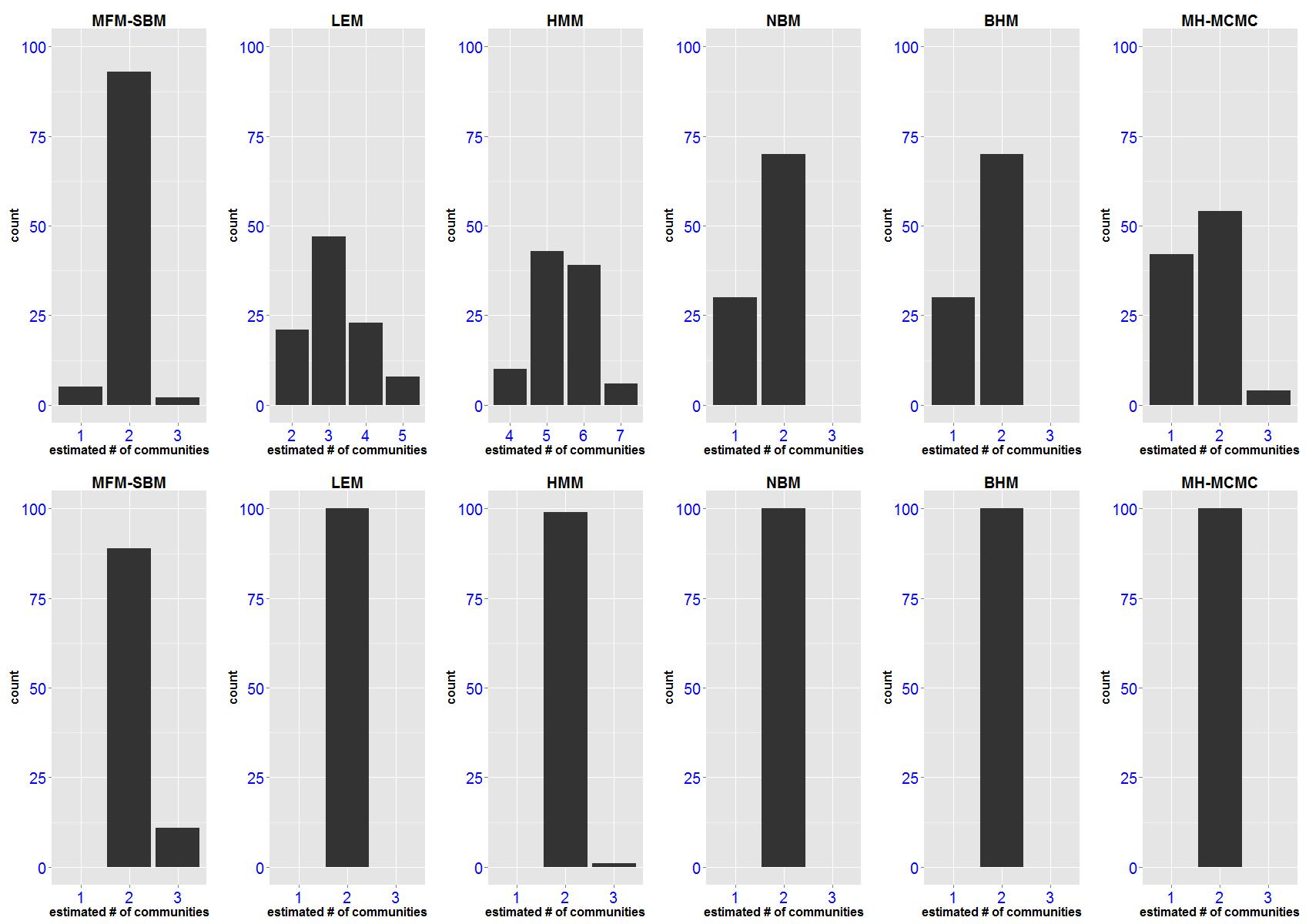}
	\caption{{\em Balanced {\bf degree-corrected} network with $100$ nodes and $2$ communities. Histograms of estimated number of communities across $100$ replicates. The lower panel is the case when the community structure in the network is prominent (p = 0.5); the top panel is for a vague block structure (p = 0.24). From left to right: our method (MFM-SBM), leading eigenvector method (LEM), hierarchical modularity measure (HMM), non back-tracking matrix (NBM) , Bethe Hessian matrix (BHM) \& MH-MCMC.} }
	\label{fig:misspec1}	
\end{figure}

\begin{figure}[htbp]
	\centering
	\includegraphics[width=1\textwidth]{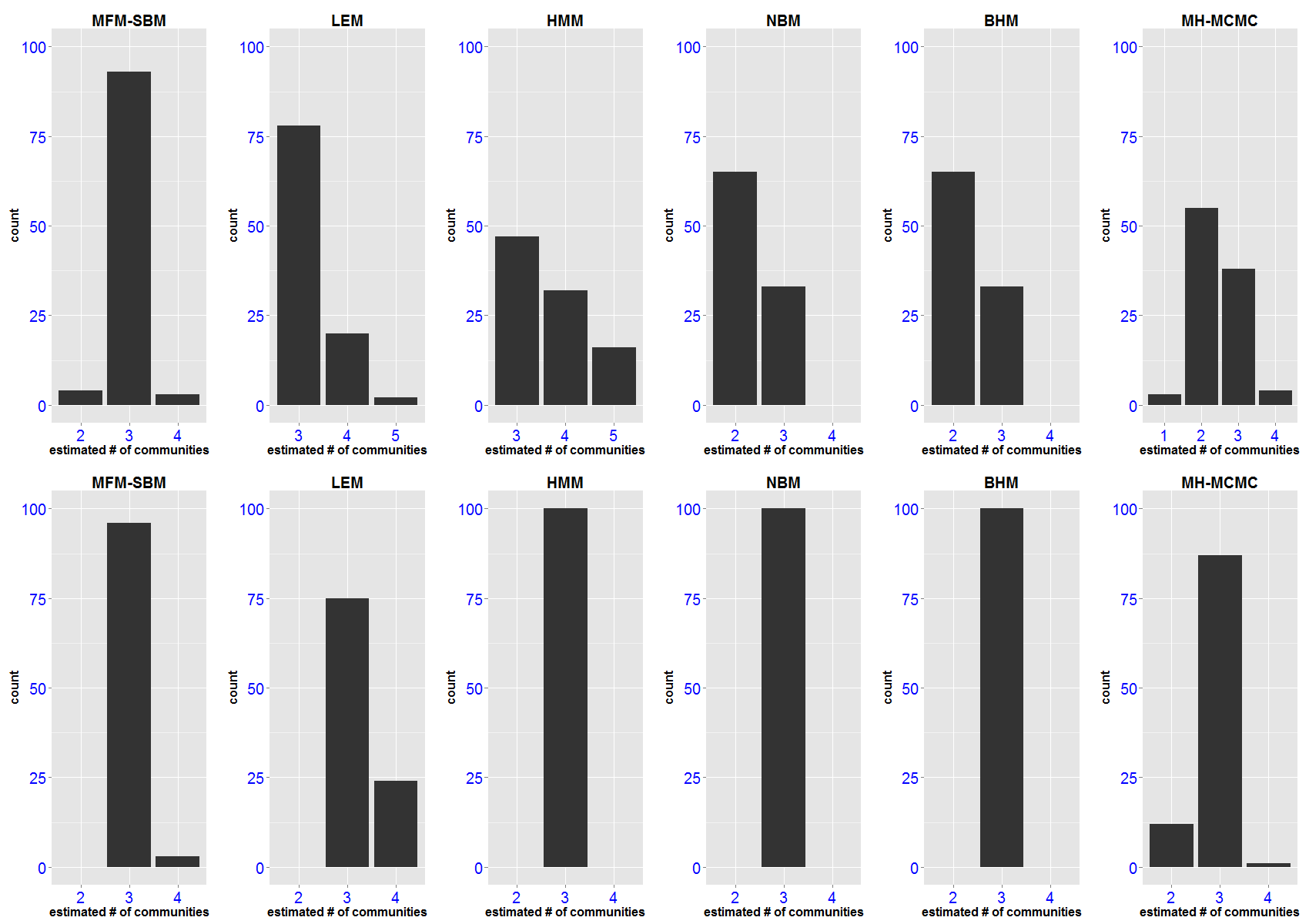}
	\caption{{\em Balanced {\bf degree-corrected} network with $100$ nodes and $3$ communities. Histograms of estimated number of communities across $100$ replicates. The lower panel is the case when the community structure in the network is prominent (p = 0.5); the top panel is for a vague block structure (p = 0.33). From left to right: our method (MFM-SBM), leading eigenvector method (LEM), hierarchical modularity measure (HMM), non back-tracking matrix (NBM) , Bethe Hessian matrix (BHM) \& Bayesian competitor (MH-MCMC). }}
\label{fig:misspec2}	
\end{figure}

As mentioned at the end of \S~5.1 of the main document, we investigate the robustness of MFM-SBM to deviations from the block model assumption. To this end, we generate data from a degree-corrected block model
\begin{align}\label{eq:DCsBm}
A_{ij} \sim \mbox{Bernoulli}(\theta_{ij}), \quad \theta_{ij} = w_i w_j Q_{z_i z_j}, \quad 1 \leq i < j \leq n, 
\end{align}
with node specific weights $w_i$s. If all $w_i$s are one, this reduces to the usual block model. We randomly set $30 \%$ of the $w_i$s to $0.8$ and the remaining to one. We generate 100 datasets for the same choices of $(n, K, p)$ as in \S~5.1. Performance in estimating the number of communities is summarized in Figures \ref{fig:misspec1} and \ref{fig:misspec2}, while Table \ref{tab:2} reports estimation accuracy of the cluster configurations. As in \S~5.1 of the main document, MFM-SBM continues to have superior performance when the block structure is vague. These simulations indicate that MFM-SBM can handle mild deviations from the block model assumption without degrading performance, though certainly there will be a breakdown point if the true model is very different from an SBM.   

\begin{table}[!htbp]
\begin{center}
\begin{tabular}{|c|cccc|}
\hline
$(k, p)$ & {\bf MFM-SBM} & {\bf LEM} &  {\bf HMM} &  {\bf MH-MCMC}\\ \hline
$k=2, p=0.50$ & 0.89 (1.00) & 1.00 (1.00) & 0.99 (1.00) & 1.00 (1.00)\\ \hline
$k=2, p=0.24$ & 0.93 (0.75) & 0.21 (0.73) & NA (NA) & 0.54 (0.57)\\ \hline
$k=3, p=0.50$ & 0.96 (0.99) & 0.75 (0.94) & 1.00 (0.99) &  0.87 (0.99)\\ \hline
$k=3, p=0.33$ & 0.93 (0.88) & 0.78 (0.73) & 0.47 (0.80) & 0.38 (0.82)\\ \hline 
\end{tabular}
\end{center}
\caption{ {\em Cluster membership estimation under degree-corrected model. The value outside the parenthesis denotes the proportion of correct estimation of the number of clusters out of 100 replicates. The value inside the parenthesis denotes the average Rand index value when the estimated number of clusters is true. NA's indicate no correct estimation of the number of clusters out of all replicates.} }
\label{tab:2}
\end{table}

\section{{\bf Convergence diagnostics}} \label{sec:cond}

Our first set of simulations investigate the algorithmic performance of MFM-SBM relative to other available Bayesian methods for different choices of the number of nodes $n$, number of communities $K$, the within-community edge probability $p$, and the relative community sizes.

\begin{figure}[htp!]
	\centering
	\includegraphics[width=0.9\textwidth]{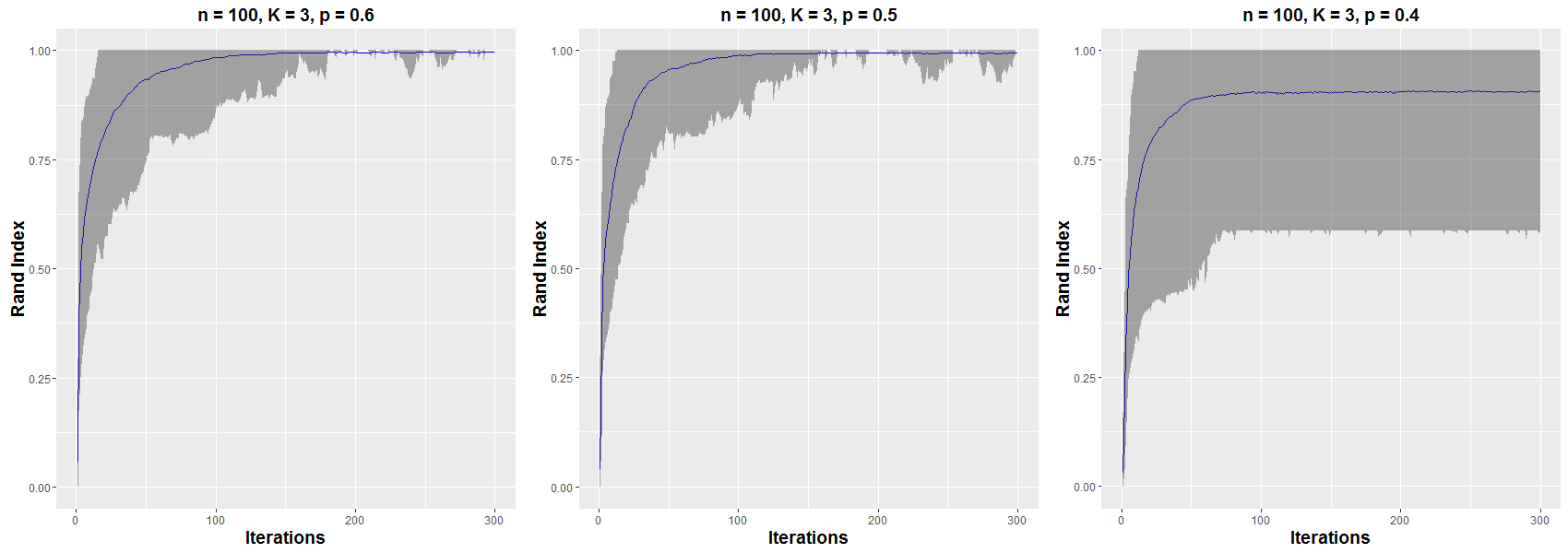}\\
	\caption{{\em Average Rand index (solid blue line) vs. MCMC iteration for MFM-SBM for 100 different starting configurations in a balanced network. $n = 100$ nodes in $K = 3$ communities of sizes 33, 33 and 34. The  shaded regions correspond to the variation of the Rand index obtained from MFM-SBM due to random initializations. }}
	\label{fig:s1}	
\end{figure}

\begin{figure}[htp!]
	\centering
	\includegraphics[width=0.9\textwidth]{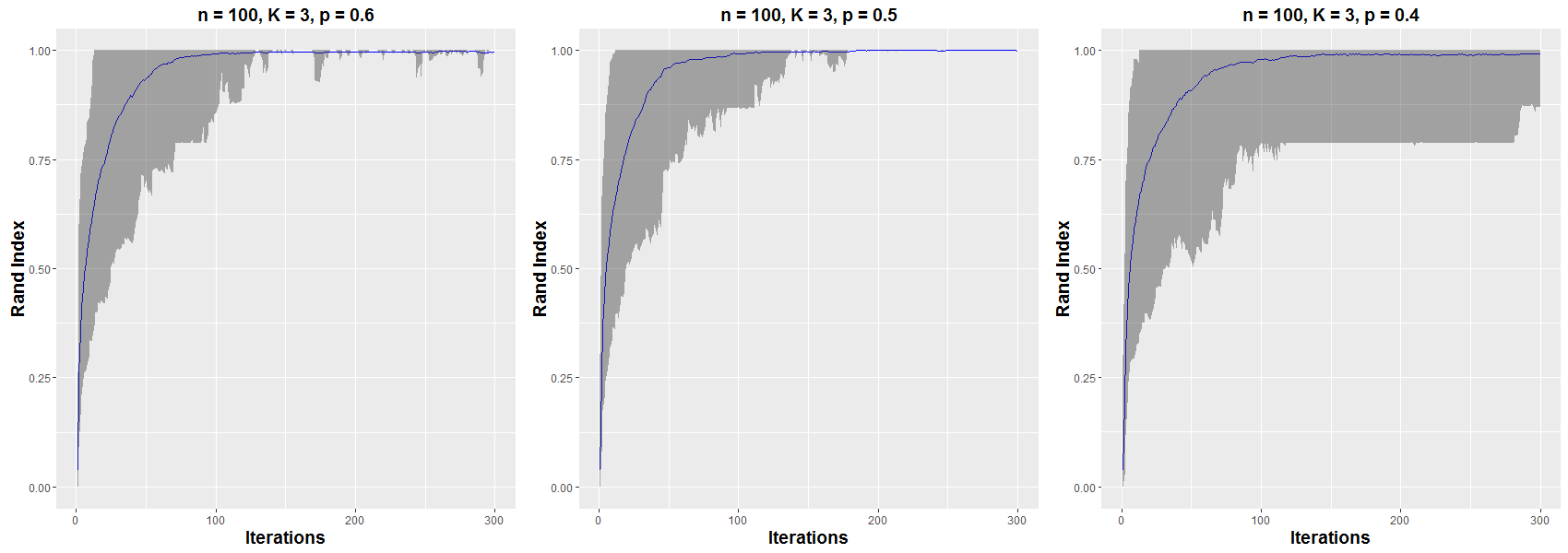}
	\caption{{\em Average Rand index (solid blue line) vs. MCMC iteration for MFM-SBM with 100 different starting configurations in an unbalanced network. $n = 100$ nodes in $K = 3$ communities of sizes 22, 33 and 45. }
 }
\label{fig:s2}
\end{figure}

\begin{figure}[htp!]
	\centering
	\includegraphics[width=0.9\textwidth]{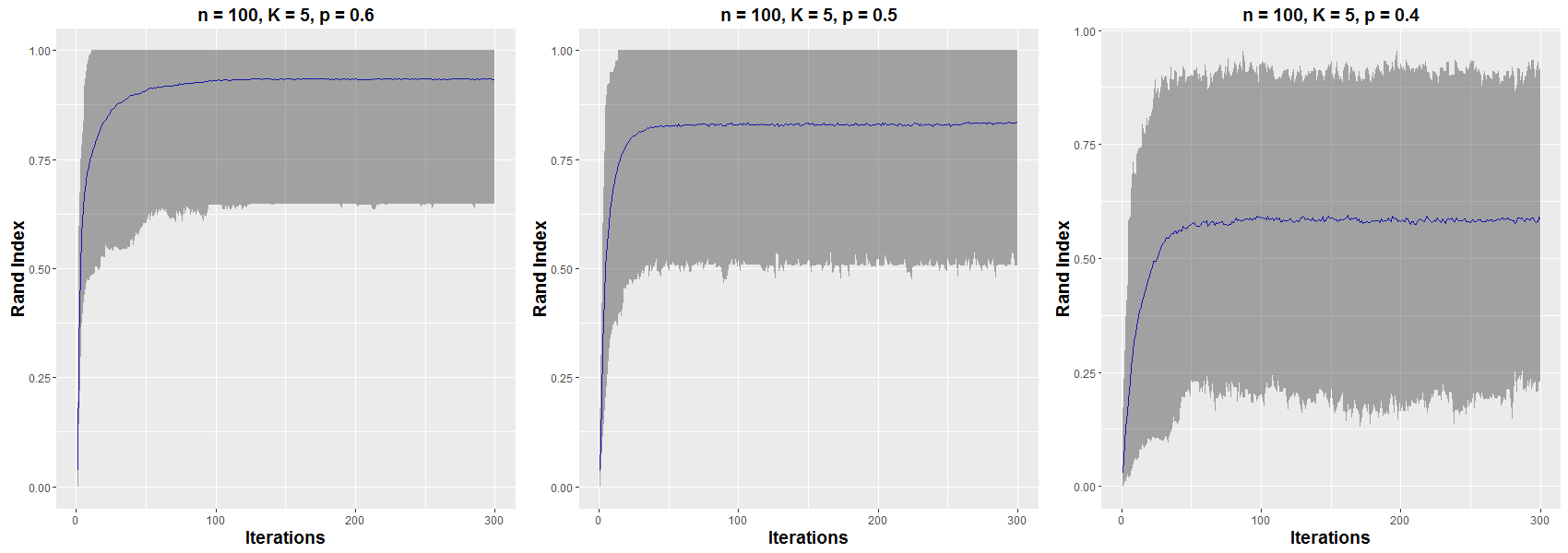}
	\caption{{\em Average Rand index (soild blue line) vs. MCMC iteration for MFM-SBM with 100 different starting configurations in a balanced network. $n = 100$ nodes in $K = 5$ communities of size 20 each. }
}
\label{fig:s3}	
\end{figure}

\begin{figure}[htp!]
	\centering
	\includegraphics[width=0.9\textwidth]{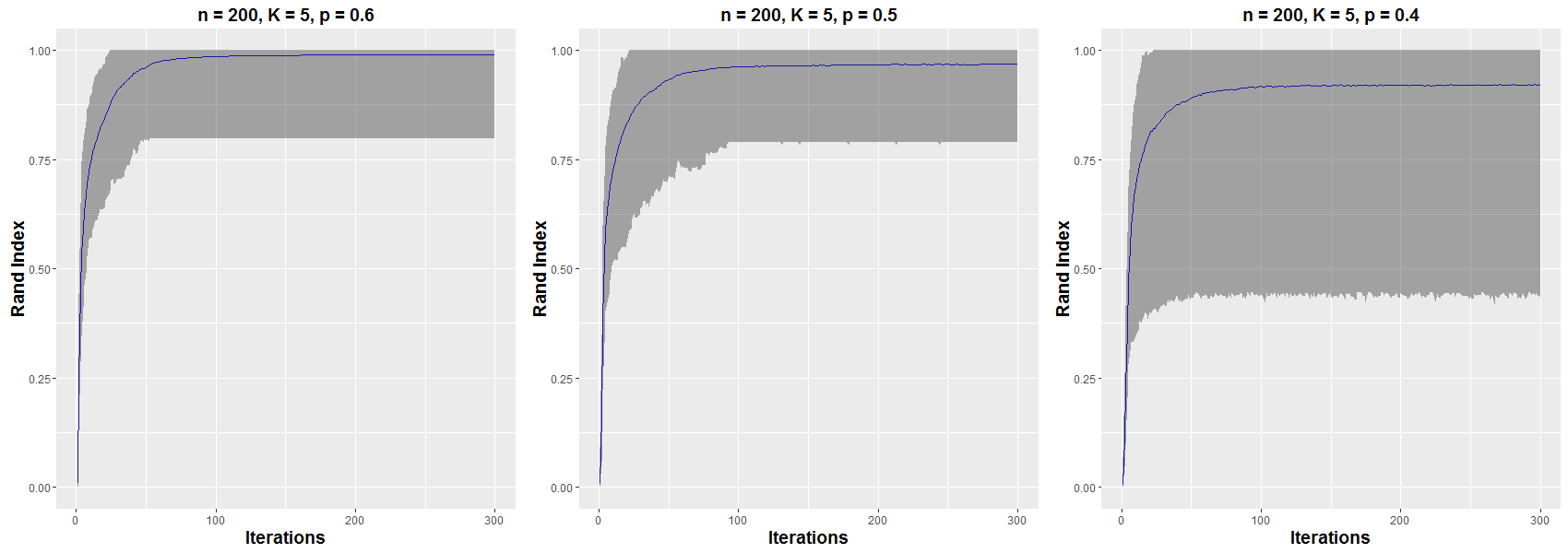}
	\caption{{\em Average Rand index (solid blue line) vs. MCMC iteration for MFM-SBM with 100 different starting configurations in a balanced network. $n = 200$ nodes in $K = 5$ communities of size 40 each.}
}
\label{fig:s4}
\end{figure}

\begin{figure}[htp!]
	\centering
	\includegraphics[width=0.9\textwidth]{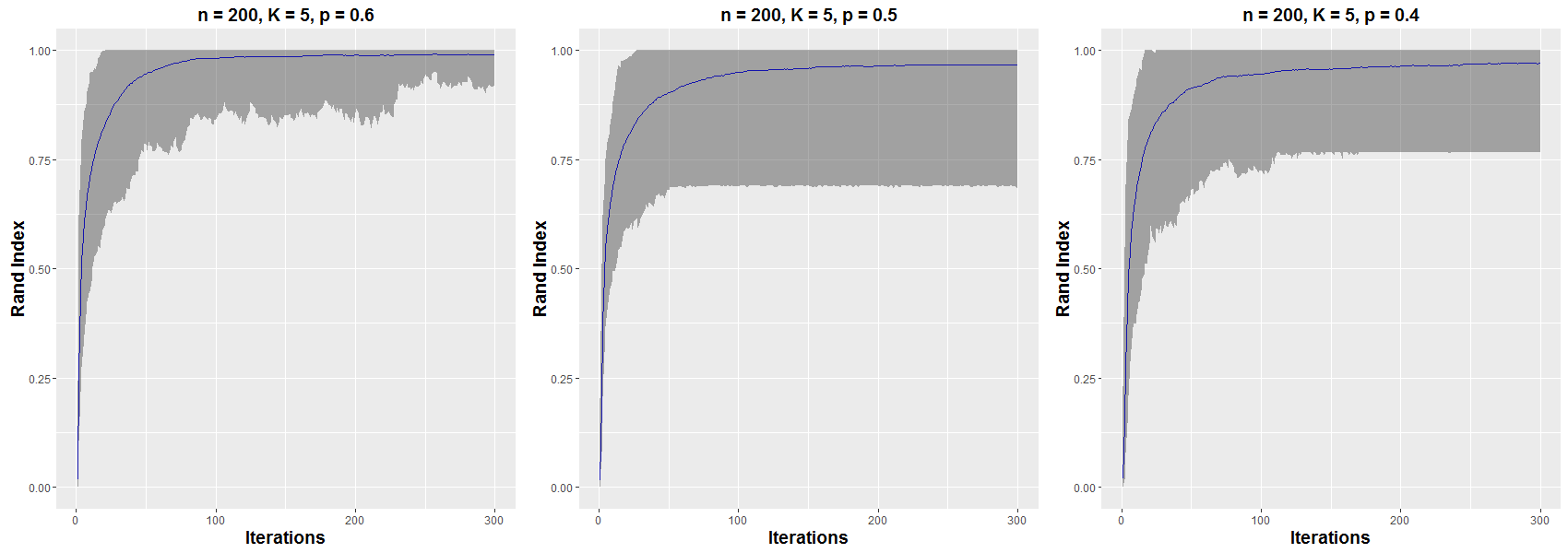}
	\caption{{\em Average Rand index (solid blue line) vs. MCMC iteration for MFM-SBM with 100 different starting configurations in an unbalanced network. $n = 200$ nodes in $K = 5$ communities of sizes 20, 30, 40, 50 and 60. }
}
\label{fig:s5}
\end{figure}

Figures \ref{fig:s1} -- \ref{fig:s5} show average value of $\mathrm{RI}(z, z_0)$ for the first 300 MCMC iterations from 100 randomly chosen starting configurations for the MFM-SBM algorithm.   In each figure, the block structure gets increasingly vague as one moves from the left to the right. It can be readily seen from Figures \ref{fig:s1} and \ref{fig:s4} that for balanced networks with sufficient number of nodes per community, the Rand index rapidly converges to $1$ or very close to $1$ within $300$ MCMC iterates, indicating rapid mixing and convergence of the chain. 
The convergence is somewhat slowed down if the network is unbalanced and the block structure is vague; see for example, the right-most panel of Figures \ref{fig:s3}. However, with a clearer block structure or more nodes available per community, the convergence improves; see the left two panels of Figures \ref{fig:s2} and \ref{fig:s3} and the right most panel of Figure \ref{fig:s5}. We additionally conclude from Figure \ref{fig:s3} - \ref{fig:s5} that as the number of community increases, we need more nodes per community to get precise recovery of the community memberships. 


\begin{figure}[htp!]
	\centering
	\includegraphics[width=0.9\textwidth]{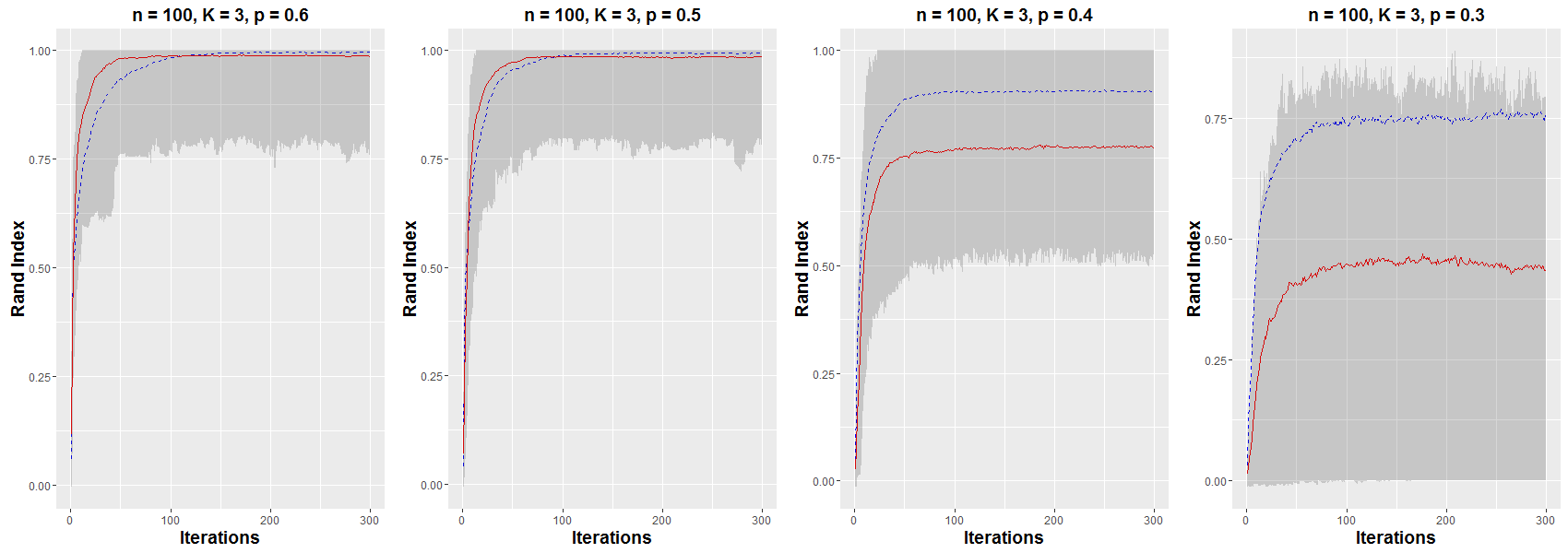}
	\caption{{\em Average Rand index vs. MCMC iteration for the MH-MCMC of \citep{newman2016estimating} with 100 different starting configurations in a balanced network (solid red line). $n = 100$ nodes in $K = 3$ communities of sizes 33, 33 and 34. 
	The  shaded regions correspond to the variation of the Rand index obtained from MH-MCMC due to random initializations. The average Rand index for MFM-SBM with 100 different starting configurations is additionally provided for comparison (dashed blue line).  }}
\label{fig:newman}
\end{figure}

We also found in the more complicated cases (e.g., right panels of Figure \ref{fig:s1}), MH-MCMC (Figure \ref{fig:newman}) does not converge as fast as our approach.

\subsection{ {\bf Mixing of the MCMC chain for $Q$ }} \label{sec:Q}

We report the results based on the simulated datasets in Figure \ref{fig:s1} with 100 nodes, 3 communities in equal sizes and different diagonal values $p$ for $Q$. The average effective sample sizes for the 250 MCMC iterations (leaving out first 50 MCMC iterations as burn-in) across 100 randomly chosen starting configurations are 252 for $p = 0.4$; 243 for $p = 0.5$ and 235 for $p = 0.6$. The reported effective sample size here is an average of element-wise effective sample sizes for all terms in matrix $\theta$. The effective sample sizes are very close to the number of MCMC iterations.    We also display the trace plots for several representative elements of the matrix $\theta$ based on simulated datasets in Figure \ref{fig:s1}. 
\begin{figure}[htp!]
	\centering
	\includegraphics[width=0.9\textwidth]{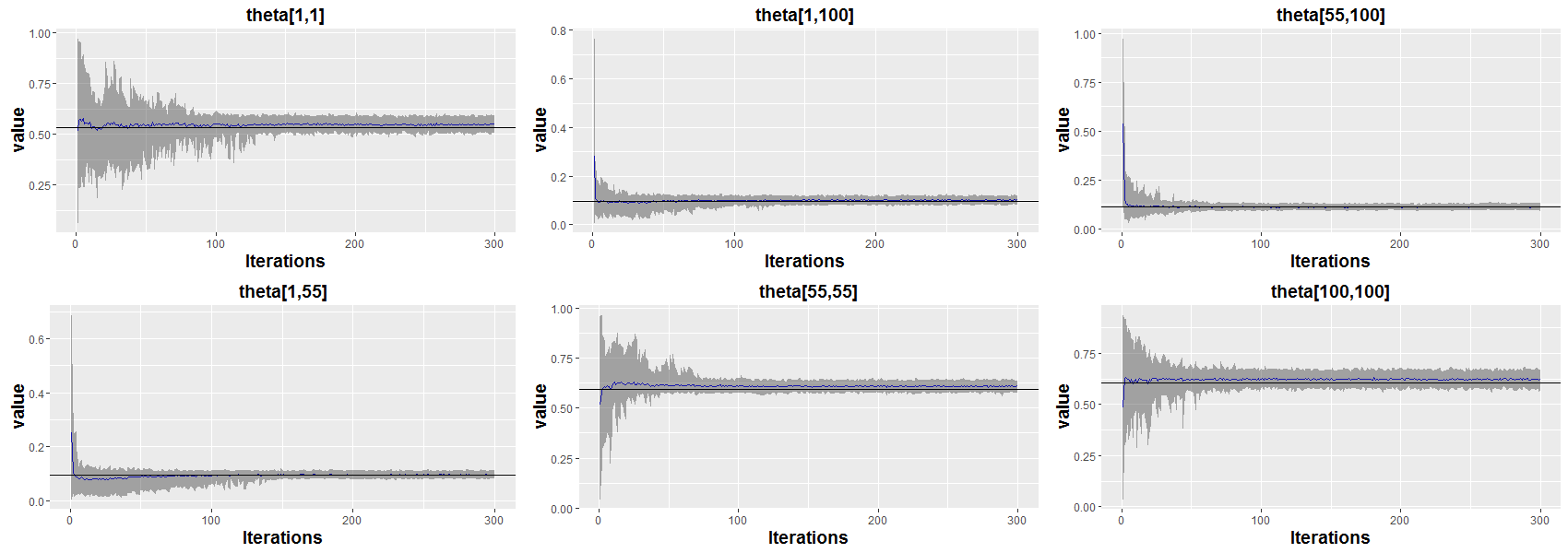}\\
	\caption{ {\em $\theta_{ij}$'s averaged across 100 different initializations vs. MCMC iteration for MFM-SBM  in a balanced network. $n = 100$ nodes in $K = 3$ communities of sizes 33, 33 and 34; $ p = 0.6$. The  shaded regions correspond to the variation of the MCMC sample due to random initializations. }}
	\label{fig:q1}	
\end{figure} 

\begin{figure}[htp!]
	\centering
	\includegraphics[width=0.9\textwidth]{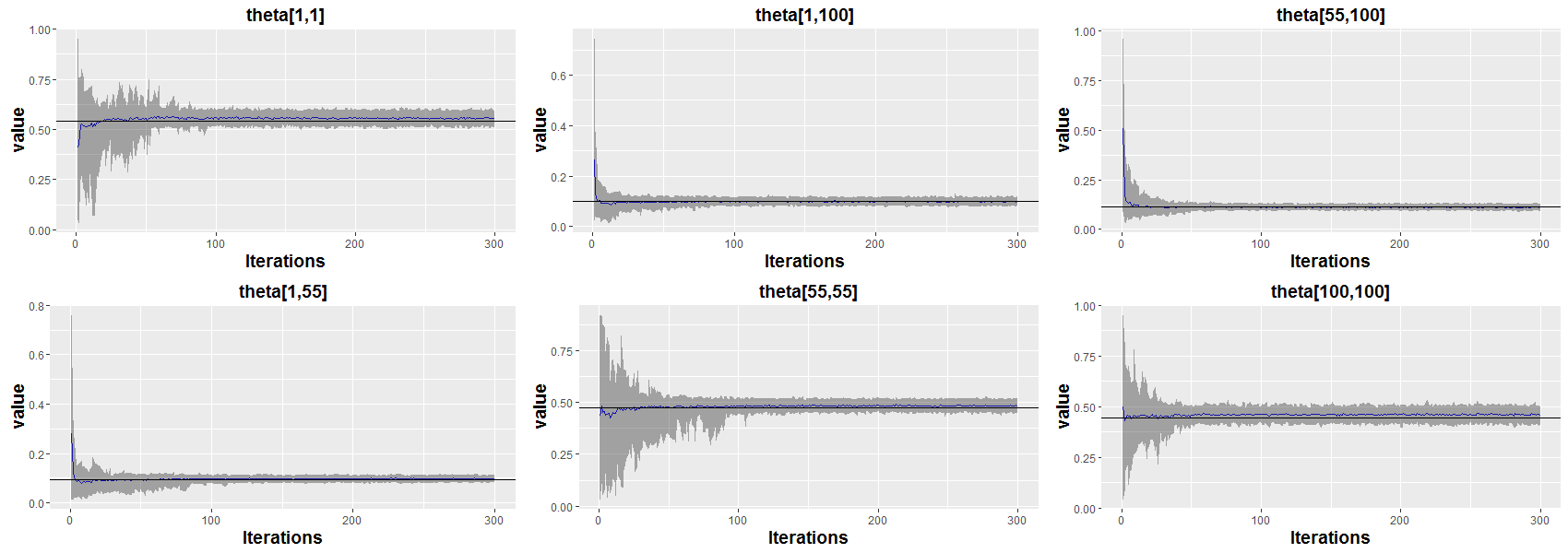}\\
	\caption{{\em $\theta_{ij}$'s averaged across 100 different initializations vs. MCMC iteration for MFM-SBM in a balanced network. $n = 100$ nodes in $K = 3$ communities of sizes 33, 33 and 34; $ p = 0.5$. }}
	\label{fig:q2}	
\end{figure} 

\begin{figure}[htp!]
	\centering
	\includegraphics[width=0.9\textwidth]{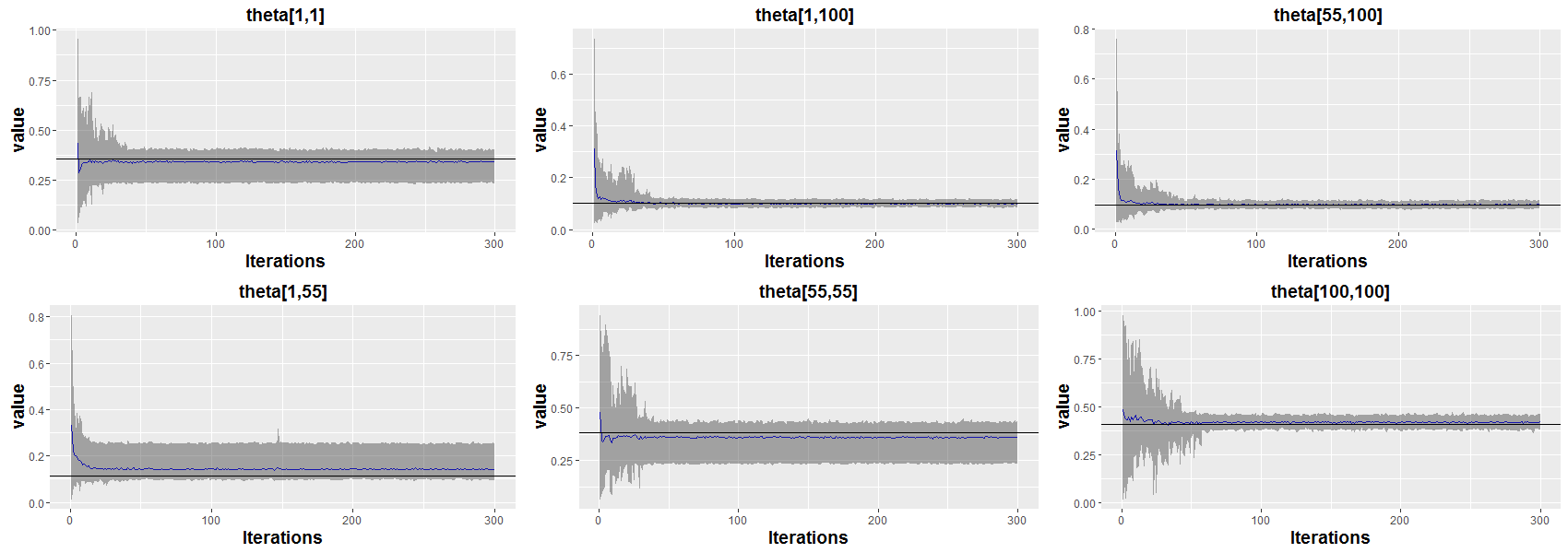}\\
	\caption{ {\em $\theta_{ij}$'s averaged across 100 different initializations vs. MCMC iteration for MFM-SBM in a balanced network. $n = 100$ nodes in $K = 3$ communities of sizes 33, 33 and 34; $ p = 0.4$. } }
	\label{fig:q3}	
\end{figure} 
Figures \ref{fig:q1} to \ref{fig:q3} depict traceplots for some representative $\theta_{ij}$s averaged over 100 initializations  for the first 300 MCMC iterations. The reference line in each subplot is the true value of the representative element based on the true clustering configuration. It is evident that $\theta_{ij}$s rapidly converge to the stationary distributions tightly centered around the true values.

\section{{\bf Community detection in Books about US Politics data}} \label{sec:uspb}
We now provide details of the second real dataset mentioned in \S6 of the main document. We consider a network of books about US politics sold by the online bookseller Amazon.com \citep{Newman2006politics}.  In this network the vertices represent 105 recent books on American politics bought from Amazon, and edges join pairs of books that are frequently purchased by the
same buyer. Books were divided according to their stated or apparent political alignment, liberal or conservative, except for a small number of books that were explicitly bipartisan or centrist, or had no clear affiliation. This is a undirected network data with 105 nodes. 
   
Results from MFM-SBM is again based on 10,000 MCMC iterations leaving out a burn-in of 4,000, initialized at a randomly generated configuration with 9 clusters. Both $\text{Beta}(2,2)$ and $\text{Beta}(1,1)$ priors on the elements of $Q$ are investigated here.
\begin{table}
\begin{center}
\begin{tabular}{|c|c|c|c|c|c|c|}
\hline
Method & {\bf MFM-SBM} & \bf{NBM} & {\bf BHM} &  {\bf LEM} & {\bf HMM} & {\bf MH-MCMC} \\ \hline
   Number of clusters & 5 & 3  & 3  & 4 & 4 & 6   \\ \hline
\end{tabular}
\end{center}
\caption{ {\em Estimated number of clusters for US Politics data} }
\label{tab:pol}
\end{table}
From Table \ref{tab:pol2} and Table \ref{tab:pol3},  both LEM and HMM find two large clusters consisting of mainly liberal or conservative books respectively (refer to cluster 3$\&$4 in table \ref{tab:pol2} and cluster 2$\&$4 in table \ref{tab:pol3}). The remaining nodes of the two clusters in these two clustering configurations consist of books from different categories.  

Among two prior choices in MFM-SBM, $\text{Beta}(2,2)$ prior on the elements of $Q$ provide a more interpretable result. From Table \ref{tab:pol1} (MFM-SBM), we find one cluster (cluster 5) consisting of books from different categories. The remaining four clusters form two large clusters consisting of mainly liberal (cluster 1$\&$3) or conservative (cluster 2$\&$4) books respectively. It is also interesting to observe ``core-periphery'' structure \cite{borgatti2000models} in those four clusters.  From that heatmap of $Q$ in Figure \ref{fig:pol2}, it is evident that there are two core clusters  surrounded by another cluster with sparse within group connections.  This structure reveals that the books in the core parts are popular books most frequently purchased by the same buyer; while the books in the peripheral region are more likely to be purchased by the same buyer more specific to his interests.  Both MFM-SBM with $\text{Beta}(1,1)$ prior and MH-MCMC reveals 6 clusters with  similar ``core-periphery'' structure.  
\begin{figure}[htp!]
	\centering
	\includegraphics[width=1\textwidth]{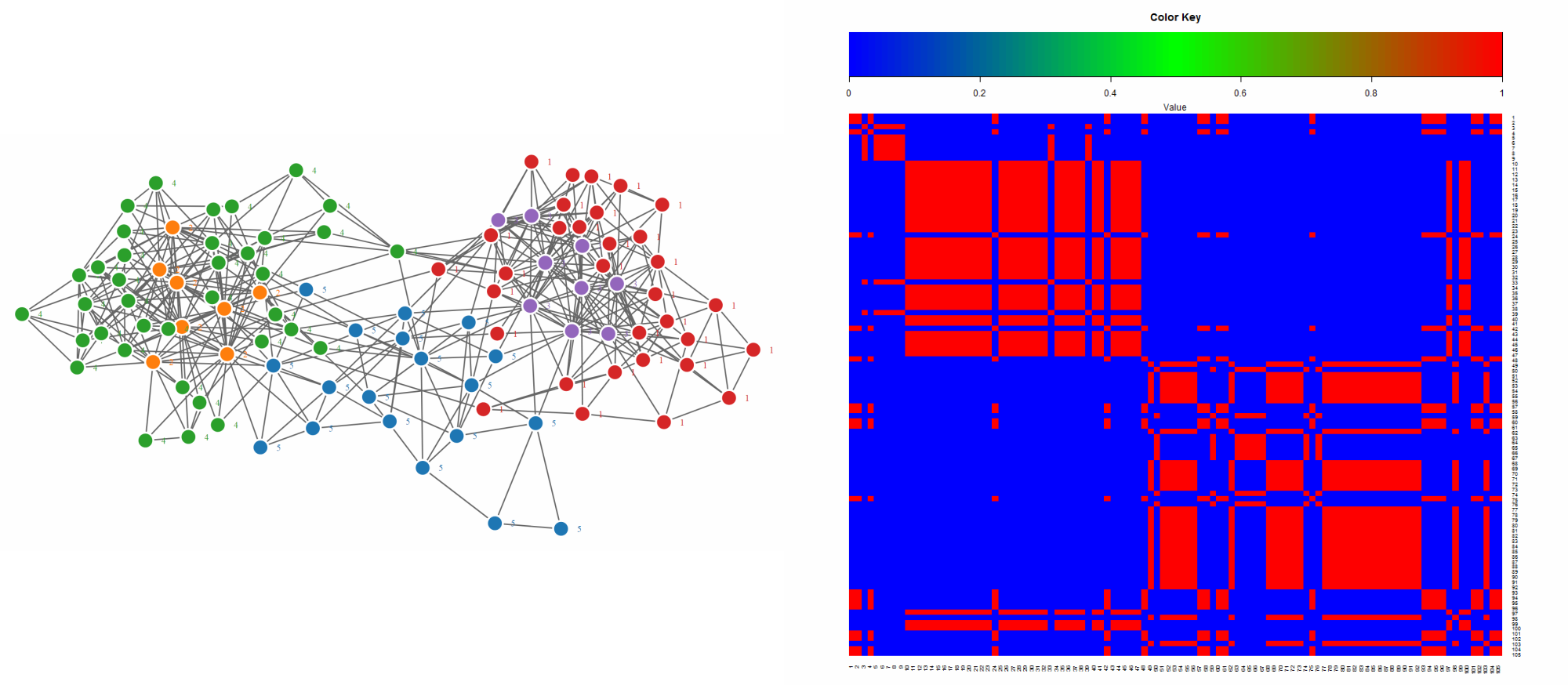}
	\caption{ {\em Estimated configuration for the US Politics books data using MFM-SBM with $\text{Beta}(2,2)$ prior on the elements of $Q$ }}
\label{fig:pol1}
\end{figure}

\begin{figure}[htp!]
	\centering
	\includegraphics[width=0.7\textwidth]{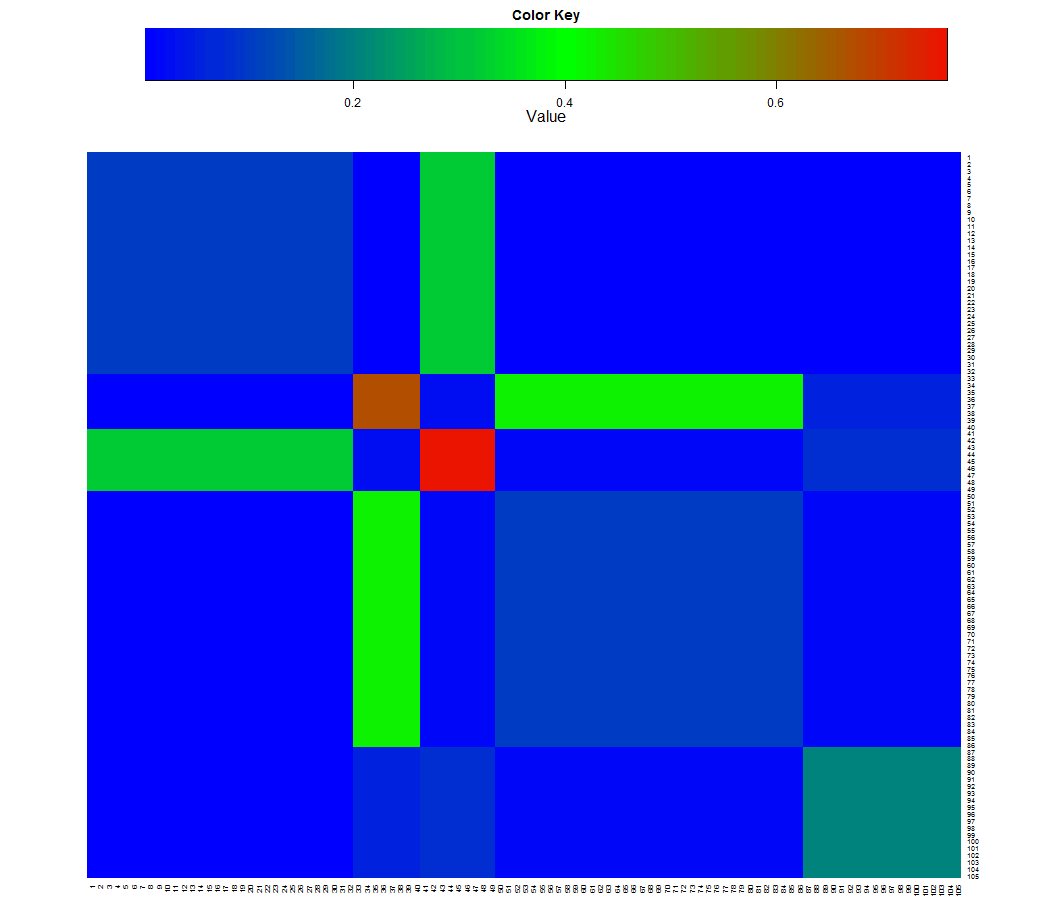}
	\caption{{\em Heatmap for $Q$ matrix for the US Politics books data using MFM-SBM with $\text{Beta}(2,2)$ prior on the elements of $Q$} }
\label{fig:pol2}
\end{figure}

\begin{figure}[htp!]
	\centering
	\includegraphics[width=1\textwidth]{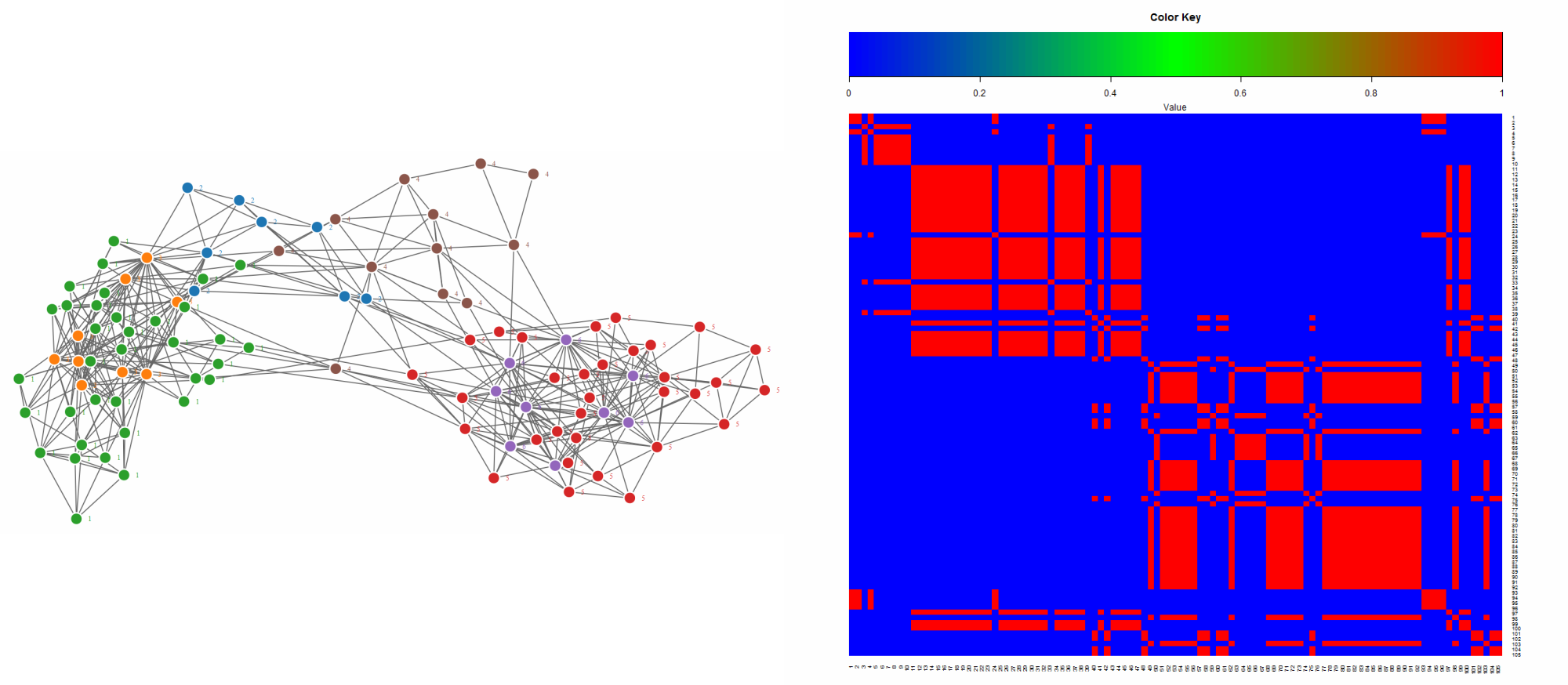}
	\caption{ {\em Estimated configuration for the US Politics books data using MFM-SBM with $\text{Beta}(1,1)$ prior on the elements of $Q$ }}
\label{fig:pol11}
\end{figure}

\begin{figure}[htp!]
	\centering
	\includegraphics[width=0.7\textwidth]{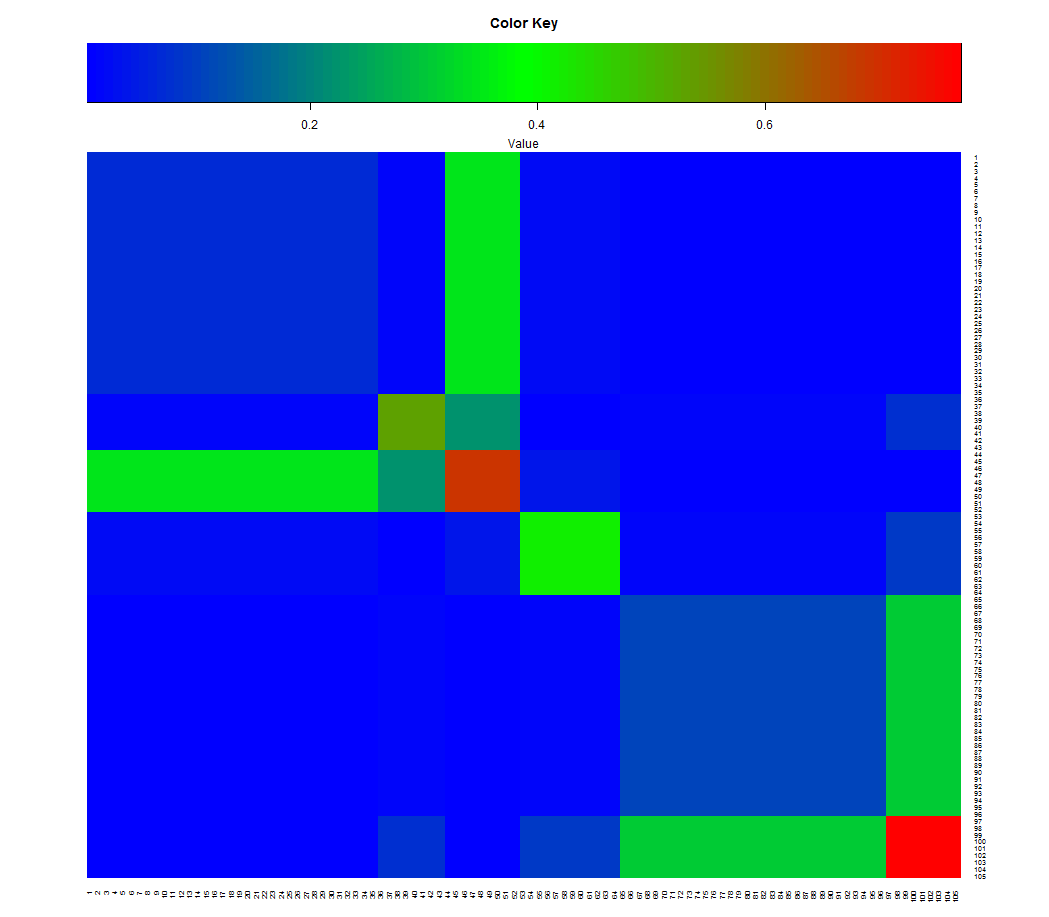}
	\caption{ {\em Heatmap for $Q$ matrix for the US Politics books data using MFM-SBM with $\text{Beta}(1,1)$ prior on the elements of $Q$}}
\label{fig:pol12}
\end{figure}

\begin{figure}[htp!]
	\centering
	\includegraphics[width=1\textwidth]{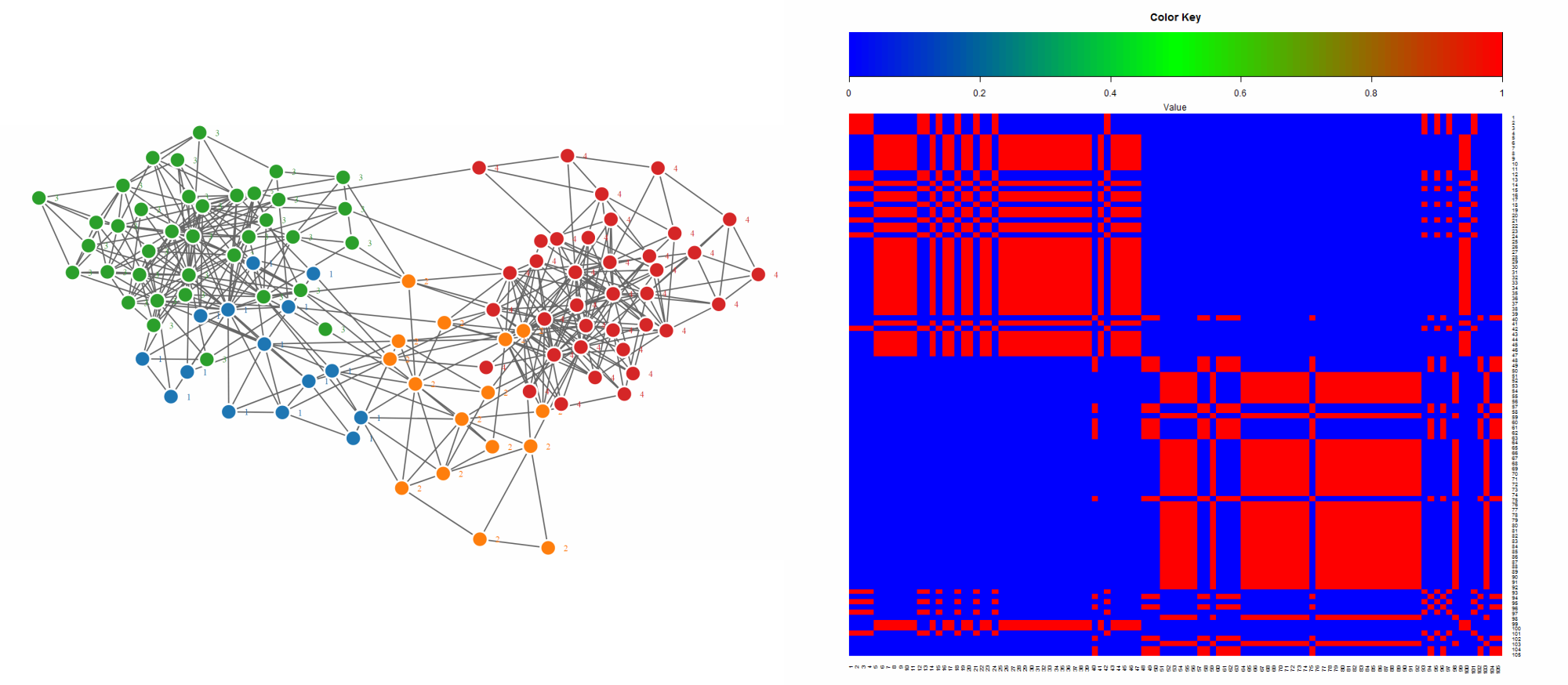}
	\caption{{\em Estimated configuration for the US Politics books data using LEM}
}	
\label{fig:polcomp1}
\end{figure}

\begin{figure}[htp!]
	\centering
	\includegraphics[width=1\textwidth]{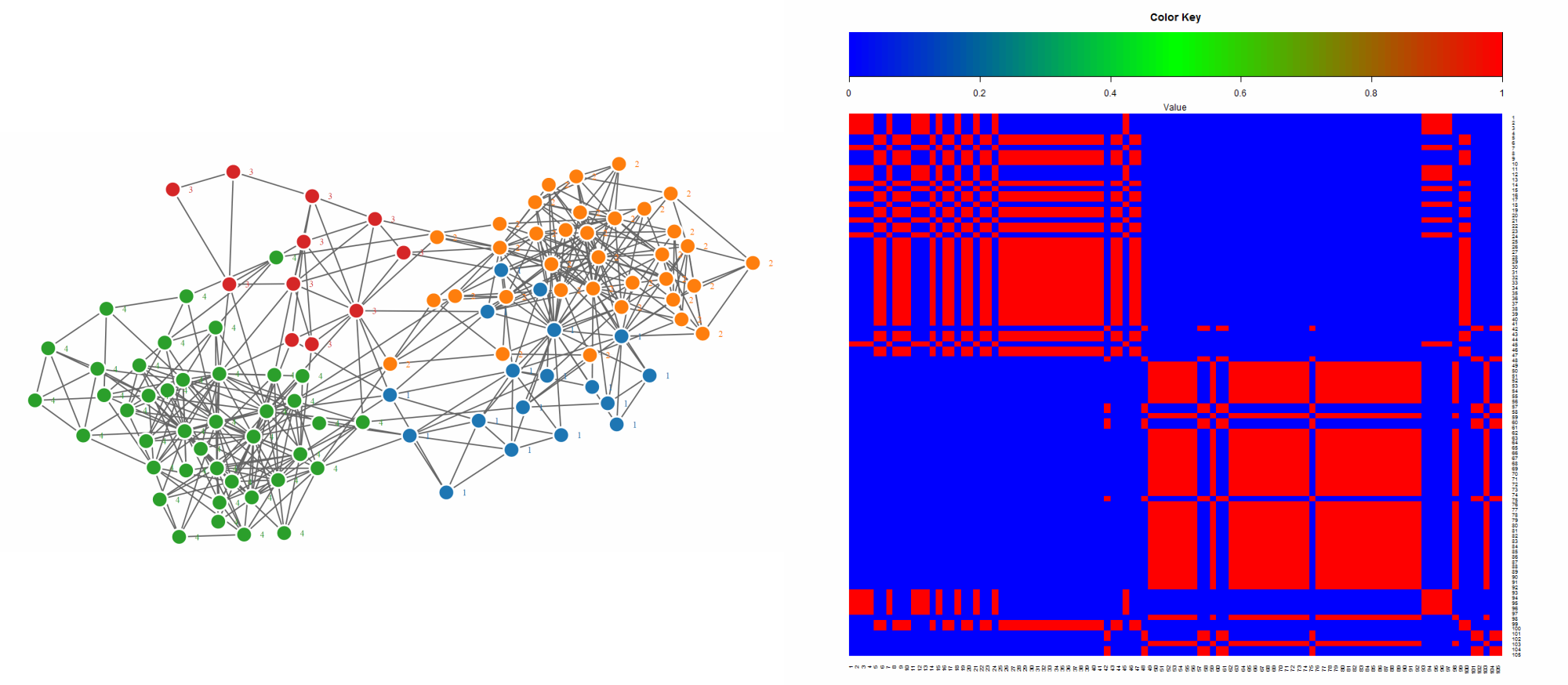}
	\caption{{\em Estimated configuration for the US Politics books data using HMM}
}	
\label{fig:polcomp2}
\end{figure}

\begin{figure}[htp!]
	\centering
	\includegraphics[width=1\textwidth]{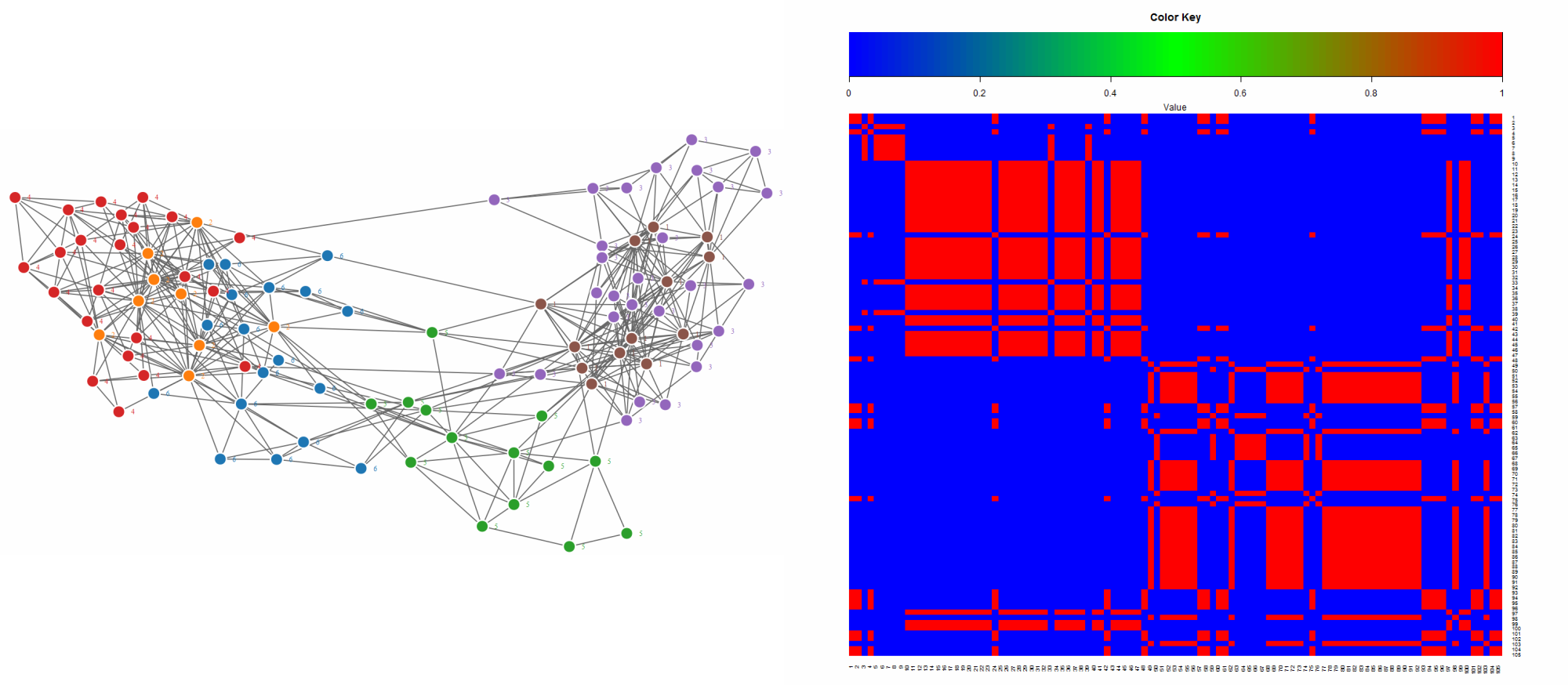}
	\caption{{\em Estimated configuration for the US Politics books data using MH-MCMC}}
\label{fig:polcomp3}
\end{figure}

The modularity based approaches (LEM and HMM) in the \texttt{igraph} package could not  find the core-periphery structure as shown in Figure \ref{fig:polcomp1} and Figure \ref{fig:polcomp2} respectively.  The heatmaps in Figures \ref{fig:pol1}, \ref{fig:pol11}, \ref{fig:polcomp1}, \ref{fig:polcomp2} and \ref{fig:polcomp3} are obtained after rearranging the nodes in order of the clusters corresponding to conservatives, liberal and neutral.

\begin{table}
\begin{center}
\begin{tabular}{|c|c|c|c|c|c|}
\hline
MFM-SBM & {\bf Cluster 1} & \bf{Cluster 2} & {\bf Cluster 3} &  {\bf Cluster 4} & {\bf Cluster 5} \\ \hline
liberal & 29 & 0  & 9  & 0 & 5   \\ \hline
conservative & 1 & 8  & 0  & 34 & 6   \\ \hline
neutral & 2 & 0  & 0  & 3 & 8   \\ \hline

\end{tabular}
\end{center}
\caption{{\em Contingency table of cluster index and book categories using MFM-SBM with $\text{Beta}(2,2)$ priors on the elements of $Q$}  }
\label{tab:pol1}
\end{table}

\begin{table}
\begin{center}
\begin{tabular}{|c|c|c|c|c|c|c|}
\hline
MFM-SBM & {\bf Cluster 1} & \bf{Cluster 2} & {\bf Cluster 3} &  {\bf Cluster 4} & {\bf Cluster 5}  & {\bf Cluster 6}\\ \hline
liberal & 0 & 0  & 0  & 5 & 29 &  9 \\ \hline
conservative & 32 & 4  & 9  & 3 & 1  & 0 \\ \hline
neutral & 3 & 4  & 0  & 4 & 2 & 0 \\ \hline

\end{tabular}
\end{center}
\caption{{\em Contingency table of cluster index and book categories using MFM-SBM with $\text{Beta}(1,1)$ priors on the elements of $Q$}  }
\label{tab:pol11}
\end{table}

\begin{table}
\begin{center}
\begin{tabular}{|c|c|c|c|c|c|}
\hline
LEM & {\bf Cluster 1} & \bf{Cluster 2} & {\bf Cluster 3} &  {\bf Cluster 4} \\ \hline
liberal & 0 & 8  & 0  & 35   \\ \hline
conservative & 11 & 3  & 35  & 0   \\ \hline
neutral & 4 & 5  & 2  & 2   \\ \hline
\end{tabular}
\end{center}
\caption{{\em Contingency table of cluster index and book categories using LEM}  }
\label{tab:pol2}
\end{table}

\begin{table}
\begin{center}
\begin{tabular}{|c|c|c|c|c|c|}
\hline
HMM & {\bf Cluster 1} & \bf{Cluster 2} & {\bf Cluster 3} &  {\bf Cluster 4}\\ \hline
liberal & 0 & 0  & 5  & 38    \\ \hline
conservative & 13 & 33  & 2  & 1    \\ \hline
neutral & 5 & 2  & 4  & 2   \\ \hline
\end{tabular}
\end{center}
\caption{{\em Contingency table of cluster index and book categories using HMM}  }
\label{tab:pol3}
\end{table}

\begin{table}
\begin{center}
\begin{tabular}{|c|c|c|c|c|c|c|c|}
\hline
MH-MCMC & {\bf Cluster 1} & \bf{Cluster 2} & {\bf Cluster 3} &  {\bf Cluster 4} &  {\bf Cluster 5} &  {\bf Cluster 6}\\ \hline
liberal & 12 & 0  & 26  & 0 & 5 & 0   \\ \hline
conservative & 0 & 9  & 1  & 22 & 3 & 14    \\ \hline
neutral & 1 & 0  & 1  & 1 & 6 & 4 \\ \hline
\end{tabular}
\end{center}
\caption{{\em Contingency table of cluster index and book categories using MH-MCMC} }
\label{tab:pol4}
\end{table}

\newpage
\section{\textbf{Proof of Theorem 4.1}}\label{sec:pfmt} 

\noindent {\bf Marginal likelihood approximation and prior-ratio bound:} \\[1ex]
The posterior expected risk $E [ d(z, z_0) \mid \m A ] = \sum_{r} r P[ d(z, z_0) = r \mid \m A]$. Recall that $\m Z_{n, K}$ denotes the space of all cluster configurations of $n$ objects into $K$ groups, $\Pi$ denotes a prior distribution on $\m Z_{n, K}$, and $z_0$ denotes the true configuration. We have 
\begin{align*}
P[ d(z, z_0) = r \mid \m A ] = \frac{\sum_{z: d(z, z_0) = r} \m L(\m A \mid z) \Pi(z)}{ \sum_{z \in \m Z_{n, K}} \m L(\m A \mid z) \Pi(z) } = \frac{\sum_{z: d(z, z_0) = r} \exp \{ \ell(z) - \ell(z_0) + \Pi_{\ell}(z, z_0) \}   }{ \sum_{z \in \m Z_{n, K}} \exp \{ \ell(z) - \ell(z_0) + \Pi_{\ell}(z, z_0) \}  },
\end{align*}
where recall $\ell(z) = \log \m L(\m A \mid z)$ is the log-marginal likelihood of cluster configuration $z$, and $\Pi_{\ell}(z, z_0) := \log \{ \Pi(z)/\Pi(z_0) \}$. Since $\sum_{z \in \m Z_{n, K}} \exp \{ \ell(z) - \ell(z_0) + \Pi_{\ell}(z, z_0) \} \ge 1$, we can bound
\begin{align}\label{eq:BR_bd}
E [ d(z, z_0) \mid \m A ]  \le \sum_r r \sum_{z: d(z, z_0) = r} \exp \{ \ell(z) - \ell(z_0) + \Pi_{\ell}(z, z_0) \}. 
\end{align}
%We first state a bound on $\Pi_{\ell}(z, z_0)$. \tcr{TBD.} 
%\\[4ex]
Next, we approximate the log-marginal likelihood $\ell(z)$ by a more manageable quantity, quantifying the approximation error. Recall the expression for $\m L(\m A \mid z)$ from \eqref{eq:marglik} in the main document. To handle the combinatorial term, we use the well-known approximation $\log {N \choose s} \approx - N \m H(s/N)$ (see, e.g., Chapter 1 of \cite{mackay2003information}), where $\m H:[0, 1] \to \mb R$ is the (negative) Binary entropy function given by $\m H(x) = x \log x + (1-x) \log (1-x)$.  

In fact, using the two-sided Stirling bound $\sqrt{2 \pi} N^{N+1/2} e^{-N} \le N ! \le e N^{N+1/2} e^{-N}$, it is straightforward to verify that 
$$
\abs{ \log {N \choose s} - \big(- N \m H(s/N) \big) } \le C \log N,
$$
where $C$ is a global constant independent of $s$ and $N$. Note that $\m H(x) < 0, \m H'(x) = \log \{x/(1-x)\} = \mbox{logit}(x)$ and $\m H''(x) = [x(1-x)]^{-1}$ for all $x \in (0, 1)$. In particular, the positivity of the second derivative of $\m H$ implies that $\m H$ is convex over $(0, 1)$, a fact which is crucial to our subsequent derivations. 

Using the above approximation and that $n_{\up}(z), n_{\down}(z) \le n^2$, we can write $\ell(z) = \wt{\ell}(z) + \ell_R(z)$, where 
\begin{align}\label{eq:ltilde}
\wt{\ell}(z) = n_{\up}(z) \m H \bigg\{ \frac{A_{\up}(z)}{n_{\up}(z)} \bigg\} 
+ n_{\down}(z) \m H \bigg\{ \frac{A_{\down}(z)}{n_{\down}(z)} \bigg\}, 
\end{align}
with the remainder term $|\ell_R(z)| \le C \log n$ for a global constant $C$ independent of $z$ and $n$. 

Putting together the various approximations, we have from \eqref{eq:BR_bd} that 
\begin{align}\label{eq:BR_bd1}
E [ d(z, z_0) \mid \m A ]  \le \sum_r r \sum_{z: d(z, z_0) = r} \exp \{ \wt{\ell}(z) - \wt{\ell}(z_0) + \Delta(z, z_0) \}, 
\end{align}
where $\Delta(z, z_0) = \ell_R(z) - \ell_R(z_0) + \Pi_l(z, z_0)$. Since $|\Pi_l(z, z_0)| \le C K d(z, z_0)$ by assumption, we have
$|\Delta(z, z_0) | \le C \max\{K d(z, z_0), \log n\}$ for all $z$.  
%We first state a Lemma which bounds the log prior differences. 
%\begin{lemma}\label{lem:prior_diff}
%Suppose $d(z, z_0) = r$. 
%\end{lemma}
We subsequently aim to bound $\wt{\ell}(z) - \wt{\ell}(z_0)$ from above inside a large $\bbP$-probability set. {\em The following result is key to our derivations.} 
\begin{proposition}\label{prop:main}
Fix $\nu > 1$. There exists a set $\m C$ with $\bbP (\m C) \ge 1 - e^{-C (\log n)^{\nu}}$, such that for any $\m A \in \m C$, we have 
\begin{align}\label{eq:marglik_diff_main}
\wt{\ell}(z_0) - \wt{\ell}(z) \ge \frac{ C \bar{D}(p_0, q_0) \ n \ d(z, z_0)} {K}
\end{align}
for all $z \in \m Z_{n, K}$, where recall that % this needs to be defined in main document as well
\begin{align}\label{eq:bar_D}
\bar{D}(p_0, q_0) := \frac{ (p_0 - q_0)^2}{ (p_0 \vee q_0) \{1 - (p_0 \wedge q_0)\}}.
\end{align}
%with $\vee$ and $\wedge$ denoting maximum and minimum respectively. 
\end{proposition}

Proposition \ref{prop:main} quantifies the difference between the (approximate) log-marginal likelihood of the true configuration $\wt{\ell}(z_0)$ and that of any other configuration $\wt{\ell}(z)$ in terms of $d(z, z_0)$, the sample size $n$, the number of communities $K$, and the quantity $\bar{D}(p_0, q_0)$. The proof of Proposition \ref{prop:main} is long and hence deferred to the next subsection. Substituting the bound \eqref{eq:marglik_diff_main} from Proposition \ref{prop:main} in \eqref{eq:BR_bd1} and using the crude bound $| \{ z \in \m Z_{n, K} : d(z, z_0) = r \} | \le K^r {n \choose r}$, we obtain, inside the set $\m C$, 
\begin{align*}
E [ d(z, z_0) \mid r ] \le \sum_r r {n \choose r} K^r \exp \bigg\{ - \frac{C \bar{D}(p_0, q_0) \ n r}{K} + C \max\{K r, \log n\} \bigg\} \le e^{-  \frac{C \bar{D}(p_0, q_0) n }{K} },
\end{align*}
where the second inequality uses the crude bound ${n \choose r} \lesssim e^{r \log n}$ and the geometric sum formula. This establishes Theorem \ref{thm:clus_cons}. 
\\[4ex]
\subsection*{{\bf Proof of Proposition \ref{prop:main}} }
%\\1[ex]
We now provide a running proof of Proposition \ref{prop:main}. We break the proof up into several parts which are somewhat independent of each other for improved readability. We first introduce some useful notation and collect some concentration inequalities. The concentration inequalities are used to define the large $\bbP$-probability set $\m C$ in \eqref{eq:bigset}. The final part of the proof bounds $\wt{\ell}(z_0) - \wt{\ell}(z)$ inside $\m C$. Readers primarily interested in the bound for the log-marginal likelihood difference can skip directly to the final part after familiarizing with the new notations. 
\\[2ex]
\noindent {\bf Additional Notation:} \\[1ex]
%Recall the quantities $A_{\up}(z), A_{\down}(z), n_{\up}(z), n_{\down}(z)$ from the main document. 
%Let us denote $q_{\uparrow} = p_0$ and $q_{\downarrow} = q_0$. 
For $z, z' \in \m Z_{n, K}$, define
\begin{eqnarray*}
%\alpha(z; z') = | \{ (i, j) : i < j, z_i \ne z_j, z'_i = z'_j \} |, \quad \gamma(z; z') =  | \{ (i, j) : i < j, z_i = z_j, z'_i \ne z'_j \} |.
n_{\up \up}(z,z') = \sum_{i < j} \ind(z_i = z_j, z_i' = z_j'), \quad A_{\up \up}(z, z') = \sum_{i < j} a_{ij} \ind(z_i = z_j, z_i' = z_j'), \\
n_{\up \down}(z,z') = \sum_{i < j} \ind(z_i = z_j, z_i' \ne z_j'), \quad A_{\up \down}(z, z') = \sum_{i < j} a_{ij} \ind(z_i = z_j, z_i' \ne z_j'), \\
n_{\down \up}(z,z') = \sum_{i < j} \ind(z_i \ne z_j, z_i' = z_j'), \quad A_{\down \up}(z, z') = \sum_{i < j} a_{ij} \ind(z_i \ne z_j, z_i' = z_j'), \\
n_{\down \down}(z,z') = \sum_{i < j} \ind(z_i \ne z_j, z_i' \ne z_j'), \quad A_{\down \down}(z, z') = \sum_{i < j} a_{ij} \ind(z_i \ne z_j, z_i' \ne z_j').
\end{eqnarray*}
To simplify notation, we shall subsequently use $\dagger$ and $\dagger'$ as dummy variables taking values in the set $\{\uparrow, \downarrow\}$.\footnote{For example, $\sum_{\dagger} n_{\dagger}(z)$ is shorthand for $n_{\uparrow}(z) + n_{\downarrow}(z)$.} With this notation, $n_{\dagger}(z) = \sum_{\dagger, \dagger'} n_{\dagger \dagger'}(z, z')$ and $A_{\dagger}(z) = \sum_{\dagger, \dagger'} A_{\dagger \dagger'}(z, z')$ for any $z, z' \in \m Z_{n, K}$. Denoting $\xi_{\uparrow} = p_0$ and $\xi_{\downarrow} = q_0$, we have %for any $\dagger, \dagger'$, 
\begin{align}\label{eq:dag_dist}
A_{\dagger \dagger'}(z, z_0) \sim \mbox{Binomial}(n_{\dagger \dagger'}(z, z_0), \xi_{\dagger'}),
\end{align}
independently across $\dagger, \dagger'$. 
For any $\dagger, \dagger'$, additionally denote
\begin{align}
& X_{\dagger} = \frac{ A_{\dagger}(z) }{ n_{\dagger}(z) }, \quad Y_{\dagger} = \frac{ A_{\dagger}(z_0) }{ n_{\dagger}(z_0) }, \quad W_{\dagger \dagger'} = \frac{ A_{\dagger \dagger'}(z, z_0) }{ n_{\dagger \dagger'}(z, z_0) } \label{eq:dag_def1} \\
& \omega_{\dagger \dagger'} = \frac{ n_{\dagger \dagger'}(z, z_0) }{ n_{\dagger}(z) }, \quad \wt{\omega}_{\dagger \dagger'} = \frac{ n_{\dagger \dagger'}(z, z_0) }{ n_{\dagger'}(z_0) }. \label{eq:dag_def2}
\end{align}
It is straightforward to verify that 
\begin{align*}
& \sum_{\dagger'} \omega_{\dagger \dagger'} = 1, \quad X_{\dagger} = \sum_{\dagger'} \omega_{\dagger \dagger'} W_{\dagger \dagger'}, \\
& \sum_{\dagger} \wt{\omega}_{\dagger \dagger'} = 1, \quad Y_{\dagger'} = \sum_{\dagger} \wt{\omega}_{\dagger \dagger'} W_{\dagger \dagger'}. 
\end{align*}
It is evident from \eqref{eq:dag_dist} that $\bbE W_{\dagger \dagger'} = \xi_{\dagger'}, \bbE Y_{\dagger'} = \xi_{\dagger'}$ and $\bbE X_{\dagger} = \bar{\xi}_{\dagger} := \sum_{\dagger'} \omega_{\dagger \dagger'} \xi_{\dagger'}$. Further, since the random variables involved are sub-Gaussian, they concentrate around their mean with large probability. We collect some useful concentration bounds next. 
\\[2ex]
{\bf Concentration bounds:} Fix $z \ne z_0 \in \m Z_{n, K}$ with $d(z, z_0) = r$. %Define $r^* = \min\{r, n/2 - r\}$. 
For a constant $\nu > 1$, let
\begin{align}
& \m C_{X}(z) = \bigg\{ |X_{\dagger} - \bar{\xi}_{\dagger} | \le \frac{ (\log n)^{\nu/2} \sqrt{r} }{ \sqrt{n_{\dagger}(z)} }, \, \forall \, \dagger \bigg\} \label{eq:CX_z} \\
& \m C_{Y}(z) = \bigg\{ |Y_{\dagger} - \xi_{\dagger} | \le \frac{ (\log n)^{\nu/2} \sqrt{r} }{ \sqrt{n_{\dagger}(z_0)} }, \, \forall \, \dagger \bigg\}. \label{eq:CY_z}
\end{align}
For $T_i \sim \mbox{Bernoulli}(p_i)$ independently for $i = 1, \ldots, N$, it follows from Hoeffding's inequality that $P( | \bar{T} - \bar{p} | > t) \le 2 e^{- 2n t^2}$ for any $t > 0$, where $\bar{p} = N^{-1} \sum_{i=1}^N p_i$. Combining with the union bound, it follows that 
\begin{align}\label{eq:conc_C1}
\bbP \big[\m C_{X}(z) \cap \m C_{Y}(z) \big] \ge 1 - 8 \, e^{- r (\log n)^{\nu}}.
\end{align}
We additionally need control on another set of random variables that appear inside Taylor expansions subsequently. Define, for each $\dagger$, 
\begin{align}\label{eq:Ldag}
L_{\dagger} = \sum_{\dagger'} \omega_{\dagger \dagger'} Y_{\dagger'} - X_{\dagger} = \sum_{\dagger'} \omega_{\dagger \dagger'} (Y_{\dagger'} - W_{\dagger \dagger'}). 
\end{align}
For any $\dagger$, define $\ddagger$ to be the reverse spin of $\dagger$, that is, $\ddagger  = \downarrow$ if $\dagger = \uparrow$ and vice versa. With this notation, $Y_{\dagger'} - W_{\dagger \dagger'} = \wt{\omega}_{\dagger \dagger'} W_{\dagger \dagger'} + \wt{\omega}_{\ddagger \dagger'} W_{\ddagger \dagger'} - W_{\dagger \dagger'} = \wt{\omega}_{\ddagger \dagger'} (W_{\ddagger \dagger'} - W_{\dagger \dagger'})$, since $1 - \wt{\omega}_{\dagger \dagger'} = \wt{\omega}_{\ddagger \dagger'}$. 
Substituting in \eqref{eq:Ldag}, 
\begin{align}\label{eq:Ldag1}
L_{\dagger} = \sum_{\dagger'} \omega_{\dagger \dagger'}  \wt{\omega}_{\ddagger \dagger'} (W_{\ddagger \dagger'} - W_{\dagger \dagger'}). 
\end{align}
Observe that $W_{\ddagger \dagger'}$ and $W_{\dagger \dagger'}$ are independent random variables with $\bbE W_{\ddagger \dagger'} = \bbE W_{\dagger \dagger'} = \xi_{\dagger'}$, implying $\bbE L_{\dagger} = 0$. Define 
\begin{align}\label{eq:CL_z}
\m C_L(z) = \bigg\{ \abs{ L_{\dagger} } \le \frac{ C (\log n)^{\nu/2} \sqrt{r} \, \sqrt{\mathbf{n}(z, z_0)} }{ n_{\dagger}(z) }, \, \forall \ \dagger \bigg\}, 
\end{align}
where 
\begin{align}\label{eq:mota_n}
\mathbf{n}(z, z_0) = \frac{ n_{\up \up}(z, z_0) n_{\down \up}(z, z_0) }{ n_{\up}(z_0) } +  \frac{ n_{\up \down}(z, z_0) n_{\down \down}(z, z_0) }{ n_{\down}(z_0) }.
\end{align}
Using a sub-Gaussian concentration inequality, we prove below that 
\begin{align}\label{eq:conc_C2}
\bbP \big[ \m C_L(z) \big] \ge 1 - 6 e^{- r (\log n)^{\nu} }. 
\end{align}
The main idea to establish \eqref{eq:conc_C2} is to recognize $L_{\dagger}$ as a weighted sum of centered Bernoulli variables in \eqref{eq:Ldag1} and use a rotation invariance property of sub-Gaussian random variables to bound the sub-Gaussian norm of the aforesaid random variable. 

Let us recall some useful facts about sub-Gaussian random variables from \S 5.2.3 of \cite{vershynin2010introduction}. A mean zero random variable $Z$ is called sub-Gaussian if $E( e^{t Z}) \le e^{C t^2 \norm{Z}_{\psi_2}^2}$ for all $t \in \mb R$, where $\norm{Z}_{\psi_2} = \sup_{s \ge 1} s^{-1/2} ( E |Z|^s)^{1/s}$ is the sub-Gaussian norm of $Z$ and $C$ is an absolute constant. Sub-Gaussian random variables satisfy Gaussian-like tail bounds: $P(|Z| > t) \le C e^{- c t^2/\norm{Z}_{\psi_2}^2}$, with $C < 3$. The following rotation invariance property is useful: if $Z_1, \ldots, Z_N$ are independent sub-Gaussian random variables, then $Z = \sum_{i=1}^N a_i Z_i$ is also sub-Gaussian, with 
$$
\norm{Z}_{\psi_2}^2 \le C \sum_{i=1}^N a_i^2 \norm{Z_i}_{\psi_2}^2, 
$$
for some absolute constant $C$. %Thus, up to constants, 

Any centered Bernoulli random variable is sub-Gaussian, with sub-Gaussian norm bounded by $1$. Since $L_{\dagger}$ is a weighted sum of Bernoulli random variables, $L_{\dagger}$ is also sub-Gaussian. Let us attempt to bound the sub-Gaussian norm of $L_{\dagger}$. First,  in \eqref{eq:Ldag1}, write $W_{\ddagger \dagger'} - W_{\dagger \dagger'} = (W_{\ddagger \dagger'} - \xi_{\dagger'}) - (W_{\dagger \dagger'} - \xi_{\dagger'})$ as a weighted sum of centered Bernoulli random variables. By rotation invariance, 
$$
\norm{ W_{\ddagger \dagger'} - W_{\dagger \dagger'} }_{\psi_2}^2 \le C \bigg( \frac{1}{n_{\ddagger \dagger'}} + \frac{1}{n_{\dagger \dagger'}} \bigg). 
$$
Another application of rotation invariance yields, 
\begin{align*}
\norm{L_{\dagger}}_{\psi_2}^2 
& \le C \sum_{\dagger'} \omega^2_{\dagger \dagger'}  \wt{\omega}^2_{\ddagger \dagger'} \bigg( \frac{1}{n_{\ddagger \dagger'}(z, z_0)} + \frac{1}{n_{\dagger \dagger'}(z, z_0)} \bigg) \\
& = \frac{C}{n_{\dagger}^2(z)} \sum_{\dagger'} \frac{  n_{\dagger \dagger'}(z, z_0) \ n_{\ddagger \dagger'}(z, z_0) }{ n_{\dagger'}(z_0) } = \frac{C \mathbf{n}(z, z_0)}{ n_{\dagger}^2(z) },
\end{align*}
using the definitions in \eqref{eq:dag_def1} and \eqref{eq:dag_def2} from the first to the second line, and noting that the summation in the penultimate line equals $\mathbf{n}(z, z_0)$ defined in \eqref{eq:mota_n}. 

% Since $n_{\dagger \dagger'}(z, z_0) + n_{\ddagger \dagger'}(z, z_0)  = n_{\dagger'}(z_0)$ and $ab/(a+b) \le \min\{a, b\}$ for $a, b > 0$, we can bound the term inside the sum in the above display by $\min\{ n_{\dagger \dagger'}(z, z_0), n_{\ddagger \dagger'}(z, z_0) \}$. Since $n_{\up \up}(z, z_0) \ge n_{\up \down}(z, z_0)$ and $n_{\down \down}(z, z_0) \ge n_{\down \up}(z, z_0)$, we further obtain
%$$
%\sum_{\dagger'} \frac{  n_{\dagger \dagger'}(z, z_0) \ n_{\ddagger \dagger'}(z, z_0) }{ n_{\dagger'}(z_0) }  \le n_{\up \down}(z, z_0) + n_{\down \up}(z, z_0). 
%$$
%A third application of rotation invariance finally implies 
%\begin{align}
%\nnorm{ \sum_{\dagger} n_{\dagger}(z) L_{\dagger} }_{\psi_2}^2 \le C \big(n_{\up \down}(z, z_0) + n_{\down \up}(z, z_0) \big). 
%\end{align}
From the general tail bound for sub-Gaussian random variables mentioned previously (see paragraph after equation \eqref{eq:conc_C2} ), we have $\bbP( | L_{\dagger} | > t) \le 3 e^{- C t^2/\norm{L_{\dagger}}_{\psi_2}^2}$ for any $t > 0$. Set $t^* = C (\log n)^{\nu/2} \sqrt{r} \, \sqrt{\mathbf{n}(z, z_0)}/n_{\dagger}(z)$ for an appropriate $C$ and use that $e^{-1/x}$ is increasing in $x$ to obtain $\bbP( | L_{\dagger} | > t^*) \le 3 e^{- r (\log n)^{\nu} }$. The inequality \eqref{eq:conc_C2} follows from an application of the union bound over $\dagger$. 
%
%
%The inequality \eqref{eq:conc_C2} now follows from the general tail bound for sub-Gaussian random variables mentioned previously (see paragraph after equation \eqref{eq:conc_C2} ) by choosing an appropriate $C$ in \eqref{eq:CL_z}.
\\[2ex]
{\bf Constructing large probability set:}\\[1ex]
We use the concentration bounds above to create the large probability set $\m C$ in Proposition \ref{prop:main} within which the log-marginal likelihood differences can be appropriately bounded. Define, 
\begin{align}\label{eq:C_r}
\m C_r = \cap_{z: d(z, z_0) = r} \big[ \m C_X(z) \cap \m C_Y(z) \cap \m C_L(z) \big], \quad \m C = \cap_{r=1}^n \m C_r. 
\end{align}
We have, 
\begin{align*}
\bbP\big[ \m C_r^c \big] \le C \abs{z: d(z, z_0) = r} \, e^{- r (\log n)^{\nu}} \le C {n \choose r} K^r e^{- r (\log n)^{\nu}} \le e^{ - C r (\log n)^{\nu} }. 
\end{align*}
For the first inequality in the above display, we used the union bound to \eqref{eq:conc_C1} and \eqref{eq:conc_C2}. The second inequality uses the crude upper bound $\abs{z: d(z, z_0) = r} \le {n \choose r} K^r$, whereas the last inequality uses the bound ${n \choose r} \le e^{r \log n}$ and the fact that $\nu > 1$. Another application of the union bound yields
\begin{align}\label{eq:bigset}
\bbP \big(\m C \big) \ge 1 - e^{-C (\log n)^{\nu}}. 
\end{align}
%We work inside $\m C$ subsequently. 
\\[0.5ex]
{\bf Bounding the log-marginal likelihood differences:} \\[1ex]
Fix $z$ with $d(z, z_0) = r$. Recall the approximation $\wt{\ell}(\cdot)$ to the log-marginal likelihood from  \eqref{eq:ltilde}. We now proceed to bound $\wt{\ell}(z_0) - \wt{\ell}(z)$ from below inside the set $\m C$. Using the notation introduced in \eqref{eq:dag_def1} and \eqref{eq:dag_def2}, we can write 
\begin{align*}
\wt{\ell}(z) = \sum_{\dagger} n_{\dagger}(z) \m H(X_{\dagger}), 
\end{align*}
and 
\begin{align*}
\wt{\ell}(z_0) 
= \sum_{\dagger'} n_{\dagger'}(z_0) \m H(Y_{\dagger'}) 
= \sum_{\dagger'} \sum_{\dagger} n_{\dagger \dagger'}(z, z_0) \m H(Y_{\dagger'}) 
= \sum_{\dagger} n_{\dagger}(z) \bigg[ \sum_{\dagger'} \omega_{\dagger \dagger'} \m H(Y_{\dagger'}) \bigg].
\end{align*}
Thus, $\wt{\ell}(z_0) - \wt{\ell}(z) = \sum_{\dagger} n_{\dagger}(z)  \big[ \sum_{\dagger'} \omega_{\dagger \dagger'} \m H(Y_{\dagger'}) - \m H(X_{\dagger}) \big]$. To tackle the inner sum, we perform a Taylor expansion of each $\m H(Y_{\dagger'})$ around $\m H(X_{\dagger})$.  After some cancellations since $\sum_{\dagger'} \omega_{\dagger \dagger'} = 1$, we obtain
\begin{align}\label{eq:marglikdif}
\wt{\ell}(z_0) - \wt{\ell}(z) 
= \sum_{\dagger} n_{\dagger}(z) \bigg[ \sum_{\dagger'} \omega_{\dagger \dagger'} \bigg\{ (Y_{\dagger'} - X_{\dagger}) \m H'(X_{\dagger}) + \frac{ (Y_{\dagger'} - X_{\dagger})^2}{2} \m H''(U_{\dagger' \dagger}) \bigg\} \bigg],
\end{align}
where $U_{\dagger' \dagger}$ lies between $Y_{\dagger'}$ and $X_{\dagger}$. 

Since $\m H$ is convex, the quadratic term in \eqref{eq:marglikdif} is positive. We show below that the quadratic term is the dominant term and the linear term is of smaller order. To that end, we first bound the magnitude of the linear term inside $\m C$. Since from \eqref{eq:CX_z}, $X_{\dagger}$ concentrates around $\bar{\xi}_{\dagger}$, and $\bar{\xi}_{\dagger}$ lies between $p_0$ and $q_0$, $|\m H'(X_{\dagger})|$ can be bounded by a constant inside $\m C$. Hence, inside $\m C$, 
\begin{align}\label{eq:bd_lin}
\aabs{ \sum_{\dagger} n_{\dagger}(z) \sum_{\dagger'} \omega_{\dagger \dagger'} (Y_{\dagger'} - X_{\dagger}) \m H'(X_{\dagger}) } \le C \sum_{\dagger} n_{\dagger}(z) \abs{ L_{\dagger} } \le C (\log n)^{\nu/2} \sqrt{r} \, \sqrt{\mathbf{n}(z, z_0)}, 
\end{align} 
where recall from \eqref{eq:Ldag} that $L_{\dagger} = \sum_{\dagger'} \omega_{\dagger \dagger'} (Y_{\dagger'} - X_{\dagger})$. From the second to third step, we used the bound on $|L_{\dagger}|$ inside $\m C$ from \eqref{eq:CL_z}. 

Next, we bound from below the quadratic term in \eqref{eq:marglikdif}. Since $U_{\dagger' \dagger}$ lies between $Y_{\dagger'}$ and $X_{\dagger}$ which in turn concentrate around their respective means inside $\m C$, we can bound $H''(U_{\dagger' \dagger})$ from below as follows:
$$
\m H''(U_{\dagger' \dagger}) = \frac{1}{ U_{\dagger' \dagger} (1 - U_{\dagger' \dagger})} \ge \frac{1}{ (p_0 \vee q_0) \{1 - (p_0 \wedge q_0)\} },
$$
where $\vee$ and $\wedge$ respectively denote the maximum and minimum. 
Thus,  
\begin{align*}
\sum_{\dagger} n_{\dagger}(z) \sum_{\dagger'} \omega_{\dagger \dagger'} \frac{ (Y_{\dagger'} - X_{\dagger})^2}{2} \m H''(U_{\dagger' \dagger})
\ge \frac{ \sum_{\dagger} \sum_{\dagger'} n_{\dagger \dagger'}(z, z_0) (Y_{\dagger'} - X_{\dagger})^2}{ (p_0 \vee q_0) \{1 - (p_0 \wedge q_0)\}  }. 
\end{align*}
Write 
$$
(Y_{\dagger'} - X_{\dagger}) = (\xi_{\dagger'} - \bar{\xi}_{\dagger}) + (Y_{\dagger'} - \xi_{\dagger'}) + (X_{\dagger} - \bar{\xi}_{\dagger}). 
$$
The bounds on $|Y_{\dagger'} - \xi_{\dagger'}|$ and $|X_{\dagger} - \bar{\xi}_{\dagger}|$ from  \eqref{eq:CY_z} and \eqref{eq:CX_z} imply that $(\xi_{\dagger'} - \bar{\xi}_{\dagger})$ is the leading term in the above display. Since we can bound $(a+b)^2 \ge a^2/2$ if $|b| = o(|a|)$, we obtain, inside $\m C$,   
\begin{align}\label{eq:bd_quad}
\sum_{\dagger} \sum_{\dagger'} n_{\dagger \dagger'}(z, z_0) (Y_{\dagger'} - X_{\dagger})^2  \ge \frac{1}{2} \sum_{\dagger} \sum_{\dagger'} n_{\dagger \dagger'}(z, z_0)  (\xi_{\dagger'} - \bar{\xi}_{\dagger})^2.
\end{align}
We have $(\xi_{\dagger'} - \bar{\xi}_{\dagger}) = (\xi_{\dagger'} - \omega_{\dagger \dagger'} \xi_{\dagger'} - \omega_{\dagger \ddagger'} \xi_{\ddagger'}) = \omega_{\dagger \ddagger'} (\xi_{\dagger'} - \xi_{\ddagger'})$, since $\omega_{\dagger \ddagger'} = 1 - \omega_{\dagger \dagger'}$. Also, $|\xi_{\dagger'} - \xi_{\ddagger'}| = |p_0 - q_0|$. Hence  
\begin{align}
& \sum_{\dagger} \sum_{\dagger'} n_{\dagger \dagger'}(z, z_0)  (\xi_{\dagger'} - \bar{\xi}_{\dagger})^2 \\
& = \sum_{\dagger} \sum_{\dagger'} n_{\dagger \dagger'}(z, z_0) \omega_{\dagger \ddagger'}^2 (p_0 - q_0)^2 \notag \\
& = (p_0 - q_0)^2 \sum_{\dagger} \sum_{\dagger'}  n_{\dagger \dagger'}(z, z_0)  \frac{ n_{\dagger \ddagger'}^2(z, z_0)}{n_{\dagger}^2(z) } \notag \\
& = (p_0 - q_0)^2  \sum_{\dagger} \frac{ n_{\dagger \up}(z, z_0) n_{\dagger \down}(z, z_0) }{n_{\dagger}(z)}, \label{eq:bd_quad1}
\end{align}
since 
\begin{align*}
\sum_{\dagger'}  n_{\dagger \dagger'}(z, z_0)  n_{\dagger \ddagger'}^2(z, z_0) = n_{\dagger \up}(z, z_0) n^2_{\dagger \down}(z, z_0) + n_{\dagger \down} n_{\dagger \up}^2(z, z_0) = n_{\dagger \up}(z, z_0) n_{\dagger \down}(z, z_0) n_{\dagger}(z).
\end{align*}
Define
\begin{align}\label{eq:mota_n1}
\wt{\mathbf{n}}(z, z_0) = \sum_{\dagger} \frac{ n_{\dagger \up}(z, z_0) n_{\dagger \down}(z, z_0) }{n_{\dagger}(z)} = \frac{ n_{\up \up}(z, z_0) n_{\up \down}(z, z_0) }{ n_{\up}(z) } +  \frac{ n_{\down \up}(z, z_0) n_{\down \down}(z, z_0) }{ n_{\down}(z) }
\end{align}
We then have, from \eqref{eq:bd_quad1}, \eqref{eq:bd_quad}, and \eqref{eq:bd_lin}, that inside $\m C$, 
\begin{align}\label{eq:mota_diff1}
\wt{\ell}(z_0) - \wt{\ell}(z) \ge \frac{ (p_0 - q_0)^2}{ 2 (p_0 \vee q_0) \{1 - (p_0 \wedge q_0)\}} \,  \wt{\mathbf{n}}(z, z_0) - C' (\log n)^{\nu/2} \sqrt{r} \, \sqrt{\mathbf{n}(z, z_0)}.
\end{align}
We now state a Lemma to bound $\wt{\mathbf{n}}(z, z_0)$ and $\mathbf{n}(z, z_0)$ in appropriate directions. 
\begin{lemma}\label{lem:mota_n}
Suppose $K \ge 2$ and $d(z, z_0) = r$. Then, $\wt{\mathbf{n}}(z, z_0)\geq \text{min}\{C rn/K,Cn^2/K^2\}$ and $\mathbf{n}(z, z_0) \le C \{nr/K + r^2\}$ for some constant $C > 0$, where $\wt{\mathbf{n}}(z, z_0)$ and $\mathbf{n}(z, z_0)$ are defined in \eqref{eq:mota_n1} and \eqref{eq:mota_n} respectively. 
\end{lemma}
The proof of Lemma \ref{lem:mota_n} is provided in the Appendix \ref{sec:aux}. Substituting the inequalities in Lemma \ref{lem:mota_n} to \eqref{eq:mota_diff1} delivers the bound \eqref{eq:marglik_diff_main} in Proposition \ref{prop:main}. 

\section{\textbf{Proof of Theorem 4.3}}\label{sec:pfcor} 
 We first introduce a few notations. Since $d_H$ is not defined between two configurations with different values of $k$, we instead work with the Rand-Index (R) in the subsequent developments.  
Define
\begin{align*}
& n_{\alpha \beta}  = |i :z_i = \alpha, z_i^0 = \beta|, \ \alpha=1,\ldots,k, \beta = 1,2;  \quad n_{\alpha} = |i: z_i = \alpha|,  \ \alpha=1,\ldots,k, \\
& B = 2\sum_{\alpha =1}^{k}n_{\alpha1}n_{\alpha2}, \quad  R = \dfrac{n_{\up\up}(z,z_0)+n_{\down\down}(z,z_0)}{{n \choose 2}}.  
\end{align*}
Clearly $0 \leq R \leq 1$ and $R =1$ indicates perfect concordance between the configurations $z$ and $z_0$.    To find a lower bound to $\Pi(K\mid \mA)$,  it is enough to find an upper bound to the Bayes factor 
$\mathcal{L}(\mA \mid k)/ \mathcal{L}(\mA \mid K)$. Observe that 
\begin{eqnarray}\label{BF2-sp}
\dfrac{\mathcal{L}(\mA \mid k)}{\mathcal{L}(\mA \mid K)} 
  \leq  \sum_{z\in Z_{n,k}}\dfrac{\mathcal{L}(A\mid z,k)}{\mathcal{L}(\mA \mid z_0,K)}\dfrac{\Pi(z\mid k)}{\Pi(z_0\mid K)}. 
\end{eqnarray}
Straightforward calculations yield, for the Dirichlet-multinomial prior with Dirichlet concentration parameter $\gamma$, 
\begin{eqnarray}\label{PRI2-sp}
\dfrac{\Pi(z\mid k=3)}{\Pi(z_0\mid K=2)}  \leq c_1 e^{nc_2},  \quad \dfrac{\Pi(z\mid k=2)}{\Pi(z_0\mid K=3)}  \leq c_3 e^{c_4n \log n}. 
\end{eqnarray}
Since the analysis leading up to \eqref{eq:mota_diff1} does not depend on whether or not $z$ and $z_0$ share the same $k$,  we have     
\begin{align}\label{eq:mota_diffk}
\dfrac{\mathcal{L}(A\mid z,k)}{\mathcal{L}(\mA \mid z_0,K)}\leq  \exp \{ C' t_n  \, \sqrt{\mathbf{n}(z, z_0)} - \bar{D}(p_0, q_0) \wt{\mathbf{n}}(z, z_0) \}
\end{align}
with probability $1- e^{-Ct_n^2}$.  Denote by $\mathcal{C}$ the set corresponding to the high-probability event in 
\eqref{eq:mota_diffk}.   In the following, we derive a lower bound for  $\mathbf{n}(z, z_0)$ respectively for the following two cases.   In both the cases, the upper bound for $\mathbf{n}(z, z_0)$ follows trivially.    \\
\underline{1.  Overfitted case ($K = 2$ and the model is fitted with $k=3$):}  Since the true model is contained in the fitted model,  a value of $R$ close to $1$ impedes the concentration of 
$k$ around $K=2$.  We derive  lower bound for  $\wt{\mathbf{n}}(z, z_0)$  in terms of the Rand-Index $R$ and investigate the bounds for different regimes of $R$.  $R \asymp 1$ corresponds to the  case when the separation between the log-marginal likelihoods is relatively weak, but strong enough to offset the model complexity and the prior.  In this case $\wt{\mathbf{n}}(z, z_0)$ and $\mathbf{n}(z, z_0)$ both are of the order $n$; however the number of such configurations is polynomial in $n$, so that the posterior concentrates at $K=2$ with a rate $e^{-Cn}$.  \\
\underline{2. Underfitted  case ($K=3$ and the model is fitted with $k=2$):}   In the underfitted case,  $R$ can never approach $1$ which makes separation between  the log-marginal likelihoods stronger.  In this case both $\mathbf{n}(z, z_0)$ and $\wt{\mathbf{n}}(z, z_0)$ are of the order $n^2$ which is enough to offset the model complexity leading to a posterior concentration rate of $e^{-n^2}$.  

In the following, we analyze the above two cases separately.  \\
\underline{1. Overfitted case:}  Here  $K =2$ and $m = n/2$ and  
\begin{eqnarray*}
n_{\up\up}(z,z_0) &=& \dfrac{\sum_{\alpha =1}^k (n_{\alpha1}^2+n_{\alpha2}^2)}{2} - m, \quad 
n_{\up\down}(z,z_0) = \sum_{\alpha =1}^{k}n_{\alpha 1}n_{\alpha 2}\\
n_{\down\up}(z,z_0) &=& m^2 - \dfrac{\sum_{\alpha =1}^k (n_{\alpha 1}^2+n_{\alpha 2}^2)}{2}, \quad 
n_{\down\down}(z,z_0) = m^2 - \sum_{\alpha =1}^{k}n_{\alpha 1}n_{\alpha 2}.  
\end{eqnarray*}
We express $\mathbf{n}(z, z_0)$ and $\wt{\mathbf{n}}(z, z_0)$  in terms of $R$ as  
\begin{align}\label{eq:boundz0}
\mathbf{n}(z, z_0) &= \dfrac{n_{\up\up}(z,z_0)n_{\down\down}(z,z_0)(1-R) + n_{\up\down}(z,z_0)n_{\down\up}(z,z_0)R}{(m^2-m)m^2/{n \choose 2}}, \\
\wt{\mathbf{n}}(z, z_0) &= \dfrac{n_{\up\up}(z,z_0)n_{\down\down}(z,z_0)(1-R) + n_{\up\down}(z,z_0)n_{\down\up}(z,z_0)R}{n_{\up}(z)n_{\down}(z)/{n \choose 2}} \label{eq:bound2}.  
\end{align}
Lemma \ref{lem:rbound} derives upper and lower bounds for $\mathbf{n}(z, z_0)$ and $\wt{\mathbf{n}}(z, z_0)$ 
depending on 5 possible range of values for $R$.  For cases 1 and 2,  
$t_n  \, \sqrt{\mathbf{n}(z, z_0)} - \bar{D}(p_0, q_0) \wt{\mathbf{n}}(z, z_0)  \leq  
n t_n  - \bar{D}(p_0, q_0) n^2$.   For Cases 3 and 4, the bounds are  $\{n t_n \sqrt{\eta_n}  -  \bar{D}(p_0, q_0) n^2 \eta_n\}$ and $
\{t_n \sqrt{n}  -  \bar{D}(p_0, q_0) n^2 \eta_n\}$ respectively.   

Thus for each of the cases 1-4, the bound for the ratio of the marginal likelihood in \eqref{eq:mota_diffk} is faster than exponential.  
For Case 5,  the bound is  $C\{t_n \sqrt{n}  -  \bar{D}(p_0, q_0) n\}$. Note that this means the ratio of the marginal likelihood in \eqref{eq:mota_diffk} can be at the minimum $e^{-Cn}$ for Case 5.  However, for $z$ satisfying Case 5, one 
 can improve on the bound of the prior ratio  in \eqref{PRI2-sp} as
\begin{eqnarray}\label{PRI2-sp2}
\dfrac{\Pi(z\mid k=3)}{\Pi(z_0\mid K=2)} \leq C \sqrt{n}(n+2)^3.  
\end{eqnarray}
The proof of  \eqref{PRI2-sp2} is appended with the proof of Lemma \ref{lem:rbound}. 
Thus for Case 5, we have
\begin{eqnarray*}
\dfrac{\mathcal{L}(A\mid z,k=3)}{\mathcal{L}(\mA \mid z_0,K=2)}\dfrac{\Pi(z\mid k=3)}{\Pi(z_0\mid K=2)} \leq  e^{-Cn}.  
\end{eqnarray*}

Instead of a global bound on the model complexity,  we separately analyze the complexity of configurations corresponding to Cases 1-4 and 5. From  the proof of Lemma \ref{lem:rbound}, configurations corresponding to Case 5 satisfy the following:  choose a constant $a$ from $m$ observations in cluster one and a constant value $b$ from cluster two, then randomly place $a+b$ nodes into three clusters.  The number such configurations is at most polynomial in $n$, say $n^\kappa$ for some $\kappa > 0$.

For Cases 1-4, choose $t_n = o(n \sqrt{\eta_n})$ with $3^n e^{-C t_n^2} \to 0$.  For Case 5, choose $t_n = o(\sqrt{n})$ with $n^\kappa e^{-C t_n^2} \to 0$.   Then $\bbP(\mathcal{C}^c)\to 0$.   Hence the right hand side of 
\eqref{BF2-sp} can be bounded by $3^n  \exp \{-C n^2 \eta_n\} + n^\kappa \exp\{-Cn\}$ which can be upper bounded by
$\exp\{-Cn\}$. 

\begin{lemma}\label{lem:rbound}

\begin{enumerate}
\item If $1-2R \asymp \beta_n \text{ or } 1-2R \asymp Cm^{-1}$ with $\beta_n\rightarrow0$ and $m\beta_n\rightarrow 0$, $\mathbf{n}(z, z_0)  \leq  Cn^2$ and 
 $\wt{\mathbf{n}}(z, z_0) \geq C n^2$.
\item  If either $1-R$ or $1-2R$ are constants, $\mathbf{n}(z, z_0)  \leq  Cn^2$ and 
 $\wt{\mathbf{n}}(z, z_0) \geq C n^2$.
\item If $1- R \asymp \eta_n$ with $\eta_n \to 0$ and $ m \eta_n\to \infty$,  $\mathbf{n}(z, z_0)  \leq  Cn^2 \eta_n$ and 
 $\wt{\mathbf{n}}(z, z_0) \geq C n^2 \eta_n$.  
 \item  When $1-R = C/m$ and $B/m \to \infty$ and 
$B / (m^2 \eta_n) \to C$, then $\mathbf{n}(z, z_0)  \leq  Cn$,  $\wt{\mathbf{n}}(z, z_0) \geq n^2 \eta_n$.  
\item When $1-R = C/m$ for some constant $C > 0$, and $B = Cm$, then $\mathbf{n}(z, z_0)  \leq  Cn$ and $\wt{\mathbf{n}}(z, z_0) \geq Cn$ 
\end{enumerate}
\end{lemma}
\noindent \underline{2. Underfitted case:}  Assume $K = 3$ and $m = n/3$. Then $%n_{\alpha \beta}  &=& |\{i\}:z_i = \alpha, z_i^0 = \beta|; \text{ for } \alpha=1,...,k, \beta = 1,3,  n_{\alpha} = |\{i\}:z_i = \alpha|;\text{ for } \alpha=1,...,k, \\
B = 2\sum_{\alpha =1}^{k}(n_{\alpha 1}n_{\alpha 2} + n_{\alpha 1}n_{\alpha 3} + n_{\alpha 2}n_{\alpha 3})$. 
%,  R = \dfrac{n_{\up\up}(z,z_0)+n_{\down\down}(z,z_0)}{{n \choose 2}}
%\end{eqnarray*}
Also, note that 
\begin{eqnarray*}
n_{\up\up}(z,z_0) &=& \dfrac{\sum_{\alpha =1}^k (n_{\alpha1}^2+n_{\alpha2}^2 + n_{\alpha3}^2)}{2} - \frac{3m}{2}, \quad 
n_{\up\down}(z,z_0) = \sum_{\alpha =1}^{k}(n_{\alpha 1}n_{\alpha 2} + n_{\alpha 1}n_{\alpha 3} + n_{\alpha 2}n_{\alpha 3})\\
n_{\down\up}(z,z_0) &=& \frac{3m^2}{2} - \dfrac{\sum_{\alpha =1}^k (n_{\alpha1}^2+n_{\alpha2}^2 + +n_{\alpha3}^2)}{2}, \quad 
n_{\down\down}(z,z_0) =3m^2 - n_{\down\up}(z,z_0).  
\end{eqnarray*}

%To find a lower bound to $\Pi(K=3\mid \mA)$,  it is enough to find an upper bound to the Bayes factor 
%$\mathcal{L}(\mA \mid K=2)/ \mathcal{L}(\mA \mid K=3)$. Observe that 
%\begin{eqnarray}\label{BF2-sp}
%\dfrac{\mathcal{L}(\mA \mid K=2)}{\mathcal{L}(\mA \mid K=3)} 
%  \leq  \sum_{z\in Z_{n,2}}\dfrac{\mathcal{L}(A\mid z,K=2)}{\mathcal{L}(\mA \mid z_0,K=3)}\dfrac{\Pi(z\mid K=2)}{\Pi(z_0\mid K=3)}. 
%\end{eqnarray}
%From Dirichlet-multinomial calculations
%\begin{eqnarray}\label{PRI2-sp}
%\dfrac{\Pi(z\mid K=2)}{\Pi(z_0\mid K=3)}  \leq Ce^{n\log(3)}n^2 . 
%\end{eqnarray}
%From the previous calculation we have 
%\begin{align}\label{eq:mota_diff1}
%\dfrac{\mathcal{L}(A\mid z,K=2)}{\mathcal{L}(\mA \mid z_0,K=3)}\leq  \exp \{ C' t_n  \, \sqrt{\mathbf{n}(z, z_0)} - \bar{D}(p_0, q_0) \wt{\mathbf{n}}(z, z_0) \}.
%\end{align}
%with probability $1- e^{-Ct_n^2}$. 
 It is straightforward to show $n_{\up\up}(z,z_0) \geq Cn^2$. Also, 
\small
\begin{align}\label{eq:B1}
B = \dfrac{n_1^2 + n_2^2}{2} + \bigg(3 - \frac{9}{2}R\bigg)m^2 + \bigg(\frac{3}{2}R - \frac{3}{2}\bigg)m \geq \frac{9}{4}m^2 + \bigg(3 - \frac{9}{2}R\bigg)m^2 + \bigg(\frac{3}{2}R - \frac{3}{2}\bigg)m = Cn^2. 
\end{align}
\normalsize
The first inequality in \eqref{eq:B1} follows because $n_1^2 + n_2^2 \geq 2(n/2)^2$ and $n = 3m$. The last equality in 
\eqref{eq:B1} follows since $0 \leq R \leq 1$. Hence $n_{\up \up}(z, z_0) n_{\up \down}(z, z_0)/ n_{\up}(z) \geq Cn^2$ and thus $\wt{\mathbf{n}}(z, z_0)\geq Cn^2$.  Choosing $t_n = o(n)$ concludes the proof. 
\newpage 

%\appendix
 \section{Proof of a few auxiliary Lemmata}\label{sec:aux}
\subsection{\textbf{Proof of Lemma \ref{lem:mota_n}}}
We introduce some additional notations to analyze the terms $n_{\up \down}(z, z_0)$ and $n_{\down \up}(z, z_0)$.  
Set $m = n/K$ and define $a_k = \abs{\{ i:  z_i \neq k, z_i^0 = k\}}$, $b_k = \abs{\{ i:  z_i = k, z_i^0 \neq k\}}$, $n_k = \abs{\{ i:  z_i = k\}}$ and $n^0_k = \abs{\{ i:  z_i^0 = k\}} = m$ for all $k =1, \ldots, K$.  Clearly,  $\sum_{k=1}^{K}a_k = \sum_{k=1}^{K}b_k = r$ and 
$n^0_k - a_k = n_k - b_k$.  Fix $z$ with $d(z, z_0) = r$.  Then $0 \le r \le n - m$.   Defining $n_{\up \down}^{(k)}(z,z_0) = |\{ i:  z_i = z_j = k, z_i^0 \neq z_j^0\}|$ and $n_{\down \up}^{(k)}(z,z_0) = |\{ i:  z_i^0 = z_j^0 = k, z_i \neq z_j\}|$, we write 
\begin{eqnarray*}
n_{\up \down}(z,z_0) = \sum_{k=1}^{K} n_{\up \down}^{(k)}(z,z_0),  \quad n_{\down \up}(z,z_0) = \sum_{k=1}^{K} n_{\down \up}^{(k)}(z,z_0). 
\end{eqnarray*}
Observe that,  
\begin{eqnarray*}
n_{\up \down}^{(k)}(z,z_0) &\geq & \abs{\{ i:  z_i = k, z_i^0 = k\}}\abs{\{ i:  z_i = k, z_i^0 \neq k\}} =  (n_k - b_k)b_k\\
n_{\down \up}^{(k)}(z,z_0) &\geq & \abs{\{ i:  z_i = k, z_i^0 = k\}}\abs{\{ i:  z_i \neq k, z_i^0 = k\}} = (n^0_k - a_k)a_k.
\end{eqnarray*}

%We next state the following results. 
%\subsection{Proof of Lemma  \ref{lem:mota_n}}
%\begin{lemma}\label{lem:wtnub}
%For  $K \geq 2$,  $\wt{\mathbf{n}}(z, z_0)\geq Crn/K^3$ for some constant $C > 0$. 
%\end{lemma}
%\begin{lemma}\label{lem:nub}
% For any  $r \in [0, n-m]$,   $\mathbf{n}(z, z_0) \leq  C \{nr/K + r^2\}$ for some constant $C > 0$. 
%\end{lemma}
\noindent {\bf Proof of lower bound on $\wt{\mathbf{n}}(z, z_0)$:}\\
Note that 
\begin{eqnarray*}
\sum_{\dagger} \frac{ \prod_{\dagger'} n_{\dagger \dagger'}(z, z_0)}{n_{\dagger}(z)} &=& \frac{n_{\up \up}(z, z_0)n_{\up \down}(z, z_0)}{n_{\up \up}(z, z_0)+n_{\up \down}(z, z_0)} + \frac{n_{\down \down}(z, z_0)n_{\down \up}(z, z_0)}{n_{\down \down}(z, z_0)+n_{\down \up}(z, z_0)}\\
&=&  \frac{n_{\up \down}(z, z_0)}{1 + \frac{n_{\up \down}(z, z_0)}{n_{\up \up}(z, z_0)}} + \frac{n_{\down \up}(z, z_0)}{1 + \frac{n_{\down \up}(z, z_0)}{n_{\down \down}(z, z_0)}} := T_1 + T_2.
\end{eqnarray*}
The proof is based on the following three inequalities:
\begin{align}
& n_{\up \down}(z,z_0) + n_{\down \up}(z,z_0)   \geq C rm,\label{em1} \\
& n_{\up \up}(z,z_0)  \geq C m^2,  \label{lem2}\\
& n_{\down \down}(z,z_0) \geq 2n_{\down \up}(z,z_0)-n_{\up \down}(z,z_0)\label{lem3}. 
\end{align}  
Hence $C >0$ denotes a generic constant.  By \eqref{em1}, either $n_{\up \down}(z,z_0) \geq Crm/2$ or  $n_{\down \up}(z,z_0) \geq Crm/2$.   If $n_{\up \down}(z,z_0) \geq Crm/2$, then  $T_1 \geq Crm$ since $n_{\up \up}(z,z_0) \geq Cm^2$ by \eqref{lem2}.  If $n_{\down \up}(z,z_0) \geq Crm/2$ and $n_{\up \down}(z,z_0) < Crm/2$,  $n_{\down \up}(z,z_0)/ n_{\up \down}(z,z_0) > 1$. Then by \eqref{lem3},  $n_{\down \down}(z,z_0) \geq 2n_{\down \up}(z,z_0)-n_{\up \down}(z,z_0)$ and hence  
\begin{eqnarray*}
\frac{n_{\down \up}(z, z_0)}{1 + \frac{n_{\down \up}(z, z_0)}{n_{\down \down}(z, z_0)}} &\geq & \frac{n_{\down \up}(z, z_0)}{1 + \frac{n_{\down \up}(z, z_0)}{2n_{\down \up}(z,z_0)-n_{\up \down}(z,z_0)}} > \frac{n_{\down \up}(z,z_0)}{2}.
\end{eqnarray*}
Thus $T_2 \geq Crm$. The lower bound on $\wt{\mathbf{n}}(z, z_0)$ then follows immediately.

We next turn our attention to proving \eqref{em1} - \eqref{lem3}. We first show \eqref{lem2}. Defining  $n_{\up \up}^{(k)}(z,z_0) = |\{ (i,j):  z_i = z_j = k, z_i^0 = z_j^0\}|$, observe that 
\begin{eqnarray}
n_{\up \up}(z,z_0) &=& \sum_{k=1}^{K}n_{\up \up}^{(k)}(z,z_0) \nonumber\\
&\geq & \sum_{k=1}^{K} {{n_k-b_k} \choose 2} =  \sum_{k=1}^{K} (n^0_k - a_k)\frac{n_k - b_k-1}{2} = \frac{n}{K}\sum_{k=1}^{K} \frac{n_k - b_k-1}{2} - \sum_{k=1}^{K} a_k \frac{n^0_k - a_k-1}{2}\nonumber\\
&=& \frac{n^2}{2K} - \frac{nr}{2K} - \frac{n}{2} - \frac{nr}{2K} + \frac{r}{2} + \sum_{k=1}^K \frac{a^2_k}{2} \nonumber\\
&\geq & \frac{n^2}{2K} - \frac{nr}{K} - \frac{n}{2} + \frac{r}{2} + \frac{r^2}{2K}= \frac{(n-r)^2}{2K}+\frac{r-n}{2} =Cm^2 \label{eq:last}
\end{eqnarray}
for some constant $C > 0$. The inequality in \eqref{eq:last} follows since $\sum a_k^2$ is minimized at $a_k = r/K$.

Next, we show \eqref{lem3}.  Observe that 
\begin{eqnarray*}
n_{\down \down}(z,z_0) &=& \abs{\{(i,j):z_i \neq z_j, z_i^0 \neq z_j^0\}} = \abs{\{ (i,j):  z_i^0 \neq z_j^0\}} - n_{\up \down}(z,z_0)\\
& = &{n \choose 2} - K{m \choose 2} - n_{\up \down}(z,z_0) = \frac{(K-1)K}{2} m^2 - n_{\up \down}(z,z_0).
\end{eqnarray*}
The conclusion will then follow if we can show  $2n_{\down \up}(z,z_0) \leq \frac{(K-1)K}{2} m^2$. We denote $a_{kt} = \abs{\{(i,j):z_i = t, z_i^0 = k\}}$, and we fix $a_{kk} = 0$ for all $k = 1,...K.$ Then $\sum_{t=1}^{K} a_{kt} = a_k$ and there are $K-1$ non-zero terms.
\begin{eqnarray}
n_{\down \up}(z,z_0) &=& \sum_{k=1}^{K} \{n_{\down \up}^{(k)}(z,z_0)\} \nonumber\\
 &= & \sum_{k=1}^{K} \bigg\{(n^0_k - a_k)a_k + {a_k \choose 2} - \sum_{t=1}^{K} {a_{kt} \choose 2}\bigg\} \nonumber\\
 &=& mr + \sum_{k=1}^{K}(-\frac{a_k^2}{2} - \sum_{t=1}^{K}\frac{a_{kt}^2}{2} )\nonumber\\
 &\leq & mr + \sum_{k=1}^{K}\bigg\{ -\frac{a_k^2}{2} - \frac{a_k^2}{2(K-1)} \bigg\} = mr - \frac{K}{2(K-1)} \sum_{k=1}^K a_k^2  \label{eq:min}\\
&\leq & mr - \frac{r^2}{2(K-1)}. \label{eq:nupdownub}
\end{eqnarray}
\eqref{eq:min} follows since $\sum_{t=1}^{K}a_{kt}^2/2$ is minimized at $a_{kt} = a_k/(K-1)$ for $t = 1,...,K$ and $t \neq k$. \eqref{eq:nupdownub} follows since $\sum_{k=1}^K a_k^2$ is minimized at $a_k = r/K$ for $k = 1,...,K$. Observe that $r \mapsto mr - r^2/2(K-1)$  is maximized at $r = (K-1)m$. Then the upper bound in \eqref{eq:nupdownub} becomes 
\begin{eqnarray*}
m(K-1)m - \frac{(K-1)m^2}{2} = \frac{(K-1)m^2}{2}. 
\end{eqnarray*}
It is easy to see that $2n_{\down \up}(z,z_0) \leq (K-1)m^2 \leq \frac{(K-1)K}{2} m^2$ when $K \geq 2$.   

We finally prove \eqref{em1}. We  split the proof into two cases.\\
\underline{{\bf Case 1:}} When $r/m \rightarrow 0$ as $m \rightarrow \infty$, we  want to show that $n_{\up \down}(z,z_0) + n_{\down \up}(z,z_0)   \geq C rm $. Observe that 
\begin{eqnarray*}
n_{\up \down}(z,z_0) + n_{\down \up}(z,z_0) &=& \sum_{k=1}^{K} \big\{n_{\up \down}^{(k)}(z,z_0) + n_{\down \up}^{(k)}(z,z_0)\big\} \geq  \sum_{k=1}^{K} (n_k - b_k)b_k  + \sum_{k=1}^{K} (n^0_k - a_k)a_k
\\
&=& \sum_{k=1}^{K} (n^0_k - a_k)(a_k + b_k) = m\sum_{k=1}^{K}(a_k + b_k) - \sum_{k=1}^{K}(a_k^2 + a_k b_k),
\end{eqnarray*}
which implies 
\small
\begin{eqnarray}\label{eq:nupdown}
n_{\up \down}(z,z_0) + n_{\down \up}(z,z_0)&\geq &2mr - \sum_{k=1}^{K}a_k (a_k+b_k) \geq 2mr - \bigg\{\sum_{k=1}^{K}a_k\bigg\}\bigg\{\sum_{k=1}^{K}(a_k+b_k)\bigg\} \nonumber\\
&=& 2mr - 2r^2 = 2rm(1-r/m) \geq Crm. 
\end{eqnarray}
\normalsize
\eqref{eq:nupdown} follows from the fact that $\sum_{k=1}^{K}a_k (a_k+b_k) < \{\sum_{k=1}^{K}a_k\}\{\sum_{k=1}^{K}(a_k+b_k)\}$.

%We next return to the proof of Lemma \ref{lem:mota_n}.   Note that 
%\begin{eqnarray*}
%\sum_{\dagger} \frac{ \prod_{\dagger'} n_{\dagger \dagger'}(z, z_0)}{n_{\dagger}(z)} &=& \frac{n_{\up \up}(z, z_0)n_{\up \down}(z, z_0)}{n_{\up \up}(z, z_0)+n_{\up \down}(z, z_0)} + \frac{n_{\down \down}(z, z_0)n_{\down \up}(z, z_0)}{n_{\down \down}(z, z_0)+n_{\down \up}(z, z_0)}\\
%&=&  \frac{n_{\up \down}(z, z_0)}{1 + \frac{n_{\up \down}(z, z_0)}{n_{\up \up}(z, z_0)}} + \frac{n_{\down \up}(z, z_0)}{1 + \frac{n_{\down \up}(z, z_0)}{n_{\down \down}(z, z_0)}} := T_1 + T_2.
%\end{eqnarray*}
%
%By \eqref{em1}, we may have either $n_{\up \down}(z,z_0) \geq Crm/2$ or 
%$n_{\down \up}(z,z_0) \geq Crm/2$.  
%
%If $n_{\up \down}(z,z_0) \geq Crm/2$, then  $T_1 \geq Crm$ since $n_{\up \up}(z,z_0) \geq Cm^2$ by \eqref{lem2}.
%
%
%If $n_{\down \up}(z,z_0) \geq Crm/2$ and $n_{\up \down}(z,z_0) < Crm/2$ which means $\frac{n_{\down \up}(z,z_0)}{n_{\up \down}(z,z_0)} > 1$ , then by \eqref{lem3},  $n_{\down \down}(z,z_0) \geq 2n_{\down \up}(z,z_0)-n_{\up \down}(z,z_0)$ and hence  
%\begin{eqnarray*}
%\frac{n_{\down \up}(z, z_0)}{1 + \frac{n_{\down \up}(z, z_0)}{n_{\down \down}(z, z_0)}} &\geq & \frac{n_{\down \up}(z, z_0)}{1 + \frac{n_{\down \up}(z, z_0)}{2n_{\down \up}(z,z_0)-n_{\up \down}(z,z_0)}} > \frac{n_{\down \up}(z,z_0)}{2}
%\end{eqnarray*}
%Thus $T_2 \geq Crm$. As a summary of $T_1$ and $T_2$, we can see $\wt{\mathbf{n}}(z, z_0)\geq C rn/K$ for some constant $C > 0$.

\noindent \underline{{\bf Case 2:}} When $r = am$, where $a$ is a constant that satisfies $0 < a \leq K-1$, 
\begin{eqnarray}\label{em1case2}
n_{\up \down}(z,z_0) + n_{\down \up}(z,z_0)   \geq C rm 
\end{eqnarray}
for some $C > 0$. 
Observe that 
\begin{align}
n_{\up \down}(z,z_0) + n_{\down \up}(z,z_0) &= (n_{\up}(z) - n_{\up \up}(z,z_0)) + (n_{\up}(z_0) - n_{\up \up}(z,z_0)) \notag\\
&= \sum_{k=1}^{K}{n_k \choose 2} + \sum_{k=1}^{K}{m \choose 2} - 2\sum_{\alpha =1}^{K}\sum_{\beta =1}^{K}{n_{\alpha \beta}  \choose 2} \notag\\
&= \sum_{k=1}^{K}(\dfrac{n_k^2 + {n^0_k}^2}{2}) - \sum_{\alpha =1}^{K}\sum_{\beta =1}^{K}n_{\alpha \beta} ^2 \notag\\
&= \sum_{\alpha =1}^{K}\dfrac{[(\sum_{\beta =1}^{K}n_{\alpha \beta} )^2 + (\sum_{\beta =1}^{K}n_{\beta \alpha} )^2]}{2} - \sum_{\alpha =1}^{K}\sum_{\beta =1}^{K}n_{\alpha \beta} ^2 \notag\\
&= \sum_{k=1}^{K}\sum_{a>b}n_{k\alpha}n_{kb} + \sum_{k=1}^{K}\sum_{\alpha>\beta}n_{\alpha k}n_{\beta k}. \label{eq:case2}
\end{align}
In the preceding display, $n_{\up \down}(z,z_0) + n_{\down \up}(z,z_0)$ are the sum of squares of all column sums and row sums minus the sum of squares of each term in matrix $N = \{n_{\alpha \beta}: \alpha = 1,\ldots, K, \beta = 1,\ldots,K\}$. This quantity is essentially the sum of interaction terms within each column and row. The matrix $N$ satisfies the following requirements:
\begin{itemize}
\item
For diagonal terms of $N$, we have $\sum_{k=1}^{K} n_{kk} \geq m$. 
\item
For all $k$ in ${1,\ldots,K}$, $\sum_{\alpha =1}^{K}n_{\alpha k} = m$.  
\end{itemize}
For each column, if there is no term in that column which satisfies $n_{k\alpha} \geq Cm$, from the second requirement above, we can see that there must be at least one term $n_{k\alpha}$ which satisfied $n_{k\alpha} \geq Cm/K$. Then it is straightforward to see for each column $k$, $\sum_{\alpha>\beta}n_{\alpha k}n_{\beta k} \geq \frac{Cm}{K}(m - \frac{Cm}{K}) \geq Cm^2/K$. When $r = am$, it is easy to show 
$n_{\up \down}(z,z_0) + n_{\down \up}(z,z_0) \geq \frac{Cm^2}{K}K = Crm$.  If there is at least one column or row in which there are more than one term that is $Cm$ (say $n_{k1}$ and $n_{k2}$ are $Cm$),  then  from  \eqref{eq:case2} and $r = am$, it follows that $n_{\up \down}(z,z_0) + n_{\down \up}(z,z_0) \geq Cm^2 = Crm$.  If there is only one term that is $Cm$ in all columns and rows and all other terms are $o(m)$, one can switch labels to make $r$ satisfy $r/m \to 0$ by putting all the $Cm$ terms into diagonal terms of the matrix $N$.  This phenomenon is exemplified in Appendix \ref{app:example} for $K =4$.

%
%
%We next return to the proof of Lemma \ref{lem:mota_n}.   Note that 
%\begin{eqnarray*}
%\sum_{\dagger} \frac{ \prod_{\dagger'} n_{\dagger \dagger'}(z, z_0)}{n_{\dagger}(z)} &=& \frac{n_{\up \up}(z, z_0)n_{\up \down}(z, z_0)}{n_{\up \up}(z, z_0)+n_{\up \down}(z, z_0)} + \frac{n_{\down \down}(z, z_0)n_{\down \up}(z, z_0)}{n_{\down \down}(z, z_0)+n_{\down \up}(z, z_0)}\\
%&=&  \frac{1}{\frac{1}{n_{\up \down}(z, z_0)} + \frac{1}{n_{\up \up}(z, z_0)}} + \frac{n_{\down \up}(z, z_0)}{1 + \frac{n_{\down \up}(z, z_0)}{n_{\down \down}(z, z_0)}} := T_1 + T_2.
%\end{eqnarray*}
%
%By \eqref{em1case2}, we may have either $n_{\up \down}(z,z_0) \geq Crm/2$ or 
%$n_{\down \up}(z,z_0) \geq Crm/2$.  
%
%
%If $n_{\up \down}(z,z_0) \geq Crm/2$, then  $T_1 \geq \text{min}\{Cm^2,Crm\}$ since $n_{\up \up}(z,z_0) \geq Cm^2$ by \eqref{lem2}.
%
%
%If $n_{\down \up}(z,z_0) \geq Crm/2$ and $n_{\up \down}(z,z_0) < Crm/2$ which means $\frac{n_{\down \up}(z,z_0)}{n_{\up \down}(z,z_0)} > 1$ , then by \eqref{lem3},  $n_{\down \down}(z,z_0) \geq 2n_{\down \up}(z,z_0)-n_{\up \down}(z,z_0)$ and hence  
%\begin{eqnarray*}
%\frac{n_{\down \up}(z, z_0)}{1 + \frac{n_{\down \up}(z, z_0)}{n_{\down \down}(z, z_0)}} &\geq & \frac{n_{\down \up}(z, z_0)}{1 + \frac{n_{\down \up}(z, z_0)}{2n_{\down \up}(z,z_0)-n_{\up \down}(z,z_0)}} > \frac{n_{\down \up}(z,z_0)}{2}
%\end{eqnarray*}
%Thus $T_2 \geq Crm$. As a summary of $T_1$ and $T_2$, we can see $\wt{\mathbf{n}}(z, z_0)\geq \text{min}\{C rn/K,Cn^2/K^2\}$ for some constant $C > 0$. \\
\noindent {\bf Proof of upper bound on $\mathbf{n}(z, z_0)$:}
From \eqref{eq:nupdownub}, $n_{\down \up}(z,z_0) \leq C rm$.  In the following, we show that $n_{\up \down}(z,z_0) \leq C \{rm +r^2\}$.  We proceed similar to \eqref{eq:nupdownub}. Observe that 
\begin{eqnarray}
n_{\up \down}(z,z_0) &=& \sum_{k=1}^{K} \bigg\{(n^0_k - a_k)b_k + {b_k \choose 2} - \sum_{t=1}^{K} {b_{kt} \choose 2}\bigg\} \nonumber\\
 &=& mr  + \sum_{k=1}^K \bigg\{-a_k b_k + b_k^2/2  - \sum_{t=1}^{K} b_{kt}^2/2 \bigg\}\nonumber \\
 &\leq & mr + C r^2
 \end{eqnarray}
 for some constant $C > 0$.  Since $n_{\up \up}(z, z_0) \leq n_{\up}(z_0)$ and  $n_{\down \down}(z, z_0) \leq n_{\down}(z_0)$,  the upper bound for $\mathbf{n}(z, z_0)$ in Lemma  \ref{lem:mota_n} follows.

 \subsection{\textbf{Example in the proof of Lemma \ref{lem:mota_n}}}\label{app:example}
 Let $N = (n_{\alpha \beta})_{1\leq \alpha, \beta \leq 4}$
%\begin{eqnarray*}
%N= \begin{bmatrix}
%    n_{11} & n_{12} & n_{13} & n_{14} \\
%    n_{21} & n_{22} & n_{23} & n_{24} \\
%    n_{31} & n_{32} & n_{33} & n_{34} \\
%    n_{41} & n_{42} & n_{43} & n_{44} 
% \end{bmatrix} 
% \end{eqnarray*} 
and  $n_{11} = Cm$ without loss of generality. A particular instance of occurrence of only  $Cm$ term in each of the columns and rows is the following:
 \begin{eqnarray*}
\begin{bmatrix}
    Cm & n_{12} & n_{13} & n_{14} \\
    n_{21} & n_{22} & Cm & n_{24} \\
    n_{31} & n_{32} & n_{33} & Cm \\
    n_{41} & Cm & n_{43} & n_{44} 
 \end{bmatrix} 
 \end{eqnarray*}
in which $n_{11}, n_{42}, n_{23}$ $\&$ $n_{34}$ are $Cm$ and all other terms are $O(m)$.  
Then if we switch the labels as $4\rightarrow 2$, $2\rightarrow 3$ and $3\rightarrow 4$ for $z$, the matrix $N$ becomes 
\begin{eqnarray*}
\begin{bmatrix}
    Cm & n_{12} & n_{13} & n_{14} \\
    n_{21} &Cm & n_{23} & n_{24} \\
    n_{31} & n_{32} & Cm & n_{34} \\
    n_{41} & n_{42} & n_{43} & Cm
 \end{bmatrix}.
\end{eqnarray*}

Then we have $n_{\up \down}(z,z_0) + n_{\down \up}(z,z_0) \geq \sum_{k=1}^{K}n_{kk}(n_k - n_{kk}) \geq Cm\sum_{k=1}^{K}(n_k - n_{kk}) = Crm$.

 \subsection{\textbf{Proof of Lemma \ref{lem:rbound}}}
 Expressing the denominator for (\ref{eq:bound2}) in terms of $B$, $R$ and $m$: 
\small \begin{eqnarray}\label{eq:denom}
\dfrac{(2R-1)(3-2R)m^4 - (6R-4R^2-1)m^3 + (4-4R)Bm^2 + B(2R-1)m - B^2 + o(m^3)}{2m^2-m}. 
\end{eqnarray}
\normalsize
\eqref{eq:denom} shows that the denominator is smaller than $Cm^2$. Since we are interested in finding a lower bound to \eqref{eq:bound2}, we henceforth assume the denominator to be $Cm^2$.  The numerator for \eqref{eq:bound2} is expressed as:
\begin{eqnarray*}
(1-R)Bm^2 + (R^2-R)m^3 + (2R-1)(1-R)m^4 - \frac{B^2}{4}. 
\end{eqnarray*}
The order of the numerator is decided by the order of $B$, $1-R$ and $1-2R$. It is straightforward to show $B \leq Cm^2$. Observe that
\begin{eqnarray*}
R &=& \dfrac{n_{\up\up}(z,z_0)+n_{\down\down}(z,z_0)}{{n \choose 2}} = \dfrac{m^2-m + \sum_{\alpha =1}^{k}(n_{\alpha 1}-n_{\alpha 2})^2/2}{2m^2-m}.
\end{eqnarray*}
The minimum value for $R$ is achieved when $n_{\alpha 1} = n_{\alpha 2}$ for all $\alpha$. Then $R_{\text{min}} \asymp 0.5 - 1/4m$.   The maximum value of $R$ is achieved when $\sum_{\alpha =1}^{k}(n_{\alpha 1}-n_{\alpha 2})^2$ is the largest. The constraint here is at least one of $n_{\alpha 1}$ and $n_{\alpha 2}$ will be non-zero for all $\alpha$. Also $\sum_{\alpha =1}^{k}n_{\alpha 1} = \sum_{\alpha =1}^{k}n_{\alpha 2} = m$. Under these constraints, the maximum value will be achieved at $n_{11} = m$, $n_{21} = ...= n_{k1} = 0$, $n_{12} = 0$, $n_{22} = m-(k-2)$ and there are $k-2$ 1's in $n_{\alpha 2}$ for $\alpha >2$. Then we have 
\begin{eqnarray*}
R_{\text{max}} &=& \dfrac{m^2-m + \{(m-k+2)^2+m^2+(k-2)\}/2}{2m^2-m} \asymp 1 - \frac{k-2}{2m} + \frac{k^2-3k}{4m^2}. 
\end{eqnarray*}
Hence $1-R \geq Ck/m$.  Define a sequence $\eta_n \rightarrow0$ and $m\eta_n\rightarrow\infty$ as $m \rightarrow\infty$.  Define another sequence $\beta_n$, which satisfies $\beta_n\rightarrow0$ and $m\beta_n\rightarrow 0$ as $m \rightarrow\infty$.  We split into five different cases. 
%We  want to find a lower bound for (\ref{eq:bound2}). Thus when combining with (\ref{PRI2}) and (\ref{likek2}), the upper bound for (\ref{BF2}) will converge to 0 as $n$ goes to infinity. For one part of case 4, we need to find a specific bound for (\ref{PRI2}) so as to make the upper bound for (\ref{BF2}) converge to 0.

\noindent \underline{{\bf Case 1:}} If $R$ is close to 0.5 and $1-2R \asymp \beta_n \text{ or } 1-2R \asymp Cm^{-1}$, then we show the lower bound of (\ref{eq:bound2}) is $Cm^2/k$. We provide the justification below. 

Note that  $2B = \sum_{\alpha =1}^k n_{\alpha}^2 - (4R-2)m^2 + (2-2R)m \geq Cm^2/k$.  
Then observe that the first term of (\ref{eq:bound2}) can be lower-bounded as
\begin{eqnarray*}
\dfrac{n_{\up\up}(z,z_0)n_{\up\down}(z,z_0)}{n_{\up}(z)} = \dfrac{\bigg\{\dfrac{\sum_{\alpha =1}^k (n_{\alpha 1}^2+n_{\alpha 2}^2)}{2} - m\bigg\}B/2}{\dfrac{\sum_{\alpha =1}^k (n_{\alpha 1}^2+n_{\alpha 2}^2)}{2} - m+B/2} \geq C\frac{m^2}{k}.
\end{eqnarray*} 

\noindent \underline{{\bf Case 2:}} If $R$ is between 0.5 and 1 and both $1-R$ and $1-2R$ are constants, we provide the justification below. 

If $B/m^2 \rightarrow 0$ as $m \rightarrow\infty$, the numerator for (\ref{eq:bound2}) is greater than $Cm^4$. Thus we have the lower bound for (\ref{eq:bound2}) as $Cm^2$. If $B/m^2 \rightarrow C$ as $m \rightarrow\infty$, we have the lower bound of (\ref{eq:bound2}) to be $Cm^2/k$ from the same justification as in Case 1.

\noindent \underline{{\bf Case 3:}} If $R$ is close to 1 and $1-R \asymp \eta_n$, we provide the justification below. 

If $\frac{B}{m^2\sqrt{\eta_n}} \rightarrow 0$ as $m \rightarrow\infty$, the numerator for (\ref{eq:bound2}) is greater than $C\eta_n m^4$. Thus we have the lower bound for(\ref{eq:bound2}) as $C\eta_n m^2$.  If $\frac{B}{m^2\sqrt{\eta_n}} \rightarrow \infty$ as $m \rightarrow\infty$, we can have the lower bound of (\ref{eq:bound2}) to be $C\frac{m^2}{k}$ or $Cm^2\sqrt{\eta_n}$ whichever is smaller, from the same justification in Case 1.

\noindent \underline{{\bf Case 4:}} If $R$ is close to 1 and $1-R \asymp Cm^{-1}$, then we show the lower bound of numerator is $Cm^{-1}$.  We provide the justification below. 

If $B/m \rightarrow \infty$ and $B/(m^2\eta_n) \rightarrow C$ as $m \rightarrow\infty$,  we can have the lower bound of (\ref{eq:bound2}) to be $Cm^2/k$ or $Cm^2\eta_n$ whichever is smaller from the same justification in case 1.  If $B/m\rightarrow C$ as $m \rightarrow\infty$, we have the lower bound of (\ref{eq:bound2}) as:
\begin{eqnarray*}
\dfrac{(2R-1)(1-R)m^4}{(2R-1)(3-2R)m^4/{n \choose 2}} &\asymp & (1-\dfrac{1}{3-2R})m^2 \geq (1-\dfrac{1}{1+\frac{k-2}{m}})m^2 \asymp  (k-2)m.  
\end{eqnarray*} 

\noindent \underline{{\bf Case 5:}}  If $1-R \asymp Cm^{-1}$ when the order of $B/m \rightarrow C$ as $m \rightarrow\infty$, the lower bound for (\ref{eq:bound2})  is $km$. However, the bound for the prior ratio in \eqref{BF2-sp} is different. If one of $n_i$ is $n-k+1$, then $B/m \rightarrow \infty$ as $m \rightarrow\infty$. If we take a look at the definition of $B = 2\sum_{\alpha =1}^{k}n_{\alpha 1}n_{\alpha 2}$, $n_{\alpha 1}n_{\alpha 2}/m \rightarrow C$ or $n_{\alpha 1}n_{\alpha 2}/m\rightarrow 0$ for all $\alpha = 1,\ldots,k$.  Under the constraint that both $n_{\alpha 1}$ and $n_{\alpha 2}$ are less than $m$, in order to maximize  $n_{\alpha} = n_{\alpha 1}+n_{\alpha 2}$,  one out of  $n_{\alpha 1}$ and $n_{\alpha 2}$ has to be $c_1m - c_2$ and the other one has to be a constant.  
Then in order to find an upper bound for the prior ratio in the right-most expression of \eqref{BF2-sp}, there are two $n_i$'s, which are of the form of $n_i = m - c_i$, where $c_i$ is at most of the order of $k$. Then  
\begin{eqnarray}
\dfrac{\Pi(z\mid K=k)}{\Pi(z_0\mid K=2)} &=& \dfrac{\dfrac{(k-1)!\prod_{i=1}^{k}n_i!}{(n+k-1)!}}{\dfrac{m!m!}{(n+1)!}} \asymp \dfrac{(k-1)!(m-c_1)!(m-c_2)!(n+1)!}{(n+k-1)!m!m!} \nonumber \\
&\asymp & C(k-1)^{k-1/2}2^k\sqrt{n}e^{-c_1k}(n+k-1)^{c_2-c_3k} \label{eq:prior_rat1}.  
\end{eqnarray}
%When combining this with (\ref{likek2}), we have the upper bound for (\ref{BF2}) is: 
%\begin{eqnarray}\label{rate2}
%C(k-1)^{k-1/2}2^k\sqrt{n}e^{-c_1k}(n+k-1)^{c_2-c_3k} k^2 n^{k-1}\log(n)e^{-Cn}
%\end{eqnarray} 
%The overall rate will be some weighted average of (\ref{rate1}) and (\ref{rate2}), which will all goes to 0.
%
%Note: for most of the cases $k\ll n$, however $k$ can be up to $\log(n)$. When this happens, $(1-R)m \rightarrow \infty$ as $m \rightarrow \infty$. Then case 4 is not valid anymore. For the first three cases, when $m\eta_n \rightarrow \infty$ as $m \rightarrow \infty$, the the overall upper bound for $\chi$ becomes $e^{\frac{-Cn^2}{k}}$. However, $k$ also fits into the definition of $\eta_n$ in this case, and the results before are still valid.    

%\newpage
\bibliographystyle{plain}
\bibliography{myrefs}
\end{document}